\documentclass{article}
\pdfoutput=1
\usepackage[margin=2cm]{geometry}
\usepackage{graphicx,bm,amsmath,amssymb,latexsym,psfrag,epsfig,euscript, multirow,color,verbatim,cancel}
\usepackage{mathrsfs,tabularx}
\usepackage{float}
\usepackage{subfigure}
\usepackage{mciteplus}
\usepackage{color}
\usepackage[normalem]{ulem}
\definecolor{darkgreen}{rgb}{0,0.5,0}

\allowdisplaybreaks
\begin{document}
\newcolumntype{C}{>{\centering\arraybackslash}X}
\newcommand{\PL}{\mathcal{P}_{\mathrm L}}
\newcommand{\PR}{\mathcal{P}_{\mathrm R}}
\newcommand{\gs}{g_{\mathrm s}}
\newcommand{\beq}{\begin{equation}}
\newcommand{\eeq}{\end{equation}}
\newcommand{\bean}{\begin{eqnarray*}}
\newcommand{\eean}{\end{eqnarray*}}
\newcommand{\be}{\begin{eqnarray}}
\newcommand{\ee}{\end{eqnarray}}
\newcommand{\MPl}{M_{\mathrm{Pl}}}
\newcommand{\Td}{T_{\mathrm{d}}}
\newcommand{\ptl}{\partial}
\newcommand{\benum}{\begin{enumerate}}
\newcommand{\eenum}{\end{enumerate}}
\newcommand{\bi}{\begin{itemize}}
\newcommand{\ei}{\end{itemize}}
\newcommand{\tb}{\tilde{B}}
\newcommand{\tw}{\tilde{W}}
\newcommand{\tg}{\tilde{g}}
\newcommand{\tH}{\tilde{H}}
\newcommand{\ct}[1]{\color{blue} #1 \color{black}}
\newcommand{\bs}[1]{\color{darkgreen} #1 \color{black}}
\begin{titlepage}

\begin{flushright}
IPPP/17/17\\
SLAC-PUB-16942
\end{flushright}

\vspace{0.2cm}
\begin{center}
\Large\bf
Phase Transitions and Baryogenesis From Decays
\end{center}

\vspace{0.2cm}
\begin{center}
{
Brian Shuve$\,^{a,b}$\footnote{E-mail:bshuve@g.hmc.edu} and Carlos Tamarit$\,^{c}$\footnote{E-mail:carlos.tamarit@durham.ac.uk}\\
\vspace{15pt}
$^{a}$Harvey Mudd College, Claremont, CA 91711, USA.\\
$^{b}$SLAC National Accelerator Laboratory, Menlo Park, CA 94025, USA.\\
 $^{c}$ Institute for Particle Physics Phenomenology, Durham University, DH1 3LE, United Kingdom.
}
\end{center} 

\setcounter{footnote}{0}
%\today
\vspace{0.2cm}
\begin{abstract}\vspace{0.2cm}
\noindent 

We study scenarios in which the baryon asymmetry is generated from the decay of a particle whose mass originates from the spontaneous breakdown of a symmetry. This is  realized in many models, including low-scale leptogenesis and theories with classical scale invariance. Symmetry breaking in the early universe proceeds through a phase transition that gives the parent particle a time-dependent mass, which provides an additional departure from thermal equilibrium that could modify the efficiency of baryogenesis from out-of-equilibrium decays. We characterize the effects of various types of phase transitions and show that an enhancement in the baryon asymmetry from decays is possible if the phase transition is of the second order, although such models are typically fine-tuned. We also stress the role of new annihilation modes that deplete the parent particle abundance in models realizing such a phase transition, reducing the efficacy of baryogenesis. A proper treatment of baryogenesis in such models therefore requires the inclusion of the effects we study in this paper.

\end{abstract}
\vfil

\end{titlepage}

\tableofcontents

\section{Introduction}

The Standard Model (SM) contains many fields but only one dimensionful parameter, since the symmetries of the theory forbid all such terms apart from the Higgs field mass. Consequently, the quarks, leptons, and gauge bosons only acquire masses via interactions with the Higgs field after electroweak symmetry breaking \cite{Glashow:1961tr,Weinberg:1967tq,Salam:1968rm}. This property of the SM was confirmed by  electroweak precision studies at LEP and SLD \cite{ALEPH:2005ab,LEP-2}, as well as the recent discovery of the Higgs boson at the LHC \cite{Aad:2012tfa,Chatrchyan:2012xdj}.

One implication of the origin of masses in the SM is that field-dependent masses were different in the early universe than today. Finite-temperature corrections to the Higgs potential confine the Higgs field to the origin at early times, and the Higgs field evolved to its present minimum via a phase transition once the universe cooled to a sufficient degree \cite{Kirzhnits:1972iw,Dolan:1973qd,Weinberg:1974hy}. The time dependence of particle masses provides an ``arrow of time'' that gives a departure from equilibrium in the early universe beyond the usual Hubble expansion, and this can have profound effects on cosmology.

A departure from  equilibrium is crucial to understanding one of the most important unsolved mysteries of particle physics:~the origin of the baryon asymmetry. In the absence of an excess of baryons over antibaryons, most of the protons and neutrons would have annihilated away in the early universe, which is in clear contradiction with the observed abundance of visible matter today. The generation of a baryon asymmetry requires, among other factors, a departure from equilibrium \cite{Sakharov:1967dj}:~the processes that create a baryon asymmetry can also destroy it when occurring in reverse and the rates exactly balance when in equilibrium, resulting in a vanishing asymmetry. The electroweak phase transition provides such a departure from equilibrium because baryon-number-violating sphaleron processes are rapid at high temperatures and slow considerably in the broken phase, resulting in electroweak baryogenesis \cite{Kuzmin:1985mm,Shaposhnikov:1986jp,Shaposhnikov:1987tw}. While the electroweak phase transition in the SM does not occur quickly enough to generate the observed baryon asymmetry \cite{Bochkarev:1987wf,Kajantie:1996mn}, extensions of the SM can modify the phase transition and make electroweak baryogenesis a viable theory (for a recent review of electroweak baryogenesis, see \cite{Morrissey:2012db}).

The baryon asymmetry can also be generated through other mechanisms, the most widely studied of which is baryogenesis through the out-of-equilibrium decay of a massive particle \cite{Weinberg:1979bt}. A heavy self-conjugate field  decays  into both baryons and antibaryons:~$CP$ violation can allow it to decay more frequently into baryons relative to antibaryons, generating an asymmetry. Inverse decay processes destroy the  asymmetry but are Boltzmann-suppressed for temperatures $T\ll M$,  falling below the expansion rate of the universe and giving a departure from equilibrium. Popular implementations of this mechanism include leptogenesis \cite{Fukugita:1986hr}, which is motivated by the see-saw mechanism for generating neutrino masses \cite{Minkowski:1977sc}, and  Grand-Unified-Theory baryogenesis \cite{Yoshimura:1978ex,Ignatiev:1978uf}. In both of these examples, the masses of the particles responsible for baryogenesis are technically natural and could arise from the spontaneous breaking of a symmetry. This is even more motivated in low-scale models where new particle masses are expected to arise dynamically, including models with classical scale invariance \cite{Khoze:2016zfi}, or weak-scale models of leptogenesis and neutrino masses \cite{Mohapatra:1986aw,Mohapatra:1986bd,Pilaftsis:2005rv,Pilaftsis:2008qt}. By analogy with the SM, this would give the decaying particles a time-dependent mass in the early universe; indeed, the phase transition typically occurs when $T\sim M$, which is the crucial era for the generation of the baryon asymmetry. In nearly all existing studies of baryogenesis in such models, however, the mass is assumed to have its zero-temperature value throughout the cosmological evolution.

The goal of this paper is to study how baryogenesis via out-of-equilibrium decays is affected by a phase transition that changes the mass of the parent particle. In particular, we are interested in determining
whether the additional departure from equilibrium during the phase transition might give rise to an enhancement of the resulting asymmetry. This is feasible in principle:~consider an illustrative example with a baryogenesis parent $X$ with zero-temperature mass $M_X$. In the conventional case where $X$ has constant mass (apart from possible thermal corrections), the asymmetry from early $X$ decays is destroyed by inverse washout processes. Asymmetry generation is only efficient for decays that occur at temperatures $T\ll M_X$ when inverse decays become Boltzmann suppressed and ineffective; however, the number density of $X$ is exponentially suppressed at this time, resulting in a small asymmetry. Let us contrast this with a scenario in which the $X$ mass suddenly turns on due to a phase transition at some temperature $T_{\rm c}\ll M$:~baryon destruction from inverse decays are completely ineffective, while the abundance of $X$ may not have time to relax to its equilibrium value during the phase transition and has an abundance equal to that of a massless field. Since all $X$ decays after the phase transition efficiently generate an asymmetry, this leads to the generation of a much larger asymmetry than with a time-independent $M_X$. 

As we demonstrate in our paper, however, there are several physical effects that complicate the above na\"ive argument. First, in realistic models where the parent particle in baryogenesis acquires a mass through a phase transition, new couplings are introduced to the fields responsible for symmetry breaking. As we show, a common feature in models with a significant departure from equilibrium is the existence of a light degree of freedom in the accompanying scalar sector. Couplings to the fields responsible for symmetry breaking therefore opens a new mode for the scattering of the heavy particle, which in turn affects the efficiency of baryogenesis. The effects of this damping of the asymmetry have been explored in the case of a time-independent parent mass \cite{Gu:2009hn,Sierra:2014sta}; we review these effects and study them in combination with a dynamical parent mass.

Secondly, the abundance of $X$ remains constant at the phase transition only if it is of the second order. In contrast, during a first-order phase transition most $X$ are reflected at the phase interface (bubble wall) in the limit $T_{\rm c}\ll M_X$, limiting the possible generation of a baryon asymmetry. Realizing the condition $T_{\rm c}\ll M_X$ for a second-order phase transition, as is necessary for the enhancement of the asymmetry, is not a generic feature of realistic models and it is challenging to find models that realize this hierarchy. This is due to radiative corrections in the symmetry-breaking scalar effective potential, which tend to spoil the properties of the phase transition in the regime in which the parametric enhancement of the asymmetry is realized at leading order. Models in which this enhancement may be consistently realized must involve extra fields with carefully adjusted couplings, and the potential shape required to give a parametric enhancement to the asymmetry requires a  relationship among parameters in the theory that lies beyond the level of precision of a full one-loop calculation.

There are not many studies in the literature on the effects of a dynamical parent particle mass on decay baryogenesis. To the best of our knowledge, the earliest studies were in the context of left-right symmetric models in which the particle responsible for lepton-number breaking undergoes a strong first-order phase transition \cite{Ganguly:1994pi}. However, an asymmetry from such a mechanism actually arises via $CP$-violating scalar dynamics generating a chemical potential for sphaleron transitions as in Ref.~\cite{McLerran:1990zh}, as opposed to a substantial asymmetry from parent decays. For models with a very strong first-order transition giving rise to a right-handed neutrino (RHN) mass, we argue that asymmetry generation through decays actually becomes hindered by an exponential suppression of the abundance due to reflection of the parent particle at the bubble wall. There has also been a discussion of baryogenesis from decays in quintessence models of dark energy \cite{Bi:2003yr}, although the evolution of the dynamical masses is so slow in this scenario that no sizeable change in the asymmetry is expected due to the dynamics of the underlying scalar field. Finally, Ref.~\cite{Pilaftsis:2008qt} considered the scenario where the right-handed neutrino responsible for leptogenesis acquire a mass coincident with the electroweak phase transition. In this case, the challenge is in generating an asymmetry before the sphalerons that couple baryon and lepton number become ineffective in the electroweak vacuum. Consequently, their results are sensitive to the time-dependence of sphaleron rates and SM gauge boson masses in a manner that is not amenable to generalization.

Additionally, the aforementioned works did not take into account the effects of scalar-RHN scattering modes on the lepton asymmetry, whose importance was recognized in
\cite{Gu:2009hn,Sierra:2014sta}; these works, however, did not address the effect of time-dependence in the masses of the heavy neutrinos.  Outside of the realm of leptogenesis from decays, other studies of cosmological implications of dynamical masses have examined, for example, leptogenesis from interactions with the bubble wall of a first-order $CP$-violating phase transition \cite{Pascoli:2016gkf}, leptogenesis in a manner akin to electroweak baryogenesis \cite{Cohen:1990it,Fornal:2017owa,Long:2017rdo}, leptogenesis from the non-thermal production of right-handed neutrinos in bubble-wall collisions \cite{Katz:2016adq}, as well as the effects of a phase transition on dark matter \cite{Cohen:2008nb} and on cosmological implications of flavour models \cite{Baldes:2016rqn}. Common to these studies, as well as our own, is the fact that the properties of particles can be very different in the early universe from the present day, and care should be exercised in interpreting and motivating experimental particle searches for such phenomena.

Our study is organized as follows:~in Sec.~\ref{sec:review}, we review out-of-equilibrium-decay baryogenesis in the absence of a phase transition, and in Sec.~\ref{sec:asym_PT} we derive the dependence of the asymmetry on a time-dependent mass in a toy example. We {consider} the effects on the asymmetry of couplings between the symmetry-breaking sector and the parent particle in Sec.~\ref{sec:damping}. In Sec.~\ref{sec:PT} we examine realistic models for phase transitions and the extent to which the phase transitions studied in earlier sections can be achieved, and we conclude with a summary and discussion of our results.

\section{Review of Asymmetry Generation from Decays}\label{sec:review}

In this section, we review the mechanism for generating an asymmetry from the out-of-equilibrium decay of a massive particle with time-independent mass \cite{Weinberg:1979bt}. We highlight the  dependence of the final baryon asymmetry on the parameters of the model, as this will facilitate a physical understanding of the asymmetry generated in our new model with a phase transition. 

 Since we ultimately wish to study the decay of a particle whose mass originates from spontaneous symmetry breaking, it is most natural to consider a particle with a technically natural mass such as a fermion. We therefore use as a specific example a toy model of thermal leptogenesis, in which a lepton asymmetry is first generated via the decays of singlet RHNs and is then transmitted to baryons via $B+L$-violating sphaleron processes \cite{Fukugita:1986hr}. However, we emphasize that the results we derive apply more generally to any model with a phase transition with asymmetry generation from out-of-equilibrium decay; in such cases, the  Yukawa couplings could deviate from the na\"ive see-saw values, and we ignore subtleties such as spectator processes and $\mathcal{O}(1)$ factors associated with transmitting the lepton asymmetry to baryons. Our findings can be trivially extended to apply to other theories, including specific models of neutrino masses. 
 
The new field content is a pair of right-handed neutrinos, $N_I$, each of which can decay into the left-handed lepton doublets, $L_\alpha$, and the Higgs field, $H$. There are many excellent reviews of this subject that provide more details than given here; see, for example, Ref.~\cite{Davidson:2008bu}. The Lagrangian is
\be
\label{eq:L}
\mathcal{L} = F_{\alpha I}\,\bar L_\alpha H^* N_I + \frac{M_I}{2} \overline{N}_I^{\rm c} N_I+ \mathrm{h.c.} + \mathcal{L}_{\rm kinetic}.
\ee
The Yukawa couplings above are specified in the basis where the $N_I$ mass is diagonal. The complex Yukawa matrix has a physical phase if there are at least two generations of $N$ and $L$, giving rise to $CP$ violation. An asymmetry in $L_\alpha$ is generated when $N_I\rightarrow L_\alpha H$ occurs at a faster rate than $N_I\rightarrow \bar L_\alpha H^*$. The difference in rates arises due to interference of tree and loop contributions to decay, and the  lepton asymmetry in flavour $\alpha$ produced per $N_I$ decay is characterized by  \cite{Covi:1996wh}
\be\label{eq:CP_source}
\varepsilon^\alpha_{I} &\equiv& \frac{\Gamma(N_I\rightarrow L_\alpha H)-\Gamma(N_I\rightarrow \bar L_\alpha H^*)}{\sum_\beta\left[\Gamma(N_I\rightarrow L_\beta H)+\Gamma(N_I\rightarrow \bar L_\beta H^*)\right]}\\
%%%
\nonumber&=& \frac{1}{8\pi}\,\frac{1}{(F^\dagger F)_{II}}\,\sum_{J\neq I}\left[\mathrm{Im}\left[(F^\dagger F)_{IJ}F^\dagger_{I\alpha}F_{\alpha J}\right]\,g(x_{IJ})+\mathrm{Im}\left[(F^\dagger F)_{JI}F^\dagger_{I\alpha}F_{\alpha J}\right]\,\frac{1}{1-x_{IJ}}\right],
\ee
where
\be
\label{eq:XIJ}
x_{IJ} = \frac{M_J^2}{M_I^2},~~~~~~g(x) = \sqrt{x}\left[\frac{1}{1-x}+1-(1+x)\log\left(\frac{1+x}{x}\right)\right].
\ee
$\varepsilon^\alpha_{I}$ is enhanced in the limit $M_I\rightarrow M_J$ due to a resonance in the self-energy contribution to the asymmetry. The above expression is valid in the limit $|M_I^2-M_J^2|\gg \mathrm{max}(M_I\Gamma_I,M_J\Gamma_J)$, while in the fully degenerate limit it is important to include effects of oscillations among mass eigenstates (see Appendix \ref{app:resonant} for generalizations of Eq.~(\ref{eq:CP_source}) in this limit) \cite{Pilaftsis:1997jf,Pilaftsis:2003gt}.

The asymmetry in $L_\alpha$ is produced via the decays of $N_I\rightarrow L_\alpha H$, while it is destroyed by the washout processes of inverse decay ($HL_\alpha\rightarrow N_I$) and off-shell $2\leftrightarrow2$ scattering (such as $HL_\alpha\leftrightarrow H^* \bar{L}_\beta$). In the semi-classical limit, the evolution of particle abundances can be modelled by Boltzmann equations. These are most simply expressed in terms of the number density normalized to the entropy density, $Y_i \equiv n_i/s$. Assuming that the interactions with $L_\alpha,\,H$ provide the only decay mode for $N_I$, then the  Boltzmann equations for the evolution of $Y_N$ and the  lepton asymmetry  in flavour $\alpha$, $Y_{\Delta L_\alpha} \equiv Y_{L_\alpha} - Y_{\bar L_\alpha} $ are\footnote{We neglect here higher-order washout terms $\sim|F|^4$ because, for the relatively small couplings we consider, these rates provide only a very minor correction to the $\mathcal{O}(|F|^2)$ terms. For example, in the limit $M_I\ll T$ we can estimate in the effective field theory $\Gamma^\alpha_{\Delta L= 2}/H\sim M_0 T \left(\sum_{I,\beta} F_{\alpha I}F_{\beta I} / M_I \right)^2 / 64\pi^3 $, where $M_0=m_{\rm P}/1.66\sqrt{g_*}$ and $m_{\rm P}$ is the Planck mass. For a typical benchmark point in the strong washout regime of $T^*/M_I\sim 0.1$, $M_1\sim10^9$ GeV, and anarchic $F\sim10^{-3}$, we obtain $\Gamma^\alpha_{\Delta L= 2}/H\sim 10^{-6}$, and so higher-order scattering is  out of equilibrium. }. 
\be
\begin{aligned}
\label{eq:Boltzmann}
\frac{dY_{N_I}}{dt} =& - \langle\Gamma_{N_I}\rangle \left(Y_{N_I}-Y_{N_I}^{\rm eq}\right),\\
\frac{dY_{{\Delta L_\alpha}}}{dt} =& \sum_{I}\,\varepsilon^\alpha_{I}\,\langle\Gamma_{N_I}\rangle\,\mathrm{Br}(N_I\rightarrow\alpha)\left(Y_{N_I}-Y_{N_I}^{\rm eq}\right) - \sum_{I}\,\langle\Gamma_{N_I}\rangle\,\mathrm{Br}(N_I\rightarrow\alpha)\,\frac{Y_{N_I}^{\rm eq}}{2Y_L^{\rm eq}}\,Y_{\Delta L_\alpha},
\end{aligned}
\ee
where $\langle\Gamma_{N_I}\rangle$ is the thermally averaged total $N_I$ width and $\mathrm{Br}(N_I\rightarrow\alpha)$ is the branching fraction of $N_I\rightarrow L_\alpha H+\bar{L}_\alpha H^*$. Note that the asymmetry is defined as the sum over the asymmetries in each doublet component.  Detailed definitions  of the various terms in the Boltzmann equation are provided in Appendix \ref{app:Tav}.

We can re-write the second Boltzmann equation as
\be\label{eq:BE_asymmetry}
\frac{dY_{\Delta L_\alpha}}{dt} &=& -\sum_{I}\,\varepsilon^\alpha_I\,\mathrm{Br}(N_I\rightarrow \alpha)\,\frac{dY_{N_I}}{dt} - \sum_{I}\,\langle\Gamma_{N_I}\rangle\,\mathrm{Br}(N_I\rightarrow\alpha)\,\frac{Y_{N_I}^{\rm eq}}{2Y_L^{\rm eq}}\,Y_{\Delta L_\alpha},
\ee
which makes clear that an asymmetry of $\varepsilon_I^\alpha$ is produced for every decay of $N_I$ (weighted by the branching fraction of $N_I$ into flavour $\alpha$). The second term destroys the asymmetry via inverse decays; because the $L_\alpha$ and $H$ must draw sufficient energy from the bath to re-constitute $N_I$, the rate is Boltzmann-suppressed for $M_I\gg T$.

The solution to the Boltzmann equation for the asymmetry in flavour $\alpha$, Eq.~(\ref{eq:BE_asymmetry}), at very late times is
\be\label{eq:BE_solution}
Y_{\Delta L_\alpha}^\infty &=& -\sum_I \,\varepsilon_I^\alpha\,\mathrm{Br}(N_I\rightarrow\alpha) \,\int_0^\infty\,dt'\,\frac{dY_{N_I}}{dt'}\,\exp\left[-\int_{t'}^\infty\,dt'' \,\Gamma^\alpha_{\rm W}(t'')\right],
\ee
where $\Gamma^\alpha_{\rm W}(t)$ is the rate of washout processes that destroy the asymmetry in $L_\alpha$,
\be
\label{eq:washout}
\Gamma^\alpha_{\rm W}(t) =  \sum_I \Gamma^\alpha_{\rm W,I}\equiv\sum_I\,\frac{Y_{N_I}^{\rm eq}}{2Y_{L_\alpha}^{\rm eq}}\,\langle\Gamma_{N_I}\rangle\,\mathrm{Br}(N_I\rightarrow\alpha).
\ee
The solution Eq.~(\ref{eq:BE_solution}) can be understood as follows:~in the time interval $[t',t'+\Delta t']$, an asymmetry in $L_\alpha$ is generated that is equal to the net number of decays of $N_I$ in this interval multiplied by $\varepsilon^\alpha_I$. This asymmetry, however, is exponentially damped by inverse decays occurring at times $t > t'+\Delta t'$. Once the integral of the washout rate falls below 1 (approximately when $\Gamma^\alpha_{\rm W}(t')$ falls below the Hubble rate $H(t')$),  the asymmetry destruction processes cease to be effective and any asymmetry generated after this time is preserved.

The dependence of the solution on the model parameters hinges on whether washout processes were ever important or not. These are classified as the strong and weak washout regimes, respectively:\\

\noindent {\bf Weak Washout:}~In this scenario, the integrated washout in the exponent of Eq.~(\ref{eq:BE_solution}) is always negligible during the epoch of net $N_I$ decays (\emph{i.e.,} $M_I>T$); this is because the Hubble expansion of the universe is always faster than inverse decays, $H(t) > \Gamma_{\rm W}(t)$, during this period. In this case,
\be
\label{eq:weakwashout}
Y_{\Delta L_\alpha}^\infty &\approx& -\sum_I \,\varepsilon^\alpha_I\,\mathrm{Br}(N_I\rightarrow\alpha) \,\int_0^\infty\,dt'\,\frac{dY_{N_I}}{dt'} = \sum_I\,\varepsilon_I^\alpha\,\mathrm{Br}(N_I\rightarrow\alpha)\,Y_{N_I}(0)
\ee
Thus, every net $N_I$ decay into $L_\alpha$ efficiently contributes to the lepton asymmetry by a factor of $\varepsilon_I^\alpha$. Note, however, that the result is sensitive to the abundance of $N_I$ at early times, $Y_{N_I}(0)$:~because the couplings to $L_\alpha$ are insufficient to bring the $N_I$ into equilibrium at early times, its initial abundance and hence the lepton asymmetry depend strongly on whatever other scattering processes could produce $N_I$ at temperatures $T>M_I$. \\

\noindent {\bf Strong Washout:}~In this scenario, there is a period of time in which the lepton-number-violating scattering processes are rapid compared to the expansion rate of the universe. This is a typical scenario due to the fact that the expansion rate is suppressed by the Planck scale and usually very small compared to reaction rates unless lepton-number-violating couplings are very small. Because of frequent decays and inverse decays, thermal equilibrium is established for the $N_I$ abundances ($Y_{N_I} \approx Y_{N_I}^{\rm eq}$) in the strong washout scenario. As $T$ cools below $M_I$, the asymmetry produced during the early decays of $N_I$ is rapidly damped away by washout processes. The importance of washout for flavour $\alpha$  is typically characterized by the following quantity:
\be
\label{eq:KI}
\mathcal{K}^\alpha_I \equiv \frac{\Gamma_{N_I}\,\mathrm{Br}(N_I\rightarrow\alpha)}{H(T=M_I)},
\ee
where $\mathcal{K}_I^\alpha$ is larger for stronger washout. 

As the universe cools below $M_I$, the rate of inverse decays (and hence the washout rate) is Boltzmann suppressed $Y_{N_I}^{\rm eq}\sim e^{-M_I/T}$. Due to the rapidly falling exponential,  the total washout rate eventually slows to equal the rate of expansion at a temperature $T_*$, defined implicitly by $\Gamma^\alpha_{\rm W}(T_*) = H(T_*)$. From this point onward, any asymmetry in $L_\alpha$ produced from $N_I$ decays is preserved, and we can estimate the final asymmetry produced from $N_I$ decays as
\be\label{eq:SW_asymmetry0}
Y_{\Delta L_\alpha}^\infty \sim \sum_I\varepsilon^\alpha_I\,\mathrm{Br}(N_I\rightarrow\alpha) \,Y_{N_I}^{\rm eq}(T_*).
\ee
In the case of a sufficiently hierarchical spectrum of $N_I$ particles, with $M_1 \ll M_{J\neq 1}$, the total washout rate decouples when only the lightest $N_1$ has an appreciable  abundance due to the stronger Boltzmann suppression for the processes involving $N_{J\neq1}$. In this case, the washout is also dominated by the rate $\Gamma_{\mathrm{W},1}$, which in turn depends on the width $\Gamma_{N_1}$. Then, using the definition of $T_*$ and the asymmetry from Eq.~\eqref{eq:SW_asymmetry0}, we may estimate the total asymmetry as
\be\label{eq:SW_asymmetry}
Y_{\Delta L_\alpha}^\infty \sim\varepsilon^\alpha_1\,\mathrm{Br}(N_1\rightarrow\alpha) \,Y_{N_1}^{\rm eq}(T_*)\sim \frac{2\varepsilon^\alpha_1\,\mathrm{Br}(N_1\rightarrow\alpha) \,Y_{L_\alpha}^{\rm eq}}{(z^*_1)^2\,\mathcal{K}^\alpha_1}.
\ee
where we have defined $z^*_1 \equiv M_1/T_*$. Due to the exponential damping of the washout rate from inverse decays, $z^*_1$ typically lies between $1-10$. Recall that $Y_{L_\alpha}^{\rm eq}$ is the abundance of the lepton, which is massless for temperatures above the weak scale and is approximately constant during leptogenesis. 

The key property to note is that the final asymmetry in $L_\alpha$ is inversely proportional to the washout strength $\mathcal{K}^\alpha_1$. A larger washout strength $\mathcal{K}^\alpha_1$ implies that washout remains effective down to a lower temperature, reducing the number of $N_1$ decays that can generate an asymmetry\footnote{Also, $z^*_1$ is larger for stronger washout;  however, this dependence is logarithmic and this is not the dominant effect on the asymmetry in the strong washout limit.}. In Eq.~(\ref{eq:SW_asymmetry}), $z^*_1$ can be determined by iteratively solving the above equation in the same manner as for the calculation of dark matter chemical decoupling in thermal freeze-out models \cite{Kolb:1990vq}.

We derived the scaling relation Eq.~(\ref{eq:SW_asymmetry}) under the assumption that the only contribution to the final $L_\alpha$ asymmetry comes from decays after washout freeze-out. This neglects earlier decays that are partially washed out, and these can contribute an $\mathcal{O}(1)$ fraction of the total. A better estimate is obtained by evaluating Eq.~(\ref{eq:BE_solution}) in the steepest-descent approximation, which gives a similar parametric dependence but without requiring the iterative solution for $z^*_1$ \cite{Abada:2006ea}:
\be\label{eq:SW_asymmetry_pheno}
Y_{\Delta L_\alpha}^\infty \approx\varepsilon^\alpha_1\,\mathrm{Br}(N_1\rightarrow\alpha)\,Y_{L_\alpha}^{\rm eq}\,\left(\frac{1}{2 \mathcal{K}^\alpha_1}\right)^{1.16}.
\ee

Our discussion so far has focused on the asymmetry in a particular flavour, $L_\alpha$. Of course, we are actually interested in the \emph{total} lepton number obtained by summing over lepton flavours. For a generic theory without any hierarchical couplings, one might expect $\varepsilon_1^\alpha$ and $\mathcal{K}_1^\alpha$ are approximately flavour independent, in which case one would get a total lepton asymmetry of the order of the individual flavour asymmetries and with the same scaling dependence. In specific cases, hierarchies in flavour couplings can give rise to non-trivial effects on the asymmetry, in which case our na\"ive scaling arguments break down \cite{Pilaftsis:2005rv}. However, the scaling derived above holds for many models and provides a useful analytic approximation for understanding the physical effects of a time-varying mass.

We may obtain an explicit analytic expression for the washout factor in terms of the parameters of the theory by substituting the temperature dependence of the Hubble rate
in a radiation-dominated universe with $g_*$ relativistic degrees of freedom,
\be
\label{eq:HT}
H\approx1.66 \sqrt{g_*} \frac{T^2}{m_{\rm P}},
\ee
where $m_{\rm P}=1.22\times10^{19}$ GeV is the Planck mass. Using the zero-temperature decay rate of the $N_I$ in the limit of massless products,
\be
\label{eq:Gamma}
\Gamma_{N_I}=\frac{(F^\dagger F)_{II} M_I}{8\pi},
\ee
this gives for the washout factor,
\be
\label{eq:KIM}
\mathcal{K}^\alpha_I  \approx \frac{|F_{\alpha 1}|^2 m_{\rm P}}{13.28\pi\sqrt{g_*} M_I}.
\ee
\\

To conclude this section, we discuss which of the above parameter regimes is most likely to be affected when $M_I$ changes due to a phase transition. The weak washout scenario already features a strong departure from thermal equilibrium, and so an additional departure from equilibrium due to a changing mass is unlikely to increase the asymmetry. Indeed, the dominant effect of a phase transition on the weak washout regime is the existence of new couplings between $N_I$ and the symmetry breaking sector that can modify its abundance prior to the epoch of leptogenesis. By contrast, the strong washout regime features lepton-number-violating processes that are in equilibrium and inhibit the generation of an asymmetry for temperatures $T>T_*$. A phase transition may provide an additional departure from thermal equilibrium, modifying the dynamics of asymmetry generation. Therefore, in what follows we focus predominantly on the strong washout regime, but comment where relevant on the effects of the new interactions on the weak washout limit.

\section{Asymmetry Generation with a Phase Transition}\label{sec:asym_PT}

\subsection{Time-Dependent Majorana Mass}

When an asymmetry is generated via the decay of a heavy particle, a departure from equilibrium occurs in two, interconnected ways. The cooling of the universe below $M_I$ results in the net decay of $N_I$ into the lighter lepton species, allowing for the generation of an asymmetry. At the same time,  the cooling also suppresses inverse decays that wash out the asymmetry, allowing the accumulation of a substantial asymmetry.

If the Majorana mass, $M_I$,  originates from a phase transition in the early universe, this can provide an additional departure from equilibrium if the mass changes on time scales that are fast relative to the Hubble expansion rate. The rapid increase of $M_I$ at the phase transition tends to suppress washout processes more quickly than the Hubble expansion alone, meaning that an asymmetry generated by the decays of $N_I$ may not have time to relax to zero as fully as in the conventional strong washout scenario that is our main focus. 

For the remainder of the paper, we replace the Majorana mass for $N_I$ with a coupling to a symmetry-breaking scalar $\Phi$:
\be
\label{eq:LMajoron}
\mathcal L_{\rm mass} = \frac{y_I}{2}\,\Phi\, \overline{N}_I^{\rm c} N_I + \mathrm{h.c.} + \mathcal{L}_\Phi.
\ee
If $N_I$ possesses an appropriate discrete or continuous symmetry, a tree-level mass term as in Eq.~\eqref{eq:L} is forbidden and the coupling to $\Phi$ provides the only contribution to the $N_I$ mass after symmetry breaking. The Majorana masses of the right-handed neutrinos are now
\be\label{eq:Majorana_SSB}
M_I(t) = y_I\langle\Phi(t)\rangle.
\ee
In this section, we are agnostic about the details of the symmetry-breaking field $\Phi$ and its dynamics. In particular, we assume that  $\Phi$ is simply a time-dependent background field and that all new states associated with this symmetry breaking are decoupled at the time of baryogenesis. While this is not a realistic assumption \cite{Gu:2009hn,Sierra:2014sta}, it does allow us to isolate the effects of the background-field phase transition from other dynamics of the new scalar interactions. With this assumption hypothesis, the generation of the baryon asymmetry is modelled by the same Boltzmann equations in Eq.~\eqref{eq:Boltzmann}, but with time-dependent $N_I$ masses. We return to the effects of $N_I$ scattering into $\Phi$ in Sec.~\ref{sec:damping}.

Our study of the effects of the phase transition on baryogenesis requires some ansatz for the form of the phase transition. For now, we restrict ourselves here to simplistic ans\"atze for $\Phi$ as a function of temperature (and hence time) that make clear the influence of the phase transition, and defer a discussion of particular models (from which the evolution of $\Phi$ during the phase transition is calculable) to Sec.~\ref{sec:PT}. While realistic phase transitions have more complicated mass profiles, our ans\"atze allow for an intuitive understanding of how the asymmetry depends on the nature of the phase transition.
We consider different scenarios:~a very fast first-order phase transition, a second-order phase transition, and a slowly evolving scalar which remains constant over the time scales of leptogenesis but differs in value between the time of leptogenesis and the present.  The slow evolution, although not providing an enhanced departure from thermal equilibrium, can be an interesting case study for thermal leptogenesis because it modifies the relation between SM neutrino masses and $M_I$ in the early universe vs.~their values today. This has been argued to allow for enhanced asymmetries \cite{Gu:2003er,Bi:2003yr}, and we include a study of this scenario for completeness. 

For the phase transitions, the time-dependent background field profiles we study in this section are:
\be
\mathbf{First\,\, order:} && \Phi(T) = \Phi_0\,\theta(T_{\rm c}-T),\label{eq:FOPT}\\
\mathbf{Second\,\,order:} && \Phi(T) = \Phi_0\sqrt{1-\frac{T^2}{T_{\rm c}^2}}\,\theta(T_{\rm c}-T)\label{eq:SOPT},
\ee
where $T_{\rm c}$ is the critical temperature of the phase transition and a free parameter of the model. These background-field profiles lead to time-dependent masses that are substituted into the Boltzmann equations.
The time-varying masses appear in the Boltzmann equations in a few places:~in the expressions for the equilibrium $N_I$ abundance, in the thermally averaged widths, and in the $CP$-violating source terms driving the creation of the asymmetry ($\varepsilon^\alpha_I$). According to Eq.~(\ref{eq:CP_source}), however, $\varepsilon^\alpha_I$ is sensitive only to the ratios among masses. Since $x_{IJ} = y_J^2/y_I^2$ is independent of time, this suggests that the $CP$-violating sources are time independent as well. This is true only under the approximations that render Eq.~(\ref{eq:CP_source}) valid, namely that the time-dependence of the $CP$ source due to coherence effects is much shorter than due to other scattering processes \cite{Buchmuller:2000nd,DeSimone:2007gkc}. We restrict ourselves to model parameters for which this is true.

\subsection{\label{sec:1PT}First-Order Phase Transition}

Prior to the start of the phase transition, all fields are massless at tree level. $N_I$ does acquire a non-trivial dispersion relation from propagating through the hot, dense plasma; however, this dispersion relation leaves intact the global lepton number symmetry. In the absence of a VEV for $\Phi$, the global symmetry prevents the accumulation of any net lepton number density. Furthermore, the $N_I$ remain relativistic as the universe cools and their number density does not depart from thermal equilibrium. Thus, no asymmetry is generated in the unbroken phase.

In the limit of a very fast phase transition, the masses of all fields suddenly turn on at $T=T_{\rm c}$. In this phase, the mass is
\be
M_I^2(T<T_{\rm c}) = y_I^2 \Phi_0^2,
\ee
where $\Phi_0$ is the zero-temperature vacuum expectation value. We neglect thermal effects on the $N_I$ propagation:~this is reasonable since the net $N_I$ number density does not change until $M_I(T) = T$ and hence no leptogenesis occurs until later than this time. For a perturbative theory, the thermal corrections to the $N_I$ mass in the high-temperature expansion are therefore subdominant to the tree-level term for $T<M_I(T)$. We define the zero-temperature mass as $M_I^0 \equiv y_I \Phi_0$. 

For $T_{\rm c}\gg M_I^0$, the effect of the phase transition on the mass or number density of $N_I$ is negligible. In other words,
the $N_I$ retain an unchanging equilibrium abundance throughout the phase transition. The largest contribution to the  asymmetry from $N_I$ decay does not occur until washout goes out of equilibrium at a scale $T_*\lesssim M_I^0\ll T_{\rm c}$, and therefore the phase transition has no effect on leptogenesis.

Conversely, for $T_{\rm c}\ll T_* \lesssim M_I^0$, the washout suddenly turns off when the mass changes. Since we are considering $N_I$ in the strong washout regime ($\mathcal{K}^\alpha_I = \Gamma_{N_I}\,\mathrm{Br}(N_I\rightarrow\alpha)/H(M_I)>1$ for some $\alpha$), the $N_I$ rapidly decay to their new equilibrium abundance, giving rise to a net asymmetry:
 \be\label{eq:firstorder_nominal}
 Y_{\Delta {L_\alpha}} = \sum_I\varepsilon^\alpha_I\,\mathrm{Br}(N_I\rightarrow\alpha)\, Y_{N_I}(T_{\rm c}).
 \ee
 In contrast, the washout processes suffer a Boltzmann suppression, $\Gamma^\alpha_W \sim \Gamma_{N_I}\mathrm{Br}(N_I\rightarrow\alpha)\,e^{-M_I^0/T_{\rm c}}$, and for $ T_{\rm c}\ll T_* < M_I^0$ the washout processes are ineffective. In this case, Eq.~(\ref{eq:firstorder_nominal}) gives the exact result for the final asymmetry. Using the simple analytic solution to the asymmetry for leptogenesis in the absence of a phase transition, Eq.~(\ref{eq:SW_asymmetry}), we find that the ratio of asymmetries is:
 \be\label{eq:firstorder_ratio}
 \frac{Y_{\Delta {L_\alpha}}^{\rm P.T.}}{Y_{\Delta L_\alpha}^{\rm no\,\,P.T.}}\approx \frac{\sum_I\varepsilon^\alpha_I\, Y_{N_I}(T_{\rm c})}{Y_{L_\alpha}^{\rm eq}\sum_I 2\varepsilon^\alpha_I\left[\mathcal{K}^\alpha_I (M_I^0)^2/T_*^2\right]^{-1}}\,\,\,\,\,\,\,\,\,\,\,\,\,\,\,\,\,\,\,\,\,(T_* > T_{\rm c}),
 \ee
In particular, we see that the ratio of asymmetries scales like $\mathcal{K}^\alpha_I$:~the larger the couplings leading to $N_I$ decay, the more pronounced the effect of a phase transition is on the asymmetry by suppressing washout. Eq.~(\ref{eq:firstorder_ratio}) is only valid in the limit $ T_{\rm c}\ll T_* < M_I^0$ such that all washout is negligible; it is possible to analytically solve for the asymmetry ratio in the intermediate case $ T_{\rm c}\sim T_*$. If $T_{\rm c}\gg T_*$, the $N_I$ equilibrate in the new phase and the asymmetry reverts to the result in the absence of a phase transition.

In order to evaluate Eq.~(\ref{eq:firstorder_ratio}), we must know the abundance of $N_I$ immediately after the phase transition, $Y_{N_I}(T_{\rm c})$. In the ideal case where the mass of $N_I$ changes instantaneously and homogeneously throughout space as described in Eq.~(\ref{eq:FOPT}), the abundance of $N_I$ does not have the opportunity to react to the change, and so $Y_{N_I}(T_{\rm c}) = Y_{L_\alpha}$ (\emph{i.e.,} the abundance is the same as for a massless species). We then find an enhancement in the asymmetry  $\propto \mathcal{K}^\alpha_I$ for the case of a first-order phase transition relative to a time-independent mass.

For realistic models, however, we know that first-order phase transitions proceed through bubble nucleation, followed by the rapid expansion of the bubble walls. The result is a mass profile possessing a spatial gradient in addition to the time dependence. Assuming the bubble nucleates at the origin, and in the thin-walled approximation, the $N_I$ mass profile at position $\vec{r}$ has the form
\be
M_I(T,\vec{r}) &\approx& M_I^0\,\theta(T_{\rm c}-T)\,\theta[v_w(t-t_{\rm c})-|\vec{r}|],
\ee
where $v_w$ is the bubble-wall velocity and $t_{\rm c}$ is the time of the phase transition. As the bubble wall expands, particles in the unbroken phase must either propagate through the wall, or be reflected at the phase boundary. If $T_{\rm c}\gtrsim M_I^0$, then the $N_I$ in the plasma have sufficient energy to penetrate the bubble wall, and an $\mathcal{O}(1)$ fraction propagate into the broken phase. In this limit, however, there is no appreciable effect of the phase transition on the asymmetry since
the $N_I$ retain a near-equilibrium abundance during the phase transition, and the bulk of the asymmetry is not generated until after washout interactions cease at a time well after the phase transition.

Conversely, if $T_{\rm c}\ll M_I$, then the vast majority of $N_I$ particles cannot penetrate the bubble wall and are reflected; conservation of energy dictates that only those with typical momentum $k\sim M_I^0$ can enter the bubble \cite{Nelson:1991ab}, and the abundance of these modes is highly Boltzmann suppressed by $e^{-M_I^0/T_{\rm c}}$. The yield of $N_I$ immediately after the passage of the bubble walls is
 \be
 Y_{N_I}(M_I=M_I^0,T_{\rm c}) \approx Y_{N_I}(M_I=0,k>M_I^0) \approx \frac{1}{2}\left(\frac{M_I^0}{T_{\rm c}}\right)^2e^{-M_I^0/T_{\rm c}}\,Y_{N_I}(M_I=0).
 \ee
 We find that the asymmetry is exponentially suppressed due to the same Boltzmann suppression of $N_I$ modes propagating through the bubble wall. Using the definition of $T_*$ as the temperature at which $\Gamma^\alpha_{\rm W}(T_*) = H(T_*)$, and considering $N_1$ as the dominant contributor to both the asymmetry and washout  to obtain the approximate scaling behaviour, we can express the baryon asymmetry ratio as 
 \be\label{eq:firstorder_bad}
  \frac{Y_{\Delta L_\alpha}^{\rm P.T.}}{Y_{\Delta L_\alpha}^{\rm no\,\,P.T.}}\sim \frac{\mathcal{K}_1^\alpha}{4}\left(\frac{M_1^0}{T_*}\right)^2\left(\frac{M_1^0}{T_{\rm c}}\right)^2e^{-M_1^0/T_{\rm c}}\,\,\,\,\,\,\,\,\,\,\,\,\,\,\,\,\,\,\,\,\,(T_* > T_{\rm c}).
 \ee
According to our analytic estimate, we see that if the phase transition occurs at $T_{\rm c}\ll T_*$, there is an exponential suppression of the asymmetry relative to the scenario with no phase transition. By contrast, if $T_{\rm c}\gtrsim T_*$, then this formula is no longer valid and instead the $N_I$ equilibrate prior to the generation of the asymmetry and there is no effect of the phase transition. In the intermediate case, $T_{\rm c}\sim T_*$, it may be possible to realize a small enhancement of the asymmetry due to the phase transition, but this is  $\mathcal{O}(1)$ because $M_I/T_*$ depends only logarithmically on the decay rate. Thus, it appears that a first-order phase transition either has no effect if it happens prior to $N_I$ going out of equilibrium, or it leads to an exponential suppression of the asymmetry if it occurs after the $N_I$ depart from equilibrium.

\begin{figure}[t]
\centering
\includegraphics[width=0.55 \textwidth ]{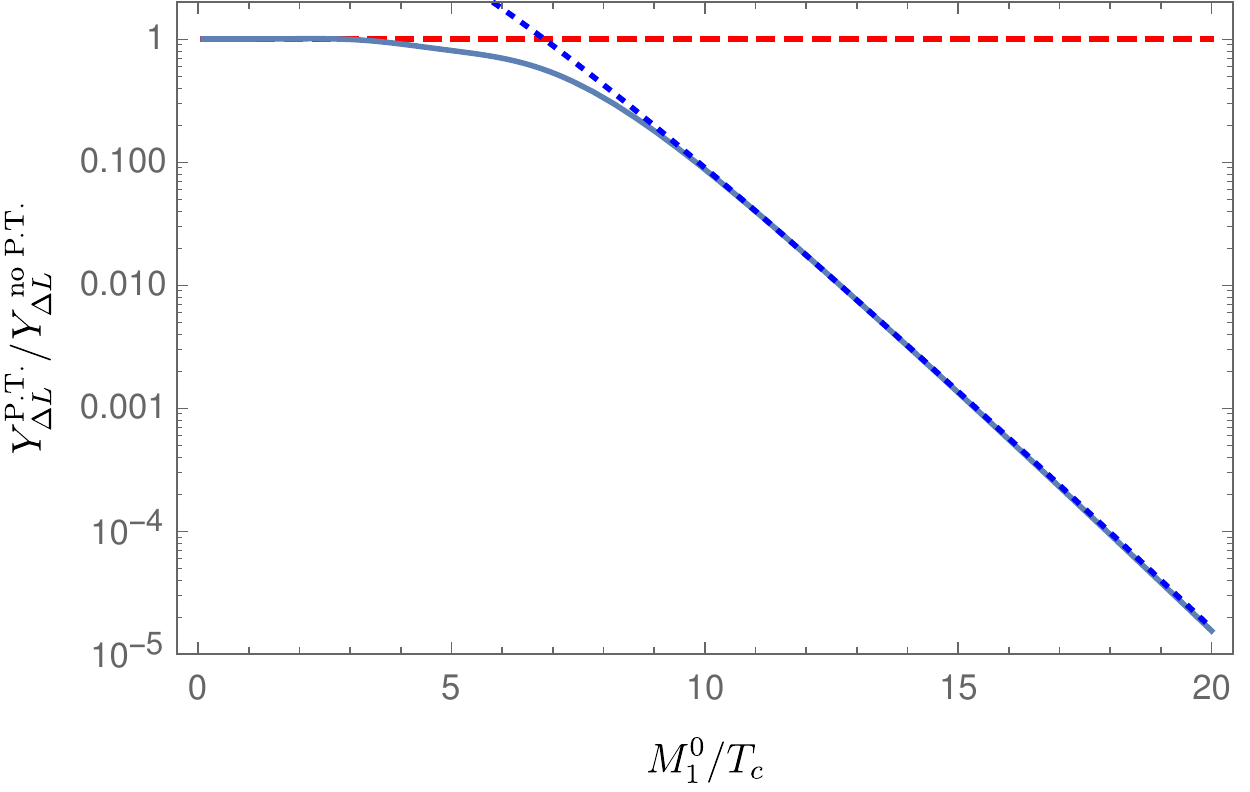}
\caption{Ratio of total lepton number asymmetry, $Y_{\Delta L}$, with a {\bf first-order phase transition} relative to the constant-mass case. Only $N_I$ modes that are sufficiently energetic to propagate through the bubble wall are considered for generating the asymmetry. The solid curve indicates the numerical solution of the Boltzmann equation, while the dotted blue line is a fit to Eq.~(\ref{eq:firstorder_bad}), and the dashed red line illustrates the ratio of unity expected when $T_{\rm c}$ is high enough that the $N_I$ equilibrate after the phase transition. For the specific numerical results plotted here, we take $M_1=M_2/2=10^9$ GeV, and the Yukawa matrices are chosen to be compatible with global fits to neutrino oscillations (see Eq.~\eqref{eq:Yuk_fig2_solid}).}
\label{fig:PT_strength}
\end{figure}

We derived the change in the asymmetry from the phase transition in Eq.~(\ref{eq:firstorder_bad}) with several simplifying assumptions, such as looking at the contribution of only one flavour of $L_\alpha$ and $N_I$. The result, however, approximately holds even when solving the full Boltzmann equations numerically. We consider a system of two flavours of $N_I$ and three flavours of $L_\alpha$ and, under the assumption of a normal neutrino hierarchy, choose a complex Yukawa matrix $F$ compatible with the latest global fits to oscillation experiments \cite{deSalas:2017kay}. We give more details in Appendix \ref{app:Fs}; the explicit choice of Yukawa matrix is given by Eq.~\eqref{eq:Yuk_fig2_solid}. We then  solve the resulting Boltzmann equations assuming initial conditions at $T_{\rm c}$ of $Y_{N_I}(T_{\rm c}) = Y_{N_I}(M=0,T_{\rm c},k > M_I^0)$ for the massive fermions in the broken phase. We compute the total lepton number asymmetry in both the case where the mass turns on suddenly and the constant-mass scenario, and their ratio is shown in Fig.~\ref{fig:PT_strength}. We see that, at large $M_I^0/T_{\rm c}$, the exponential suppression of the asymmetry predicted by Eq.~(\ref{eq:firstorder_bad}) is evident, while at large $M_I^0/T_{\rm c}$, the $N_I$ equilibrate in the broken phase and the asymmetry is essentially unchanged.

%%%%%%%%%%%%%%% FIGURE %%%%%%%%%%%%%%%%%%%%%
%%%%%%%%%%%%%%% FIGURE %%%%%%%%%%%%%%%%%%%%%

 In our discussion, we have focused exclusively on the effects of the changing mass, $M_I(T)$, on the baryon asymmetry from a first-order phase transition. There may be other effects of the bubble-wall dynamics that can lead to successful baryogenesis apart from decays  \cite{McLerran:1990zh,Cohen:1990it,Li:2001st,Katz:2016adq,Fornal:2017owa,Long:2017rdo}.

\subsection{Second-Order Phase Transition}
\label{subsec:2ndorder}
We now consider phase transitions where $\Phi$ relaxes homogeneously to its zero-temperature vacuum with no discontinuity in the order parameter. Such phase transitions give rise to a temperature-dependent mass with a profile like
\be
\label{eq:profile}
M_I(T) &=& M_I^0\left(1-\frac{T^2}{T_{\rm c}^2}\right)^n\,\theta(T_{\rm c}-T).
\ee
The mass vanishes  until a temperature $T_{\rm c}$, at which point the mass smoothly transitions to the zero-temperature value $M_I^0$. The power of $n$ depends on the details of the phase transitions:~for example, a second-order phase transition that occurs due to the competition between thermal corrections to a scalar mass and the tree-level term has $n=1/2$ in the high-temperature expansion (as with the SM Higgs potential in the high-temperature expansion; however, in the case of the SM the high-temperature expansion fails to capture the correct dynamics of the phase transition, which in fact is a cross-over). Higher-order corrections and deviations from the high-$T$ expansion modify the above form, but we use this simple ansatz to illustrate the parametric behaviour of the baryon asymmetry.

Conventionally, second-order phase transitions are believed to not provide a strong departure from thermal equilibrium beyond Hubble expansion. The reason is that the change in mass occurs continuously due to a change in $T$, which itself arises from Hubble expansion. Therefore, the change in mass is expected to be comparable to the characteristic expansion time scale. A simple example shows that this is not always true, however. If we compute the time derivative of our mass ansatz for $T<T_{\rm c}$, we find
\be\label{eq:derivative_profile}
\frac{dM_I}{dt} &=& 2Hn\,M_I(T)\, \frac{T^2}{T_{\rm c}^2-T^2}.
\ee
For $T_{\rm c}>T\ge T_{\rm c}/\sqrt{2n+1}$, the mass is changing faster than the Hubble expansion, meaning that processes sensitive to $M_I$ could deviate from equilibrium more sharply within this window of temperatures. In other words, for $T\sim T_{\rm c}$ the mass changes faster than  processes with typical timescale $\Gamma\sim H$ can respond to the change.

Washout processes go out of equilibrium when their rate becomes slower than their own rate of change,
\be\label{eq:secondO_washoutcond}
\frac{1}{\Gamma^{\alpha}_{\rm W}}\,\frac{d\Gamma^\alpha_{\rm W}}{dt}  \approx -\Gamma^\alpha_{\rm W}.
\ee
Once this is satisfied, the washout processes turn off faster than they can appreciably destroy the baryon asymmetry. We now derive an estimate for when this condition is satisfied taking Maxwell-Boltzmann statistics for simplicity in the approximate analytic results. Using Eq.~(\ref{eq:washout}) and the 
results of Appendix \ref{app:Tav}, we can write
\be
\Gamma_{\rm W}^{\alpha} &\approx& \frac{1}{32\pi T^2}\,\sum_I \,|F_{\alpha I}|^2\,M_I^3\,K_1(M_I/T),
\ee
where $K_1$ is the modified Bessel function of the first kind. According to our earlier arguments, the phase transition only has a major effect in the strong washout limit. In this case, we expect washout to decouple for $M_I/T \gg1$. Taking this limit, we find a  compact form for the derivative of the washout rate: 
\be
\frac{d\Gamma^\alpha_{\rm W}}{dt} &=& -H\,\sum_I\,\Gamma^\alpha_{\mathrm{W},I}\,\frac{M_I}{T}\left(1 - \frac{T}{M_I}\,\frac{dM_I}{dT}\right).
\ee
We evaluate this in the hierarchical case where a single species, $N_J$, dominates the washout for simplicity. We first evaluate the condition for the mass-independent scenario, and then proceed to the mass-varying case.\\

\noindent {\bf Time-Independent Mass:}~In the case of constant mass and one species, $J$, dominating the washout, we find that Eq.~(\ref{eq:secondO_washoutcond}) reduces to the usual requirement that washout decouples when the washout rate falls below the Hubble expansion rate,
\be\label{eq:washout_ooeq_usual}
\Gamma^\alpha_{\rm W}(T_*)  \approx H(T_*) \,\frac{M_J^0}{T_*},
\ee
where $T_*$ is the temperature at which this equality holds. This condition can be formulated entirely in terms of the washout factor, $\mathcal{K}^\alpha_J = \Gamma^\alpha_{N_J}/H(M_J^0)$, and the dimensionless ratio $z^*_J\equiv M_J^0/T_*$, 
\be
\frac{\mathcal{K}^\alpha_J}{4}\,{z^*_J}^3\,K_1(z_J^*) &=& 1.
\ee
There is no explicit mass dependence, and this explains why the asymmetry in each flavour can be  estimated simply in terms of $\mathcal{K}^\alpha_J$ in Eq.~(\ref{eq:SW_asymmetry}).\\

\noindent {\bf Time-Dependent Mass:}~If, instead, the mass derivative term dominates the expression for the change in washout (\emph{i.e.,} $dM_J/dT > M_J/T$), we find the condition of washout freeze-out in the single-flavour limit changes to
\be\label{eq:washout_ooeq_ours}
 \Gamma^\alpha_{\rm W}(T_*')   &=&  -H(T_*')\,\frac{dM_J(T_*')}{dT},
\ee
where $T_*'$ is defined as the temperature at which this equality holds. We see that a large derivative for the time-dependent mass results in washout decoupling at an earlier time than might otherwise be expected. Expressing this equality purely in terms of the washout factor $\mathcal{K}^\alpha_J$ and the dimensionless ratio $z_J'\equiv M_J(T_*')/T_*'$, we can write the condition of washout freeze-out as 
\be
\label{eq:washoutdec}
-\frac{\mathcal{K}^\alpha_J}{4}\,\frac{M_J^0}{T_*'}\left(\frac{dM(T_*')}{dT}\right)^{-1}{z_J'}^3\,K_1(z_J') = 1.
\ee
We therefore find that the washout condition in the mass-varying scenario is the same as for a time-independent mass, provided we substitute the factor $\mathcal{K}^\alpha_J$ for an {\bf effective washout factor}: 
\be\label{eq:effective_washout}
(\mathcal{K}_J^\alpha)^{\rm eff} &\equiv & - \mathcal{K}^\alpha_J\,\frac{M_J^0}{T_*'}\left(\frac{dM_J(T_*')}{dT}\right)^{-1}.
\ee
Because everything else in the condition for washout freeze-out depends only on the dimensionless ratio $M_J(T)/T$, the asymmetry scales like
\be
Y_{\Delta L} \sim \frac{1}{ (\mathcal{K}^\alpha_J)^{\rm eff}}.
\ee
%.
 Consequently, if the mass changes very rapidly, $(\mathcal{K}^\alpha_J)^{\rm eff} \ll \mathcal{K}^\alpha_J$, this results in an enhancement of the asymmetry over the time-independent scenario.

To  obtain an estimate of $(\mathcal{K}^\alpha_J)^{\rm eff}$ (and hence an analytic scaling for the lepton asymmetry), we must determine the temperature $T_*'$ relative to known scales in the theory. The second-order phase transition only has an effect if $M_J(T)$ varies rapidly with $T$, and from Eq.~(\ref{eq:derivative_profile}) we see that this occurs only for $T\approx T_{\rm c}$. For a typical second-order phase transition, the $T$-dependence of the mass in the high-$T$ expansion is given in Eq.~\eqref{eq:profile} with $n=1/2$ (see also Eq.~\eqref{eq:SOPT}). In this case, we find 
\be
\frac{dM_J}{dT} &\approx& -\frac{(M_J^0)^2}{z_J\,T_{\rm c}^2}
\ee
where we recall that $z_J =  M_J(T)/T$. Although $dM/dT$  nominally diverges at $T=T_{\rm c}$, no asymmetry is  generated at this time because net decays do not start occurring until $M_J(T) \approx T$. Because washout decouples exponentially quickly for $T<M(T)$, we expect that $z_{J*}\sim1-10$.  We therefore find that the transition is fastest (and the asymmetry is maximized) for $M_J^0\gg T_{\rm c}$.  

$T_*'$ becomes closer to $T_c$ for faster transitions, and the enhancement of the asymmetry grows due to the suppression of the effective washout Eq.~\eqref{eq:effective_washout}. The effective washout factor is then
\be
(\mathcal{K}^\alpha_J)^{\rm eff} &\approx&  \mathcal{K}^\alpha_J \,z_{J*}\,\frac{T_{\rm c}}{M_J^0}.
\ee

\noindent {\bf Asymmetry Ratio:}~Since $Y^\alpha_{\Delta L}\sim 1/(\mathcal{K}^\alpha)_{\rm eff}$, we find a \emph{linear enhancement} of the asymmetry for a delayed transition $M_J\gg T_{\rm c}$ in the single-flavour limit:
\be
  \frac{Y_{\Delta L_\alpha}^{\rm P.T.}}{Y_{\Delta L_\alpha}^{\rm no\,\,P.T.}}\sim \frac{M_J^0}{T_{\rm c}}\,z_J^*.
\ee
The enhancement grows linearly until the phase transition occurs so rapidly that washout reactions have no time to respond to the changing mass. At this point, we enter an ``effective'' weak washout limit, and as shown in Eq.~(\ref{eq:weakwashout}), the asymmetry is determined exclusively by the $CP$-violating sources and the relativistic abundance of the massless $N_J$ prior to the phase transition\footnote{We have assumed that the number of entropic degrees of freedom is the same as the number of radiation degrees of freedom.}:
 \be
 (Y^\alpha_{\Delta L})^{\rm max} = \varepsilon_J^\alpha\,\mathrm{Br}(N_J\rightarrow\alpha)\, Y^{\rm eq}_{N_J}(T>T_{\rm c})\approx\frac{45}{g_*\pi^4} \varepsilon_J^\alpha \,\mathrm{Br}(N_J\rightarrow\alpha) .
 \ee
For even faster phase transitions, the asymmetry  no longer has any dependence on the critical temperature and approaches a constant.\\

%%%%%%%%%%%%%%% FIGURE %%%%%%%%%%%%%%%%%%%%%
\begin{figure}[t]
\centering
\includegraphics[width=0.7\textwidth ]{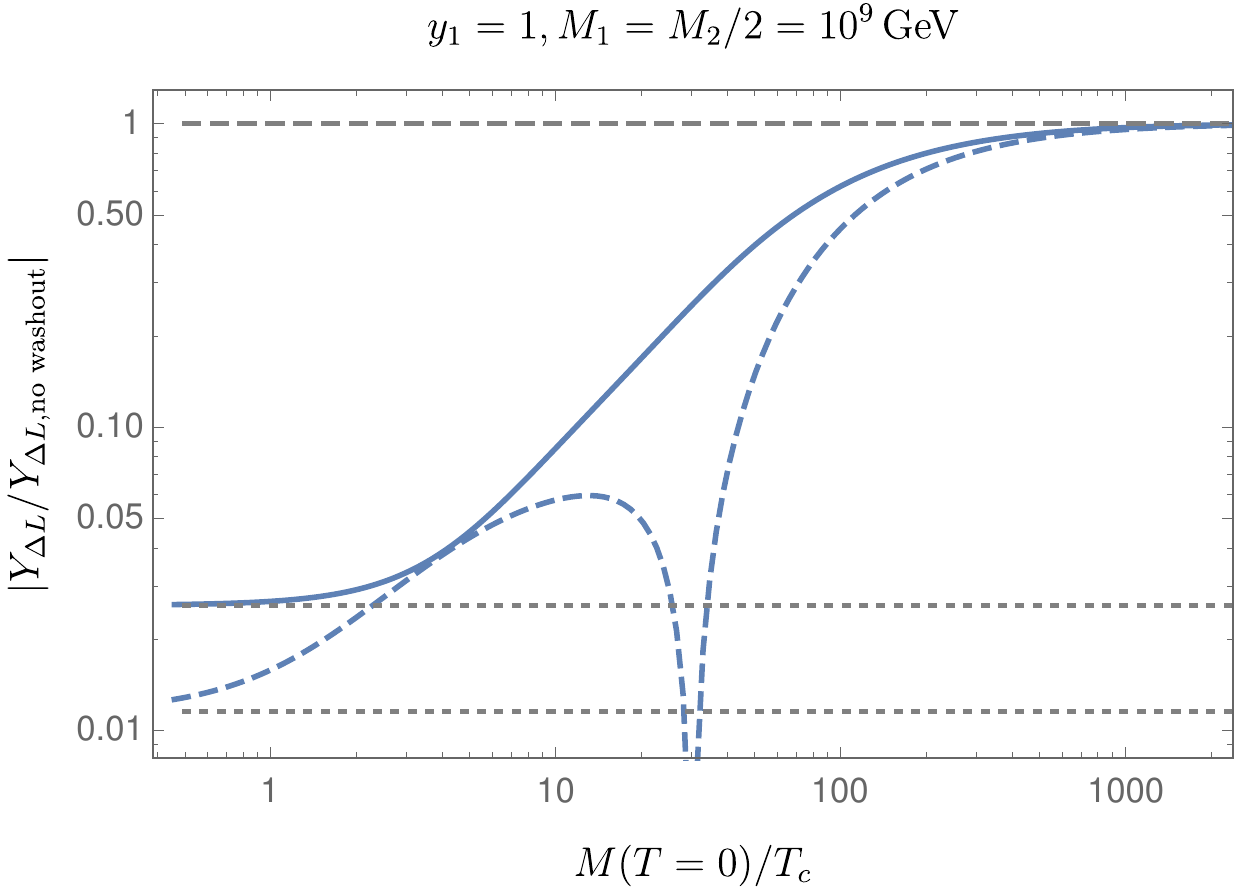}
\caption{Dependence of the asymmetry on the critical temperature of a {\bf second-order phase transition} giving rise to RHN masses, Eq.~(\ref{eq:LMaj}). Results are shown in the tree-level approximation with leading temperature corrections. The dashed-gray and dotted-gray lines show the asymptotic results expected for zero washout and constant masses, respectively. We set $y_1=1$, $M_1=M_2/2=10^9$ GeV, and calculate the asymmetry for two  choices of Yukawa matrix $F$ compatible with global fits to oscillation measurements, 
given by Eqs.~\eqref{eq:Yuk_fig2_solid} and \eqref{eq:Yuk_fig2_dashed} in Appendix \ref{app:Fs}. For the solid blue line, the sign of the asymmetry for constant $M_I$ does not change in the limit of zero washout, as opposed to the case of the dashed blue line, for which flavour effects change the sign of the asymmetry. }
\label{fig:enhancement}
\end{figure}
%%%%%%%%%%%%%%% FIGURE %%%%%%%%%%%%%%%%%%%%%

\noindent {\bf Numerical Study:}~The above analytic estimates used the single-flavour limit to obtain approximate scaling relations. However, we now examine numerically a $3\times2$ flavour system as would be expected for a model of RHNs generating the observed neutrino masses and mixings. We consider a model with right-handed neutrinos, $N_J$, and a symmetry-breaking scalar  $\Phi$ with the following interactions:
\begin{equation}
\label{eq:LMaj}
 \mathcal L_{\rm mass}\supset -\frac{y_I}{2}\,\Phi\, \overline{N}_I^{\rm c} N_I + \mathrm{h.c.}+m_\Phi^2\Phi^\dagger\Phi-\frac{\lambda}{4}(\Phi^\dagger\Phi)^2,
\end{equation}
with additional operators forbidden by means of appropriate continuous or discrete symmetries. The mass profiles for the RHNs, $M_I(T)$, can be determined from the leading order contributions to the  high-temperature expansion
of the effective potential for the $\Phi$ field (see Appendix \ref{app:VeffT} for a brief overview). The latter  gives a  VEV, $\Phi(T)$, with the same temperature dependence of Eq.~\eqref{eq:SOPT} and a critical temperature  
\begin{equation}
\label{eq:Tc}
 T^2_{\rm c}=\frac{12\lambda\Phi^2_0}{2\lambda+\sum_I y^2_I},
\end{equation}
where $\Phi_0\equiv \Phi(T=0)$. We show in Fig.~\ref{fig:enhancement} the enhancement of the asymmetry as a function of $M^0_{N_I}/T_{\rm c}$, obtained by solving Boltzmann equations numerically with the mass profiles that result from Eqs.~\eqref{eq:Majorana_SSB}, \eqref{eq:SOPT},  and \eqref{eq:Tc}. As in Fig.~\ref{fig:PT_strength}, we chose $M_1=M_2/2=10^9$ GeV, and used two choices of Yukawa matrices $F$ compatible with the latest global fits to oscillation experiments, given by Eqs.~\eqref{eq:Yuk_fig2_solid} and \eqref{eq:Yuk_fig2_dashed}. The Yukawa $y_1$ was fixed at 1, so that the value of $\lambda$ is given implicitly by Eq.~\eqref{eq:Tc} following from the choice of $M_I^0/T_{\rm c}$. Fig.~\ref{fig:enhancement} illustrates that, as argued before, the asymmetry is indeed enhanced with growing $M_I^0/T_c$, saturating at the zero-washout limit given by the upper  dashed line. The growth is approximately, although not exactly, linear with the numerical calculations including flavour effects. Since washout can be flavour dependent, it can alter the sign of the asymmetry with respect to its value in the zero washout limit. As a fast second-order phase transition tends to suppress washout effects, the sign of the asymmetry can flip as $M_I^0/T_c$ grows, as seen in the dashed line of Fig.~\ref{fig:enhancement}. For the solid line, the choice of Yukawa matrix gives no strong flavour effects, washout affects all flavours evenly, and there is no change of sign in the asymmetry.

For $M_I^0/T_c\lesssim5$, the asymmetry coincides with that in the constant mass limit, given by the dotted lines in Fig.~\ref{fig:enhancement}. This is because the temperature at the onset of the phase transition is large enough for the $N_I$ to attain near-equilibrium abundances, and so the effect of the phase transition is erased. A sizeable enhancement of the asymmetry from a large departure of equilibrium after the phase transition therefore requires $M_I^0/T_c>10$. In this analysis, we do not include the effects of scattering or decays of the particle excitations of the $\Phi$ field, to which we turn in Sec.~\ref{sec:damping}.\\
 
{\bf Assumptions:}~The previous results were obtained with several simplifying assumptions. For example, we calculated the $\Phi$ VEV using the tree-level potential supplemented with the dominant high-temperature corrections.  However, as we show in Sec.~\ref{sec:PT}, quantum corrections can spoil the shape of the potential precisely in the region of large  $M_I^0/T_c$, leaving as an open question whether large enhancements of the asymmetry may occur in realistic scenarios.

In our study of second-order phase transitions, we have also assumed that entropy is conserved such that the yield, $Y_{N_I}/s$, is constant apart from collisions that change its number density according to the Boltzmann equation. To justify this assumption, as well as address apparent violations of energy conservation, we first expand on our treatment of the $N_I$ number densities in a first-order transition, and then return to the question of a second-order phase transition. In a first-order phase transition, the relevant dynamics is described by the nucleation and rapid expansion of bubbles of true vacuum. Such a process is driven by  quantum transitions between vacua and, being distinct from the processes arising from the adiabatic Hubble expansion, lead to many kinds of non-equilibrium processes including spatial field gradients, plasma disturbances ahead of the bubble wall, and bubble-wall collisions, all of which can lead to substantial changes in entropy. In this scenario, the expansion of the universe is typically negligible during the very short duration of the phase transition, resulting in a conservation of energy density that can be used to make arguments about the flux of RH neutrinos entering the bubble (which we claimed was suppressed) or the transfer of energies to the plasma leading to reheating/entropy generation.

On the other hand, in a second-order phase transition there is a smooth, homogeneous evolution of the scalar VEV. Because the scalar is assumed to be in equilibrium, there is a single value of $T$ (which changes inversely with the scale factor), and so the phase transition is driven by the Hubble expansion itself. Due to the expansion, the energy density is not conserved; however, the energetics of the phase transition are encoded in the scalar free energy, or finite-$T$ potential, and so the effects of the energy transfers between the scalar and RHN sectors due to the changing RHN mass have already been accounted for via their contributions to the finite-$T$ potential. 

In the second-order phase transition, we expect the number density of RHNs to change via collisions and decays, as specified in the Boltzmann equation. However, we do not expect the changing $\Phi$ field itself to modify the number density beyond the indirect $N_I$ depletion due to collisions:~the change in mass is somewhat faster than Hubble, but still very slow compared to the changes associated with a first-order phase transition or the values needed to give rise to background-field-induced particle production. The departure from equilibrium comes from a small lag in the RHN number density, which is not quite able to keep up with the equilibrium value; for the reasons outlined above, however, this is not expected to change the overall entropy because it is ultimately driven by the adiabatic expansion.

Finally, we  comment that all of the results in this section are derived by including only the RHN dynamics and treating the changes to the RHN masses as resulting from the variation of a background field. We are effectively tracing over the  degrees of freedom driving the phase transition, which results in an explicitly time-dependent mass term for the RHNs. It is therefore unsurprising that energy is added to the RHN system during the phase transition since time-translation invariance has been violated via a driving term. However, in Section \ref{sec:PT} we consider the dynamics of the scalar sector on equal footing with the RHNs, and in minimizing the free energy of the combined system we are able to account for the energy transfers between the scalar and RHN sectors.  Indeed, the  fine-tuning that we find in the scalar potential points to the challenge of realizing a phase transition that proceeds in the adiabatic manner used in Section \ref{sec:asym_PT} without either occurring too early (giving rise to a small $\langle\Phi\rangle / T_{\rm c}$ and negligible effect on the asymmetry) or else producing a barrier between vacua leading to a first-order phase transition.

%%%%%%%%%%%%%%%%%%%%%%%%%%%%%%%%%%%%%%%%%%%%%%%%%%%%%%%%%%%%%%%%%%%%%%%%%%%%%%%%%%%%%%%%%
%%%%%%%%%%%%%%%%%%%%%%%%%%%%%%%%%%%%%%%%%%%%%%%%%%%%%%%%%%%%%%%%%%%%%%%%%%%%%%%%%%%%%%%%%
%%%%%%%%%%%%%%%%%%%%%%%%%%%%%%%%%%%%%%%%%%%%%%%%%%%%%%%%%%%%%%%%%%%%%%%%%%%%%%%%%%%%%%%%%

%%%%%%%%%%%%%%%%%%%%%%%%%%%%%%%%%%%%%%%%%%%%%%%%%%%
%%%%%%%%%%%%%%%%%%%%%%%%%%%%%%%%%%%%%%%%%%%%%%%%%%%
%%%%%%%%%%%%%%%%%%%%%%%%%%%%%%%%%%%%%%%%%%%%%%%%%%%
\subsection{Slowly evolving scalar field}

If the $N_I$ masses induced by the symmetry-breaking scalar are time-dependent but very slowly varying, they can be considered constant throughout baryogenesis. Thus there will not be an enhanced departure from equilibrium. However, it does mean that the $N_I$ masses in the early universe  are not directly related to their values today. In particular, if we consider the case where the $N_I$ are actual RHNs giving rise to the observed SM neutrino masses, the value of $N_I$ at the time of baryogenesis may not be directly related to the small SM neutrino masses constrained by low-energy neutrino experiments or by the cosmic microwave background. As advocated in Ref.~\cite{Bi:2003yr}, this could in principle allow for exceptions to the Davidson-Ibarra bound on $N_I$ masses that apply for non-resonant, hierarchical leptogenesis scenarios \cite{Davidson:2002qv}. Such bounds require high reheat temperatures after inflation ($\gtrsim10^9$ GeV), which can be problematic in models with new, super-weakly coupled low-mass degrees of freedom:~for example, high reheat temperatures in supersymmetric models can imply a cosmologically disfavoured over-abundance of gravitinos \cite{Ellis:1984eq}. 

To understand whether a relaxation of the Davidson-Ibarra bound is permitted in models with slowly varying $N_I$ mass, we first review the origin of the bound, which arises from a relation between the $CP$-violating sources of the asymmetry, $\varepsilon_I^\alpha$, and the light SM neutrino masses. The physical reason for the bound is that, in the hierarchical limit and absent any cancellations in matrix products, $\varepsilon^\alpha_1$ is proportional to the square of the Yukawa matrix $F_{\alpha I}$ as seen in Eq.~\eqref{eq:CP_source}. Since the typical scale of the Yukawa couplings is $F^2 \sim m_\nu M_1 / v^2$, a smaller RHN mass gives a smaller source for the asymmetry. Quantitatively, one starts with Eq.~\eqref{eq:CP_source}, and using the usual see-saw relation between the SM neutrino masses, RH neutrino masses, and Yukawa couplings, one obtains 
\begin{equation}
\label{eq:lightm}
 m^\nu_{IJ}\approx \sum_K \frac{v^2}{M_K}\,F_{IK} F_{JK},
\end{equation}
where $v$ is the Higgs VEV. The Davidson-Ibarra bound applies in the hierarchical limit where leptogenesis is dominated by $N_1$. For  $x_{J1}=M_J^2/M_1^2\gg1$ in Eq.~\eqref{eq:CP_source}, the $CP$-violating source due to the decays of $N_1$  is \cite{Davidson:2002qv}
\begin{equation}
\label{eq:epsilon_1}
 \varepsilon_1^\alpha \approx \frac{3 M_1}{16\pi v^2(F^\dagger F)_{11}} \,{\rm Im}[F_{\alpha1} (m^\nu F)_{\alpha1}].
\end{equation}
This expression demonstrates  that, for fixed SM neutrino masses $m^\nu_{IJ}$ and a lower bound on $\varepsilon_1^\alpha$ from the requirement of successful baryogenesis, there exists a lower bound for $M_1$. 

We now turn to how leptogenesis is affected in models where the $M_I$ were different in the early universe than today. The simplest way to see that the $CP$-violating source is time independent is by referring to the original formulation in Eq.~\eqref{eq:CP_source}. There, the $N_I$ masses only appear in the ratio $M_J^2/M_I^2$, and hence a universal scaling of all Majorana masses in the early universe  does not change $\varepsilon_1^\alpha$. In Eq.~\eqref{eq:epsilon_1}, the same result holds due to the fact that the re-scaling of $M_1$ in the early universe is exactly compensated by the scaling of $m^\nu$. Therefore, the $CP$-violating source is time independent, in contradiction with the claim of Ref.~\cite{Bi:2003yr}.

The most important effect of a different mass for $N_I$ in the early universe is on the efficacy of washout processes. Recall that the dimensionless washout factor is
\be
\mathcal{K}_1^\alpha &=& \frac{\Gamma_{N_1}\,\mathrm{Br}(N_1\rightarrow\alpha)}{H(M_1)}.
\ee
Considering only the scaling due to mass, the width varies linearly with $M_1$ while the Hubble scale evaluated at $T=M_1$ varies quadratically with $M_1$ in a radiation-dominated universe. We therefore have for constant Yukawa couplings and varying mass,
\be
\label{eq:Kappascaling}
\mathcal{K}_1^\alpha \propto \frac{1}{M_1}.
\ee
 In turn, we have that the asymmetry in the strong washout regime is $Y_{\Delta L_\alpha}\sim 1/\mathcal{K}_1^\alpha$, and so
\be
Y_{\Delta L_\alpha} \propto M_1.
\ee
To summarize, for $N_1$ in the strong washout limit, a larger value for $M_1$ in the early universe results in a linear enhancement of the lepton asymmetry with the $N_1$ mass.

For $M_1$ that is sufficiently large, $\mathcal{K}_1^\alpha\lesssim1$ and leptogenesis occurs instead in the weak washout limit. Here, the asymmetry is proportional simply to $\varepsilon_1^\alpha Y_{N_1}(0)$, see Eq.~\eqref{eq:weakwashout}. In the weak washout limit, the asymmetry is sensitive to the primordial $N_1$ abundance since scattering with SM leptons is insufficient to establish an equilibrium abundance.  If some other particle couples to $N_1$ at $T\gg M_1$ such that it comes into thermal equilibrium, then $Y_{N_1}(0)=Y_{N_1}^{\rm eq}$. Since both $\varepsilon_1^\alpha$ and $Y_{N_1}(0)$ are independent of the early-universe value of $M_1$, then the asymmetry no longer changes with respect to $M_1$.
If instead the Yukawa couplings between $N_1$ and $L_\alpha$ provide the dominant interactions of $N_1$, then the abundance of $N_1$ at the time it begins decaying is completely determined by the Yukawa couplings, and hence $\mathcal{K}_1^\alpha$. For smaller $\mathcal{K}_1^\alpha$, fewer $N_1$ exist and can decay to produce an asymmetry. Therefore, we find that there is a maximum value of $M_1$ in the early universe corresponding to $\mathcal{K}_1^\alpha \approx 1$, and for larger $M_1$, the abundance at the time of decay drops and the asymmetry decreases once again.

%
%%%%%%%%%%%%%%% FIGURE %%%%%%%%%%%%%%%%%%%%%
\begin{figure}[t]
\centering
\includegraphics[width=0.55 \textwidth ]{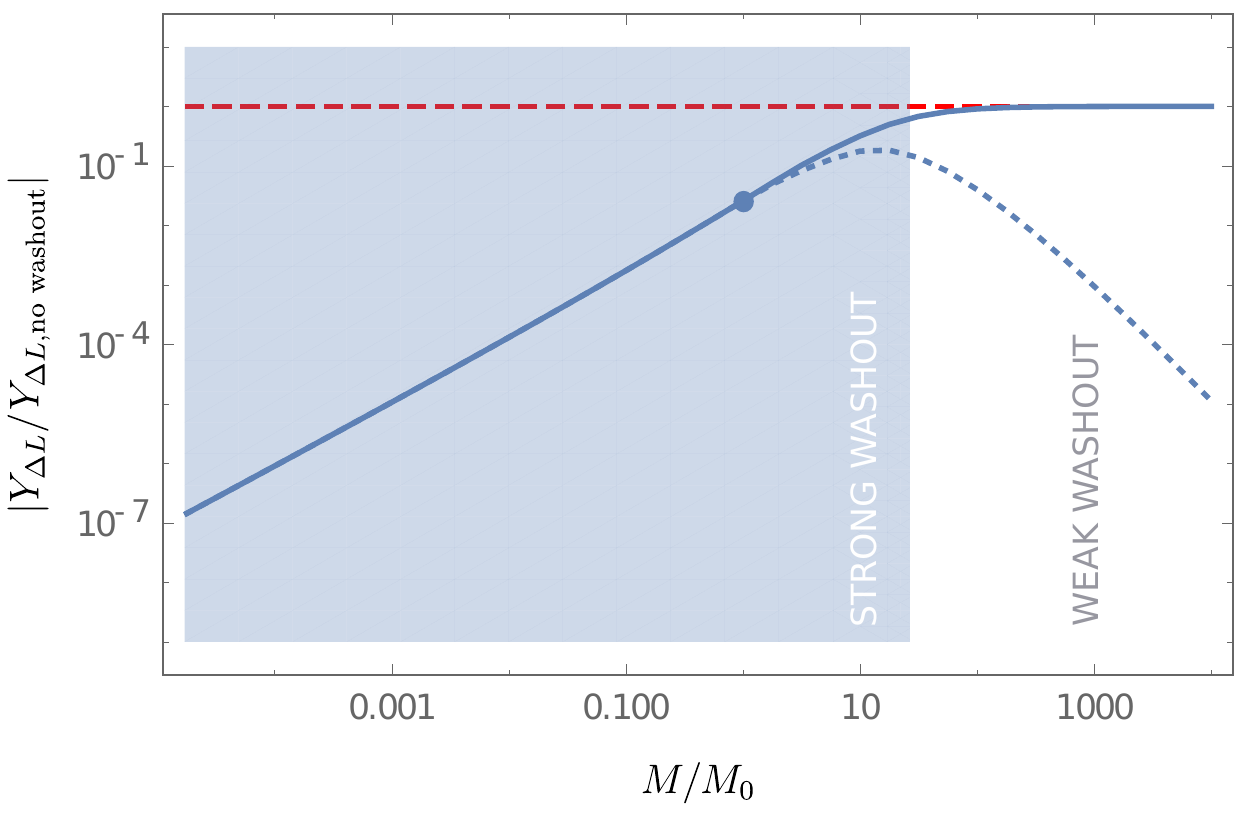}
\caption{Lepton asymmetry under a {\bf universal rescaling} of the RHN masses ($M_I$) in the early universe relative to today. The zero-temperature masses are $M_{N_1}^0=10^9$ GeV, $M_{N_2}^0=2\times10^9$ GeV, and we use a Yukawa matrix compatible with oscillation measurements, see Eq.~\eqref{eq:Yuk_fig2_solid}. The horizontal dashed line corresponds to the maximum possible asymmetry (achieved with zero washout and initial equilibrium abundances for the $N_I$).  The blue lines include washout effects:~the solid line corresponds to initial equilibrium abundances for the $N_I$, and the dotted blue line corresponding to zero initial abundances. The blue dot corresponds to no difference between the $M_I$ in the early universe and the present. The shaded blue region delimits the strong washout regime, $\mathcal{K}_I^\alpha > 1$ for some flavours $I$, $\alpha$.}
\label{fig:slowevolution}
\end{figure}
%%%%%%%%%%%%%%% FIGURE %%%%%%%%%%%%%%%%%%%%%
%

To verify this behaviour, we consider a concrete $3\times2$ flavour scenario where we fix the zero-temperature $N_I$ mass, $M_I^0$, and vary its mass at the time of leptogenesis. We use $M_{1}^0=10^9$ GeV, $M_{2}^0=2\times10^9$ GeV, and a  Yukawa matrix compatible with global fits to oscillation experiments, given by Eq.~\eqref{eq:Yuk_fig2_solid}. We then solve the Boltzmann equations numerically. We show the results in Fig.~\ref{fig:slowevolution}, and the figure clearly shows the linear dependence of the asymmetry with $M_1$ in the strong washout regime, its flattening at large $M_1$ if we assume a primordial equilibrium abundance, and its turnover and decrease at large $M_1$ if we assume the abundance originates only from scattering with SM leptons.

We note that for SM neutrino masses consistent with the Planck cosmological bound $\sum_i m^\nu_i<0.23$ eV \cite{Ade:2015xua}, and with recent fits to oscillation data yielding a largest mass difference of the order of 0.05 eV \cite{Esteban:2016qun}, the ratio $\Gamma_{N_1}\,\mathrm{Br}(N_1\rightarrow\alpha)/H(M_1^0)\gg1$, which  corresponds to the strong washout regime. This can be easily seen by substituting these numerical values into Eq.~\eqref{eq:lightm} and Eq.~\eqref{eq:KIM}.  In the strong washout regime,  an enhancement of the asymmetry requires larger Majorana masses in the early universe, $M_I>M_I^0$, which can arise if the symmetry-breaking scalar field rolls down from large to small field values. If leptogenesis occurs from a thermal abundance of $N_I$, then the asymmetry-enhanced scenario requires a larger reheat temperature than in conventional leptogenesis to populate the $N_I$ particles. This is at odds with the findings of Ref.~\cite{Bi:2003yr}.

To summarize, we reach a different conclusion  relative to Ref.~\cite{Bi:2003yr}, and the origin of the discrepancy appears to originate with an incorrect treatment of the $CP$-violating source in the referenced work. In Ref.~\cite{Bi:2003yr}, it is claimed that $\varepsilon$ is enhanced for $M_I< M_I^0$, which counteracts the effects of the stronger washout assuming $\mathcal{K}$ is not too large. Instead, we have shown that the dimensionless $\varepsilon$ is independent of the overall mass scales of $N_I$, and so a universal rescaling of all $M_I$ does not change $\varepsilon$. Indeed, in the strong washout regime expected from current measurements of SM neutrino masses \cite{deSalas:2017kay}, decreasing $M_I$ relative to $M_I^0$ only serves to enhance washout and suppress the value of the asymmetry. Therefore, having a smaller $M_I$ in the early universe does not enhance the lepton asymmetry, and cannot be used to circumvent the Davidson-Ibarra bound in the strong washout regime.

%%%%%%%%%%%%%%%%%%%%%%%%%%%%%%%%%%%%%%%%%%%%%%%%%%%%%%%%%%%%%%%%%%%%%%%%%%%%%%%%%%%%%%%%%
%%%%%%%%%%%%%%%%%%%%%%%%%%%%%%%%%%%%%%%%%%%%%%%%%%%%%%%%%%%%%%%%%%%%%%%%%%%%%%%%%%%%%%%%%
%%%%%%%%%%%%%%%%%%%%%%%%%%%%%%%%%%%%%%%%%%%%%%%%%%%%%%%%%%%%%%%%%%%%%%%%%%%%%%%%%%%%%%%%%

\section{Asymmetry Damping from New Annihilation Modes}\label{sec:damping}

To this point, we have considered the effects of a time-dependent  particle mass on the asymmetry generated from its decays. Realistically, such a time-varying mass originates from the dynamics of some scalar field(s), $\Phi$, breaking the $B-L$ symmetry. In realistic models, some or all of the components of $\Phi$ may have masses comparable or below the typical momentum scale of interactions in the plasma at the time of leptogenesis, and it is therefore important to consider the effects of $N_I$ interactions with $\Phi$ particles in determining the final lepton asymmetry. Such effects have been considered in Refs.~\cite{Gu:2009hn,Sierra:2014sta}, although our work additionally combines the effects of the time varying parent-particle mass with the effects of scattering between $N_I$ and $\Phi$.

\begin{figure}[t]
\centering
\includegraphics[width=0.7\textwidth ]{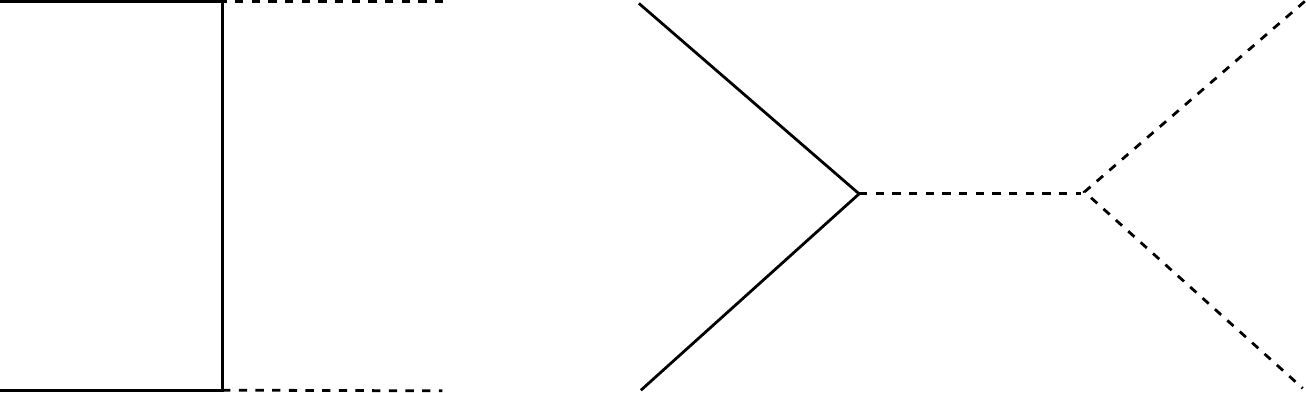}
\caption{Annihilation diagrams of the RHNs (solid lines) into scalars (dashed lines).}
\label{fig:annihilation}
\end{figure}

The Yukawa interaction between $N_I$ and $\Phi$  as specified in Eq.~\eqref{eq:LMajoron} leads to an irreducible scattering process $N_I \bar{N}_I \rightarrow \Phi^\dagger \Phi$ that can change the number density of $N_I$. This process is shown in the left pane of Fig.~\ref{fig:annihilation}. Additionally, the scalar potential in the broken phase typically contains cubic couplings of the scalar field components, and these lead to $N_IN_I\rightarrow \Phi\Phi$ scattering indicated by the diagram in the right pane of Fig.~\ref{fig:annihilation}. It is important to include these scattering processes in the Boltzmann equations, Eq.~\eqref{eq:Boltzmann}, whenever $M_\Phi \le M_I$.

We now argue that the scattering of $N_I$ into $\Phi$ is typically kinematically accessible and important if $\Phi$ is a complex scalar that breaks a continuous global symmetry, and if the conditions for asymmetry enhancement due to a ``fast'' second-order phase transition are realized (as outlined in Section~\ref{subsec:2ndorder}). Let us first assume that $\Phi$ is a complex field breaking a continuous global symmetry. There is at least one massless Goldstone mode, $\varphi$, in the broken phase and so $N_I$ annihilation into the Goldstone fields is always kinematically accessible. The only way that this process could be unimportant is if the Yukawa couplings, $y_I$, are all small. However, we have seen that a substantial modification of the asymmetry in a second-order phase transition requires
\be
\frac{M_I(T=0)}{T_{\rm c}} = \frac{y_I\Phi_0}{T_{\rm c}} \gg1.
\ee
Thus, the only way that $y_I$ can be small is if $\Phi_0\gg T_{\rm c}$. However, as we show in Section~\ref{sec:PT}, it is challenging to obtain large hierarchies in $\Phi_0/T_{\rm c}$, and so achieving $M_I^0/T_{\rm c}\gg1$ typically requires large $y_I$. We therefore find it likely that annihilations into at least some components of $\Phi$ are important during leptogenesis. Possible exceptions to this argument include scenarios with a discrete, rather than global, symmetry of a multi-field model; however, we still find that the requirement of a delayed phase transition typically leads to a relatively flat direction in the potential, which in turn suggests the existence of low-mass scalars to which $N_I$ can annihilate.

%%%%%%%%%%%%%%% FIGURE %%%%%%%%%%%%%%%%%%%%%
%%%%%%%%%%%%%%% FIGURE %%%%%%%%%%%%%%%%%%%%%

The effect of annihilations and their inverse processes was systematically studied in Ref.~\cite{Sierra:2014sta} in the scenario with a time-independent $N_I$ mass. In that work, it was pointed out that annihilations have two effects. First, when the $N_I$ are weakly coupled to the thermal bath and cannot otherwise reach a thermal abundance (such as in the weak washout regime), inverse annihilations open up new channels of production that increase the $N_I$ population and can enhance the asymmetry. On the other hand, once the $N_I$ can reach a thermal distribution at high temperatures, the generic effect of annihilations is to provide a lepton-number-preserving mode to relax the $N_I$ abundance to its equilibrium value, decreasing the asymmetry resulting from decays\footnote{Additionally, it may be possible that new modes for asymmetry generation can occur via $\Phi$ interactions with SM fields. This is highly model-dependent, however, and in the simplest scenario where $\Phi$ decays to SM fields via a small mixing with the SM Higgs, $\Phi$ effectively carries no baryon or lepton number.}.  As we have done throughout our paper, we  concentrate on the strong washout regime where the effects of a time-varying mass are most pronounced.

While we have argued that a ``fast'' phase transition typically implies light scalar degrees of freedom, more general parts of parameter space may feature $M_\Phi > M_I$. In this case, there are two effects of $N_I-\Phi$ scattering on the lepton asymmetry. First, $\Phi \Phi^\dagger\rightarrow N_I N_I$ scattering leads to an additional production mode of $N$ in the early universe as mentioned above, and so this additional production mode can lead to a larger asymmetry in the weak washout regime, Eq.~\eqref{eq:weakwashout}. Second, the inverse process $N_I N_I\rightarrow \Phi\Phi^\dagger$ can deplete the $N_I$ number density in a lepton-number-conserving manner, reducing the efficacy of leptogenesis from decays. However, the fact that the annihilation process is kinematically forbidden at zero momentum for $M_\Phi > M_I$ gives rise to a Boltzmann suppression of the annihilation rate, $\sim e^{-(M_\Phi - M_I)/T }$. Our results for the asymmetry damping from new annihilation modes in Section \ref{sec:ann_analytic} can be easily extended to the case of $M_\Phi > M_I$ by multiplying all annihilation rates by this Boltzmann factor\footnote{There are additional, non-Boltzmann-suppressed semi-annihilation modes mediated by off-shell $\Phi$, such as $N_I N_I\rightarrow N_I L_\alpha H$, but these are suppressed by the Yukawa coupling $F_{\alpha I}$ and three-body phase space, and are subdominant to the $N_I$ decays.}. In the limit $M_I \ll M_\Phi$, the $\Phi$ particles decouple and the asymmetry prediction reverts to the scenario with no additional annihilation modes studied in Section \ref{sec:asym_PT}.

\subsection{Boltzmann Equations with Annihilations}

In the presence of new annihilation modes $N_I N_I\rightarrow \phi\phi$ (where $\phi$ stands in for any of the scalar components of $\Phi$), the Boltzmann equation for the $N_I$ number density is
\be
\label{eq:BNann}
\frac{dY_{N_I}}{dt} =& -\langle \Gamma_{N_I}\rangle\left(Y_{N_I}-Y_{N_I}^{\rm eq}\right) - {2\,s(z)}\langle\sigma_{N_IN_I\rightarrow\phi\phi}v\rangle\left[Y_{N_I}^2-\left(Y_{N_I}^{\rm eq}\right)^2\right].
\ee
We use the convention where the thermally averaged cross section includes a symmetry factor of $1/2$ for identical initial states. Because the $N_I$ obtain their masses from the same $y_I$ couplings that mediate the annihilation terms, we need only concern ourselves with on-diagonal annihilations. Assuming $Y_{N_I}\approx Y_{N_I}^{\rm eq}$ (as is true in the strong washout regime and before inverse processes decouple), we can define a rate 
\be
\Gamma^{\rm ann}_{I} \equiv 4\langle\sigma_{N_IN_I\rightarrow\phi\phi}v\rangle s(z) Y_{N_I}^{\rm eq}.
\ee
In this limit, we have
\be\label{eq:dampingrate}
\frac{dY_{N_I}}{dt} &\approx& -\left[\langle \Gamma_{N_I}\rangle + \Gamma^{\rm ann}_I\right]\left(Y_{N_I}-Y_{N_I}^{\rm eq}\right).
\ee
It is clear that, if $\Gamma^{\rm ann}_I \gtrsim \langle\Gamma_{N_I}\rangle$, then the change in $N_I$ abundance is dominated by annihilations, and it is important to include the annihilation term when solving the Boltzmann equations.

The effects of annihilations on the lepton asymmetry can be found by substituting Eq.~\eqref{eq:dampingrate} into the Boltzmann equation for $Y_{\Delta L_\alpha}$, namely the second line in Eq.~\eqref{eq:Boltzmann}. We then have as our equation for the asymmetry, $Y_{\Delta L_\alpha}$,
\be\label{eq:asymmetrydamping}
\frac{dY_{\Delta L_\alpha}}{dt} &=& - \sum_I\,\varepsilon_I^\alpha\,\frac{\langle\Gamma_{N_I}\rangle\,\mathrm{Br}(N_I\rightarrow\alpha)}{\langle\Gamma_{N_I}\rangle+ \Gamma_I^{\rm ann}}\,\frac{dY_{N_I}}{dt} - \Gamma_{\rm W}^\alpha\,Y_{\Delta L_\alpha},
\ee
where again $\Gamma_{\rm W}^\alpha$ is the washout rate. The solution to this equation has an integral form analogous to Eq.~\eqref{eq:BE_solution}. Based on our earlier arguments, we know that the asymmetry is predominantly generated for times $t>t_*$, where $\Gamma_{\rm W}^\alpha(t_*) = H(t_*)$. We can identify two limits of the solution, depending on the relative magnitudes of $\langle\Gamma_{N_I}\rangle$ and $\Gamma_I^{\rm ann}$ at $t_*$:
\begin{enumerate}
\item If $\langle\Gamma_{N_I}\rangle(t_*) > \Gamma_I^{\rm ann}(t_*)$, then annihilations are subdominant to decays, and the full Boltzmann equation, Eq.~\eqref{eq:asymmetrydamping}, reduces to the simpler case with no annihilations included and the asymmetry is the same as before.
\item If $\langle\Gamma_{N_I}\rangle(t_*) < \Gamma_I^{\rm ann}(t_*)$, then we see that the lepton asymmetry is not efficiently produced because most of the $N_I$ disappear via annihilation into $\phi$ instead of decays into $L_\alpha$. We expect that the resulting lepton number asymmetry is approximately suppressed by the ratio $\langle\Gamma_{N_I}\rangle(t_*) / \Gamma_I^{\rm ann}(t_*)$. Often, it is said that ``$N_I$ is kept more in equilibrium'' by the annihilation processes.
\end{enumerate}
It should be noted that $\Gamma^{\rm ann}_I\propto Y_{N_I}^{\rm eq}$, and so the annihilation rate decreases exponentially as a function of time for $T\ll M_I$ (as is familiar, for example, from the thermal freeze-out of dark matter annihilations). Consequently, the  annihilations play less of a role in determining the final lepton asymmetry for a later decoupling of washout.

Broadly speaking, the above arguments only apply to individual flavours.  There is always the possibility of nontrivial flavour effects  which could, for instance, enhance the total asymmetry in the presence of annihilations. For example, one can consider a situation with weak washout in which the sum of $CP$-violating sources is zero due to some lepton flavour symmetries, and the total asymmetry would vanish due to a cancellation of the  asymmetries in different flavours. If annihilation rates are not flavour-universal, then the  flavoured asymmetries would be suppressed by different factors, the cancellation would be spoiled and a net asymmetry would arise.

\subsection{Analytic Estimate of Annihilation Rates}\label{sec:ann_analytic}

In order to understand when annihilations might be important for the lepton asymmetry, we may consider the leading analytic dependence of each of the washout and annihilation rates, as the importance of annihilations depends on their relative size. Considering a model where $N_I$ obtains a mass through interactions with a symmetry-breaking complex scalar as in Eq.~\eqref{eq:LMajoron}.  In the broken phase, the scalar $\Phi$ decomposes into radial and angular modes, $\Phi = (v_\phi + \phi_{\rm r})e^{i\phi_{\rm i}/v_\phi}/\sqrt 2$. When scalar masses and cubic interactions are small, the annihilation of $N_I$ proceeds mainly through  the first family of diagrams in Fig.~\ref{fig:annihilation}; for simplicity, we consider the annihilation into $\phi_{\rm r}$ for our analytic estimates but the full result is not qualitatively different. This annihilation rate of $N_I$ into two scalars is velocity suppressed  due to the negative intrinsic parity of the initial state of two Majorana fermions. In the small-velocity  and small-$\phi_{\rm r}$-mass expansions, we find:
\begin{equation}
\label{eq:sigmalows}
 \sigma_{N_I N_I\rightarrow \phi_{\rm r}\phi_{\rm r}}(s)=\frac{3y_I^4\sqrt{s-4M_I^2}}{128\pi M_I^3}+{\cal O}(s-4M_I^2)^{3/2},
\end{equation}
and the full result is given in Appendix \ref{app:XS}. 
To find $\Gamma^{\rm ann}_I$, we must compute the thermally averaged cross section. This involves an integration over $s$ with an exponentially suppressed weight for $M_I/T\gg1$ (as seen in Appendix \ref{app:Tav}), and so it is permissible to evaluate the integral using the velocity expansion. The result in the $M_I/T\gg1$ limit is
\begin{equation}
 \Gamma^{\rm ann}_{I}\approx \frac{9 \sqrt{2}\, y_I^4 \,T^{5/2}}{128 \pi^{5/2}M_I^{3/2}}e^{-M_I/T}.
\end{equation}
This can now be compared with the decay rate.

For $T \ll M_I$, the thermal averaging has no effect on the width of $N_I$ (see Appendix \ref{app:Tav}), and so we can simply use the zero-temperature width from Eq.~\eqref{eq:Gamma},
\be
\Gamma_{N_I} &=& \frac{(F^\dagger F)_{II}\, M_I}{8\pi}.
\ee
Taking the ratio then gives
\begin{equation}
\label{eq:Gammadamp}
 \frac{\Gamma^{\rm ann}_I}{\langle \Gamma_{N_I}\rangle}\sim \frac{9 \sqrt{2}\,y_I^4}{16\pi^{3/2} (F^\dagger F)_{II}}z^{-5/2}_Ie^{-z_I},
\end{equation}
where we use the usual dimensionless quantity $z_I \equiv M_I/T$. The time at which annihilations become subdominant to decays, $z_I^{\rm a}$, is defined implicitly by $\Gamma_I^{\rm ann}(z_I^{\rm a})/\Gamma_{N_I}(z_I^{\rm a}) \equiv 1$.

We need to evaluate Eq.~\eqref{eq:Gammadamp} at the time when washout interactions decouple. Let us consider for simplicity the single-flavour limit, in which the $N_I$ have a hierarchical spectrum such that only annihilations and decays of $N_1$ are relevant. Using Eqs.~\eqref{eq:washout} and \eqref{eq:Gamma}, we have
\begin{equation}
\label{eq:GammaWH}
 \frac{\Gamma_{\rm W}^\alpha(T)}{H(T)}\sim \frac{\sqrt{\pi}\,\mathcal{K}^\alpha_1}{4\sqrt{2}}z_1^{7/2} e^{-z_1}.
\end{equation}
The condition that $\Gamma_{\rm W}^\alpha(z_1^*) / H(z_1^*) = 1$ defines a time $z_1^*$. 

We can compare the time at which annihilations become subdominant to decays, $z_1^{\rm a}$, to the time at which washout decouples, $z_1^*$. In particular, let us consider the case of SM neutrino masses generated using a simple see-saw mechanism. Using the see-saw relations, the Yukawa couplings $F\sim (m^\nu M_1 / v^2)^{1/2}$ and $m^\nu \approx 0.1$ eV, it is straightforward to use Eqs.~\eqref{eq:Gammadamp} and \eqref{eq:GammaWH} to find that $z_1^* = z_1^{\rm a}$ for $M_1\sim10^6-10^7$ GeV. For higher $N_1$ masses, the Yukawa couplings are large enough that the decay dominates over annihilations at $z_1^*$, and so annihilations are not important during the epoch of asymmetry generation. By contrast, for $M_I\lesssim10^6-10^7$ GeV, the Yukawa couplings are small enough that annihilations are important during asymmetry generation and suppress the efficacy of leptogenesis. In the most conventional regime of hierarchical leptogenesis in models satisfying the Davidson-Ibarra bound \cite{Davidson:2002qv}, annihilations are never important. These overall conclusions can be different in models where the Yukawa couplings are much larger than the na\"ive expectation due to cancellations among entries \cite{Casas:2001sr}:~in this case, $F$ is larger than expected and annihilations play less of a role than in the na\"ive see-saw.

For non-hierarchical  $N_I$, one expects the individual washout rates to be comparable. Given the stronger overall washout at late times for a given value of $M_1$, washout  processes tend to decouple at later times for a degenerate spectrum, and so one expects that annihilations can have a somewhat smaller effect than in hierarchical scenarios with a similar value of $M_1$. The only way to determine precisely whether they are important is to calculate the relative annihilation, decay, and washout rates considering all flavours and see which processes decouple first.

We now turn to the question of how the above arguments change in the presence of a time-dependent mass. Specifically, we are most interested in how annihilations can affect the asymmetry with a second-order phase transition, as we saw in Section~\ref{subsec:2ndorder} that this was the type of phase transition that could lead to an enhanced asymmetry. The effect of the second-order phase transition is to make the time of washout freeze-out earlier, because the mass is changing with a rate faster than the characteristic Hubble expansion. However, we see that making $z_1^*$ occur earlier only leads to an increase in the asymmetry provided $z_1^{\rm a} < z_1^*$; otherwise, the $N_1$ abundance is damped by annihilations and the resulting asymmetry is not as enhanced as would otherwise have been anticipated. We see, therefore, that the annihilations of $N_I$ into $\phi$ tend to decrease the enhancement associated with a second-order phase transition.

\subsection{Numerical Analysis}

To illustrate the previously discussed features, we perform a numerical analysis including the effects of annihilations. We consider a $3\times2$ flavour system, with two values of Yukawa couplings to the scalar $\Phi$ ($y_1=1$ or $y_1=0.3$), and including the full cross sections for the annihilation of $N_I$ into all the complex scalar components, both in  the broken and unbroken phases. We provide details of the calculation in Appendix~\ref{app:XS}. \\

\noindent {\bf Time-Independent Masses:}~First we consider  the case where the VEV of $\Phi$ is time-independent, and we assume equilibrium boundary conditions at $T=100M_1$. 
We start from two choices of the Yukawa matrix, $F^h_7$ and $F^d_7$ --given in equations \eqref{eq:Yuk_fig5} and \eqref{eq:Yuk_fig5deg}-- which, for $M_1=10^7$ GeV, are compatible with oscillation data for hierarchical ($F^h_7$) and degenerate ($F^d_7$) scenarios. Then  we vary $M_1$, and make two different choices for $F_{\alpha I}$: either $F=\sqrt{M_1 / 10^7\,\,\mathrm{GeV}}\,F^{h/d}_7$, which keeps $m^\nu$ constant, or $F=F^{h/d}_7$. The masses of the scalars are taken to be zero (corresponding to negligible self-interactions of the scalar).

\begin{figure}[t]
\centering
\includegraphics[width=0.48\textwidth]{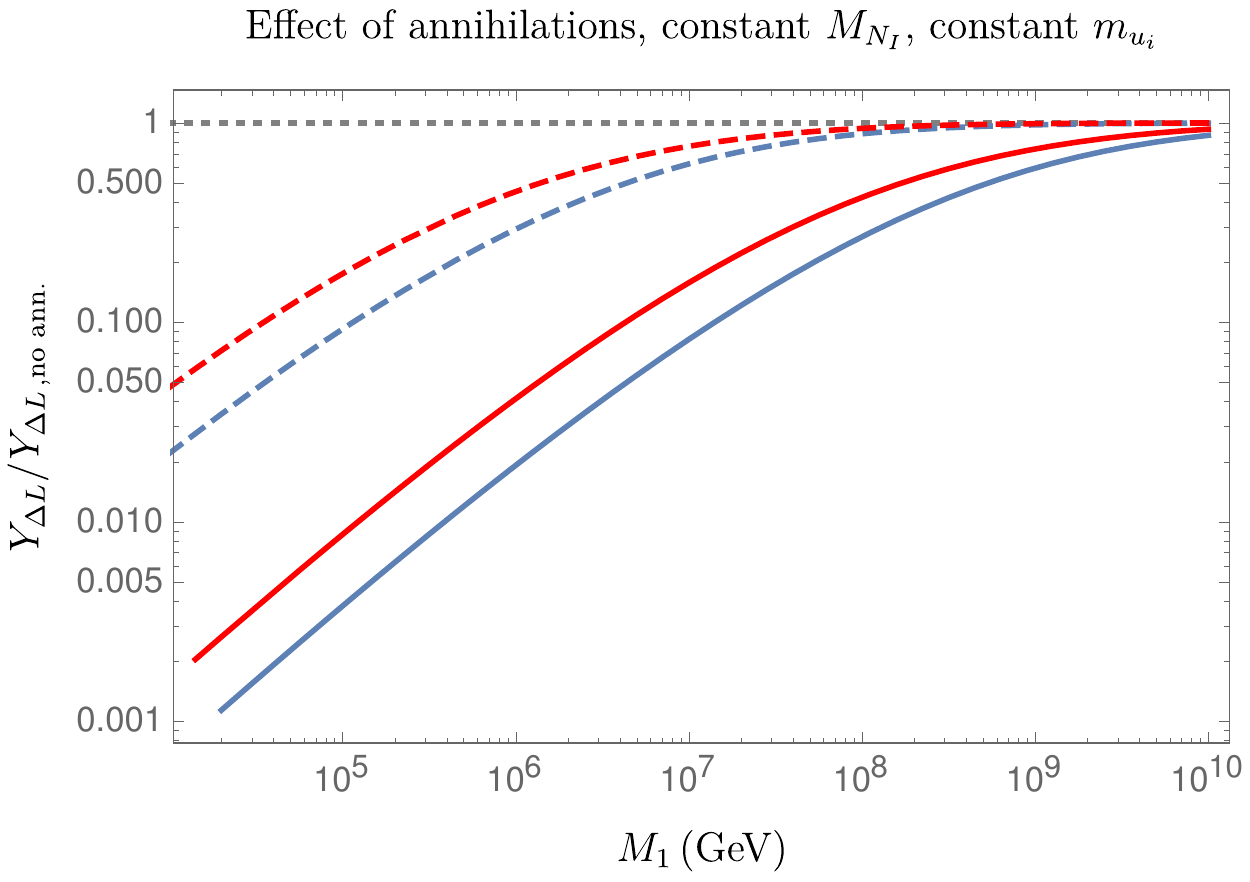}
\includegraphics[width=0.48\textwidth]{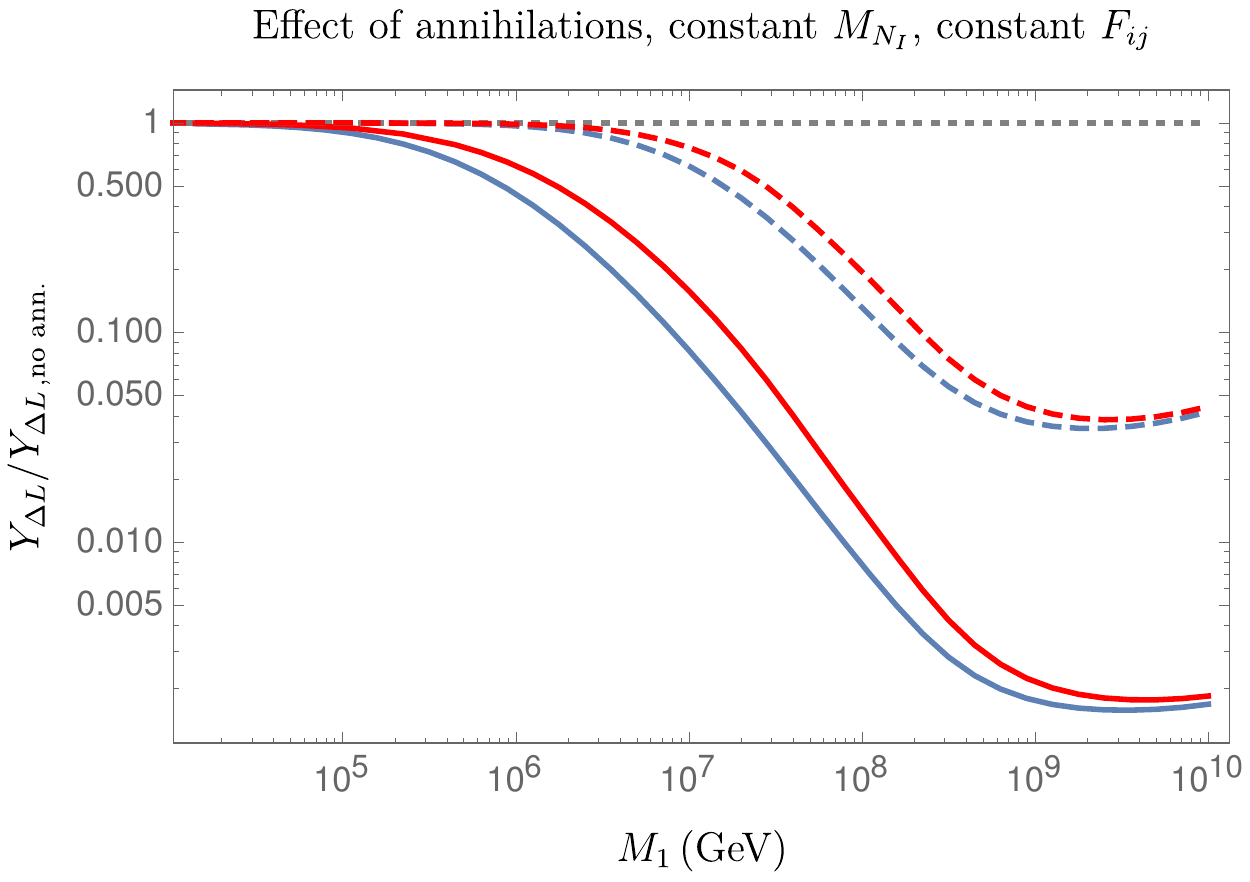}
\caption{Ratio of the total lepton asymmetry {\bf including annihilations vs.~without annihilations}, plotted as a function of $M_1$. We consider time-independent $M_I$,  two flavours of $N_I$ and three flavours of $L_\alpha$. On the left plot, the leptonic Yukawa couplings $F_{\alpha I}$ are rescaled under changes of $M_1$ to keep the  masses of the SM $\nu$ particles constant and $\lesssim 0.2$ eV. On the right plot, we kept $F_{\alpha I}$ constant as described in the text.  We assumed either $M_2=2 M_1$ (hierarchical scenario,  blue lines) or $(M_2-M_1)/M_1=10^{-6}$ (resonant scenario, red lines). Solid lines correspond to $y_1=1$, and dashed coloured lines to $y_1=0.3$.}
\label{fig:annihilationplots}
\end{figure}

Our results for the variation of the asymmetry as a function of $M_1$ and for our choice of $F^{h/d}_7$  are shown in Fig.~\ref{fig:annihilationplots}. In all cases, the effects of annihilations are less pronounced for $y_1=0.3$ (dashed coloured lines) than for $y=1$ (solid lines). Also, as anticipated earlier annihilations are more relevant for a hierarchical spectrum (blue lines, with $M_2=2 M_1$), than for a degenerate one (red lines, with $(M_2-M_1)/M_1=10^{-6}$). This is due to the larger washout rate in the degenerate case. 

In the left pane of Fig.~\ref{fig:annihilationplots}, we show the results for the scenario in which $M_1$ and $F$ are correlated to preserve the see-saw relation, which in turn affects the degree to which annihilations are important relative to decays. For smaller $M_1$, we see that annihilations are more important due to the smaller couplings $F$, and consequently the asymmetry is suppressed at these smaller masses. 

In the right pane of Fig.~\ref{fig:annihilationplots}, the lepton Yukawa couplings $F$ are kept at a constant value. In this case, increasing $M_1$ makes the washout weaker in relative terms according to Eq.~\eqref{eq:Kappascaling} and \eqref{eq:GammaWH} so that washout decoupling happens at earlier times. This increases the range of temperatures at which  annihilations can affect the asymmetry, and thus they more dramatically suppress the asymmetry. The asymmetry suppression becomes maximal when washout becomes irrelevant for all $T\lesssim M_1$, in which case the asymmetry is purely determined by the moment in which annihilations become subdominant with respect to decays; the corresponding value of $z^{\rm a}_1$ does not depend on $M_1$ for constant $F$ (see Eq.~\eqref{eq:Gammadamp}) and thus the curves in the right pane of Fig.~\ref{fig:annihilationplots} become flat. \\

%%%%%%%%%%%%%%% FIGURE %%%%%%%%%%%%%%%%%%%%%
%%%%%%%%%%%%%%% FIGURE %%%%%%%%%%%%%%%%%%%%%

\noindent {\bf Time-Dependent Masses:}~We now turn to the case in which the VEV of the mass-originating scalar $\Phi$ and the associated particle masses are temperature dependent as a consequence of a second-order phase transition. As in Section \ref{subsec:2ndorder}, we focus on the model of a single symmetry-breaking scalar with the interactions of Eq.~\eqref{eq:LMaj}, and use the values for the $\Phi$ VEV and the particle masses in the  high-temperature expansion from  Eq.~\eqref{eq:Majorana_SSB}, \eqref{eq:SOPT},  and \eqref{eq:Tc}, as well as the results for the scalar masses in Appendix \ref{subapp:ann:single}. For computing the thermally averaged annihilation cross sections, we distinguish the broken and unbroken phases, with the zero-temperature cross sections given in Appendices \ref{subapp:XS:broken} and \ref{subapp:XS:unbroken}, respectively.  The thermally averaged cross section, calculated as detailed in Appendix \ref{app:Tav}, diverges at the critical temperature when the RHNs and scalars are massless. This divergence is regularized by a resummation of thermal contributions to the $N_I$ propagator.  However, leptogenesis occurs for $T<M_I(T)$, in which case the thermal contributions to $N_I$ propagation are subdominant to the tree-level mass and can be neglected.

\begin{figure}[t]
\centering
\includegraphics[width=0.48\textwidth]{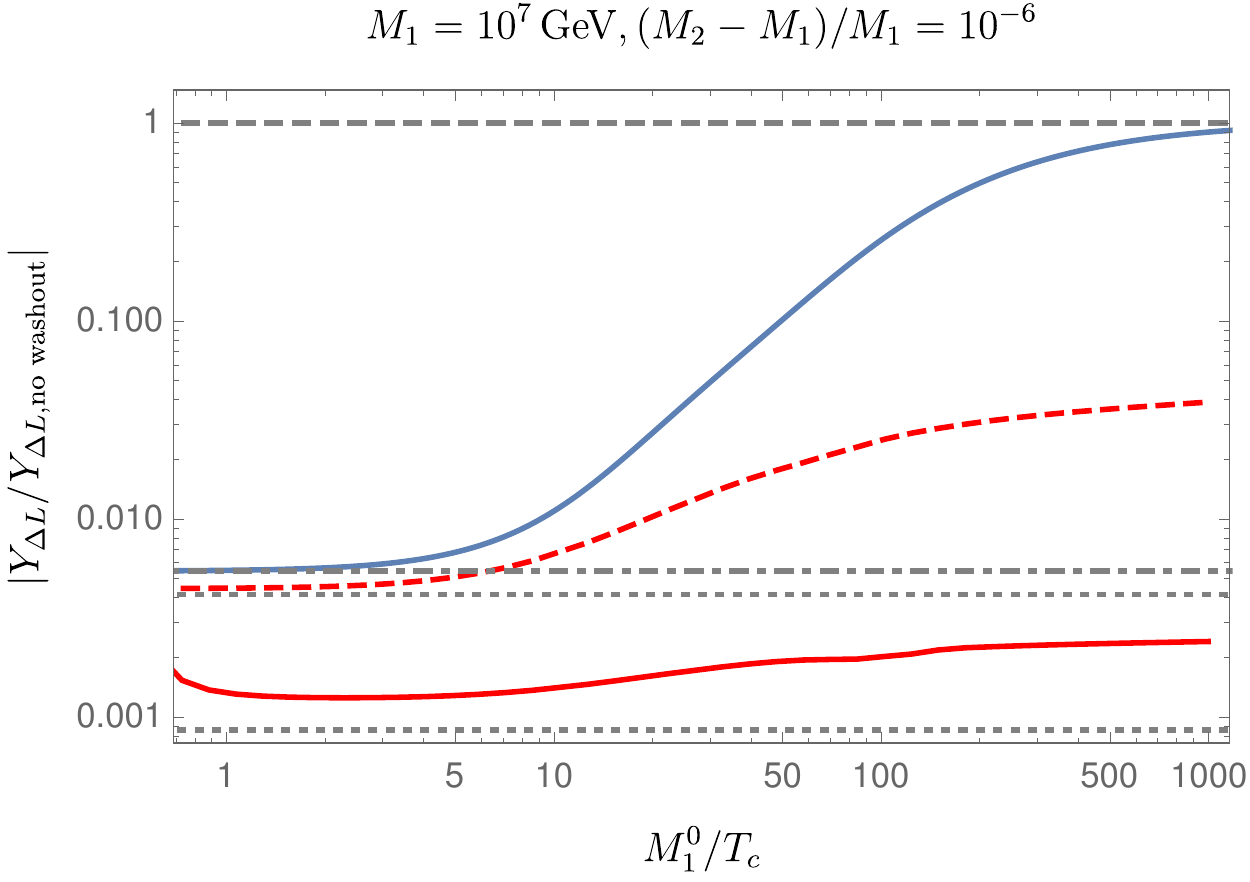}
\includegraphics[width=0.48\textwidth]{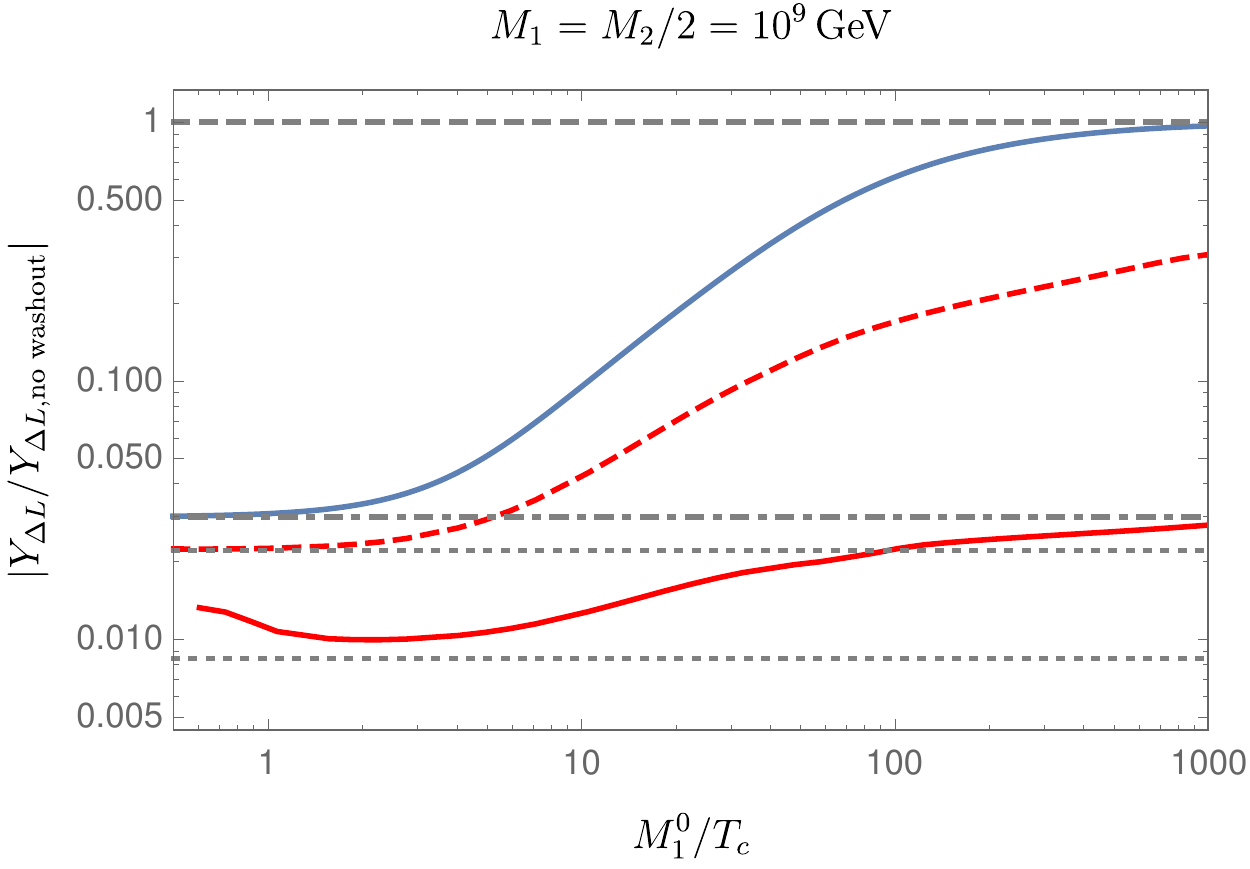}
\caption{Ratio of the total lepton asymmetry vs.~its value in the zero washout limit, plotted as a function of $M^0_1/T_c$, for two flavours of $N_I$ and three flavours of $L_\alpha$. For the left pane, $M_1=10^7$ GeV, $(M_2-M_1)/M_1=10^{-6}$, while on the right pane $M_1=M_2/2=10^9$ GeV. The leptonic Yukawa couplings are compatible with oscillation data. The solid blue lines neglect annihilations, while we include annihilations in the red lines with  $y_1=1$ (solid red), and $y_1=0.3$ (dashed red). The horizontal lines represent, from top to bottom, the zero washout limit in the absence of annihilations, the constant-mass limit in the absence of annihilations, followed by the constant-mass limits for the cases without annihilations for $y_1=0.3$ and $y_1=1$, respectively.}
\label{fig:2ndorderannihilationplots}
\end{figure}

We show the results of the numerical calculations in Fig.~\ref{fig:2ndorderannihilationplots}. On the left pane, we consider a degenerate spectrum, with $M_1=10^7$ GeV, $(M_2-M_1)/M_1=10^{-6}$, while in the right pane we have $M_1=M_2/2=10^9$ GeV. In both cases we used  Yukawa matrices consistent with oscillation data and given in Eq.~\eqref{eq:Yuk_fig5deg} (left pane) and Eq.~\eqref{eq:Yuk_fig2_solid} (right pane) in appendix \ref{app:Fs}. The matrices have the property that the asymmetry for a time-independent mass and negligible annihilations has the same sign regardless of whether or not washout processes are active (when neglecting annihilations, the asymmetry has the behaviour of the solid line in figure \ref{fig:enhancement}). The solid blue lines give the behaviour of the asymmetry when annihilations are neglected, and the red lines include the effect of annihilations for $y_1=1$ (solid red), and $y_1=0.3$ (dashed red). The horizontal lines represent, from top to bottom, the zero washout limit without annihilations, and then the constant mass limits for the case without annihilations, for $y_1=0.3$, and for $y_1=1$. Note that annihilations thwart the enhancement of the asymmetry due to the second-order phase transition, the effect being more pronounced for larger $y_I$ and lower $M_1$.

%%%%%%%%%%%%%%% FIGURE %%%%%%%%%%%%%%%%%%%%%
%%%%%%%%%%%%%%% FIGURE %%%%%%%%%%%%%%%%%%%%%

\section{Realistic Phase Transitions and Baryogenesis} \label{sec:PT}

In the previous sections, we have found that the baryon asymmetry resulting from the out-of-equilibrium decay of a particle $N_I$ can be affected by a phase transition in its mass, $M_I$. In particular, we found that an enhancement of the asymmetry is possible in models with fast second-order phase transitions, while the asymmetry may suffer a suppression due to additional annihilation modes. The baryon asymmetry was calculated assuming a particular time-dependent mass profile for $N_I$ and, in particular, taking the zero-temperature mass, $M_I^0$, and critical temperature, $T_{\rm c}$, as free parameters. We found that for second-order phase transitions to give a substantial enhancement of the baryon asymmetry, it was necessary to have $M_I^0 / T_{\rm c} \gtrsim10$ (see Fig.~\ref{fig:enhancement}). 

We now turn to addressing the question of how such time-dependent mass profiles can be obtained in realistic, perturbative models of spontaneous symmetry breaking. As we show below, the large value of $M^0_I/T_c\gg 1$ needed for an appreciable enhancement of the asymmetry is not a generic feature of scalar potentials and only results from tuned parameters in the potential which can be destabilized by quantum corrections. Below, we first consider the case of symmetry breaking in single-scalar models in Section~\ref{subsec:single}. Noting that symmetry breaking patterns can be substantially different in multi-field models, we then study two-field models in Section~\ref{subsec:multiple}.

\subsection{\label{subsec:single}Single-Field Models}

The simplest model of $N_I$ achieving a mass through spontaneous symmetry breaking is if the symmetry-breaking sector consists of a single scalar field, $\Phi$, which is responsible for giving rise to the RHN mass. The tree-level potential was given in Eq.~\eqref{eq:LMaj}:
\be
 \mathcal L_{\rm mass}\supset -\frac{y_I}{2}\,\Phi\, \overline{N}_I^{\rm c} N_I + \mathrm{h.c.}+m_\Phi^2\Phi^\dagger\Phi-\frac{\lambda}{4}(\Phi^\dagger\Phi)^2.
\ee
This tree-level potential is corrected by finite-density effects in the early universe. If we consider only the leading $T^2$ terms in the finite-temperature potential resulting from a high-temperature expansion (see Appendix~\ref{app:VeffT}), the relation between zero-temperature $N_I$ mass and $\Phi$ VEV is:
\begin{align}
\label{eq:MTc}
 \frac{M^0_I}{T_c}=\frac{y^2_I}{12\lambda}\left(2\lambda+\sum_J y^2_J\right).
\end{align}
This result suggests that, in order to achieve $M^0_I/T_{\rm c}\gg1$, one needs $y_I^4\gg\lambda$ for at least one flavour $N_I$. This limit is problematic:~for example, radiative corrections to the $\Phi$ quartic coupling from loops of $N_I$ scale like $y_I^4$, and so in this limit radiative corrections can dominate over the tree-level contributions to the potential. This suggests, at the very least, the necessity of a cancellation between tree- and loop-induced contributions to the potential that realize the relation $y_I^4 \gg \lambda$ for the renormalized couplings.

In reality, the situation is  worse than a fine tuning of parameters. The reason is that the renormalized quartic coupling can be small at only a single scale as a result of fine tuning. Renormalization group (RG) effects modify the quartic coupling at other scales in the potential, and large Yukawa couplings $y_I$ can de-stabilize the minimum of the potential under RG evolution. This effect is, for example, well-appreciated in the SM and has received renewed interest with the recent measurements of the Higgs boson and top quark masses \cite{Degrassi:2012ry,Buttazzo:2013uya}; in our case, the computation is simpler because we do not need to concern ourselves with questions of gauge invariance in the effective potential and tunnelling calculations \cite{Metaxas:1995ab,Plascencia:2015pga}.

To account for these effects, we compute the RG-improved effective potential with the RG scale set to  a VEV-dependent quantity. In order to minimize logarithmic corrections, the latter can be chosen as the largest particle mass in the $\Phi$ background \cite{Ford:1992mv}, which for $\mathcal{O}(1)$ couplings coincides with $|\Phi|$. Choosing then $\mu=|\Phi|$, in the limit of large field values the quartic coupling becomes
\be
\label{eq:lambdaeff}
V_4&=&\frac{\lambda_{\rm eff}(|\Phi|)}{4}(\Phi^\dagger\Phi)^2, \\
|\Phi|\frac{\partial\lambda_{\rm eff}(|\Phi|)}{\partial|\Phi|}&\equiv&\beta_\lambda(|\Phi|)=-\frac{1}{4\pi^2}\sum_I y^4_I+{\cal O}(\lambda y^2_I).
\ee
 For a reference scale $\mu_0$, the effective quartic coupling can be approximated as,
\begin{align}
\label{eq:lefflog}
 \lambda_{\rm eff}(|\Phi|)=\lambda(\mu_0)+\beta_\lambda\log\frac{|\Phi|}{\mu_0},
\end{align}
which can also be  directly derived from the large-field expansion of the Coleman-Weinberg potential, Eq.~\eqref{eq:V0}, in Appendix~\ref{app:VeffT}. 
For $y_I$ not too large with respect to $\lambda$, the effect of $\beta_\lambda<0$ is to make the potential negative at large field values where $\lambda_{\rm eff}(|\Phi|)$ crosses zero. The result is a local, metastable minimum for $\Phi$ at small field values, and a global minimum at large field values, much like in the metastable case of the SM Higgs potential. This is not necessarily a problem, as quantum tunnelling and thermal transitions to the unstable region are typically extremely suppressed if it happens at sufficiently large field values. Even if the origin/zero-temperature metastable ``vacuum'' are not the true minima of the potential for any $T$, the presence of the distant true vacuum can be irrelevant, with baryogenesis  proceeding as expected and the metastable vacuum surviving throughout the history of the Universe\footnote{The suppressed quantum and thermal tunnelling out of the metastable vacuum are again analogous to those of the Higgs' electroweak vacuum in the SM \cite{Degrassi:2012ry}. Although the existence of the unstable region could be problematic during inflation, due to enhanced quantum of light fields fluctuations in the presence of curvature, the field can be stabilized with nonminimal gravitational interactions that enhance the effective mass in the presence of curvature \cite{Espinosa:2007qp}.}.

  For larger tree-level values of $y_I^4/\lambda$,  $\lambda_{\rm eff}(|\Phi|)$ becomes negative even for values of $|\Phi|\sim m_\Phi$ and the effective potential may not have a  local minimum in the vicinity of $m_\Phi$ at all. Since according to Eq.~\eqref{eq:MTc} this is the coupling regime expected to give large $M_I^0/T_{\rm c}$, we expect that an upper bound on $M_I^0/T_{\rm c}$ may be derived from the requirement of the existence of a local minimum of the potential near $m_\Phi$ at zero temperature. We do this by selecting a set of Yukawa couplings, $y_I$, and finding the value of the quartic couplings, $\lambda$, at which the metastable vacuum disappears. This corresponds to a maximum value of $M_I^0/T_{\rm c}$.
  
  %%%%%%%%%%%%%%% FIGURE %%%%%%%%%%%%%%%%%%%%%
\begin{figure}[t]
\centering
\includegraphics[width=0.49\textwidth ]{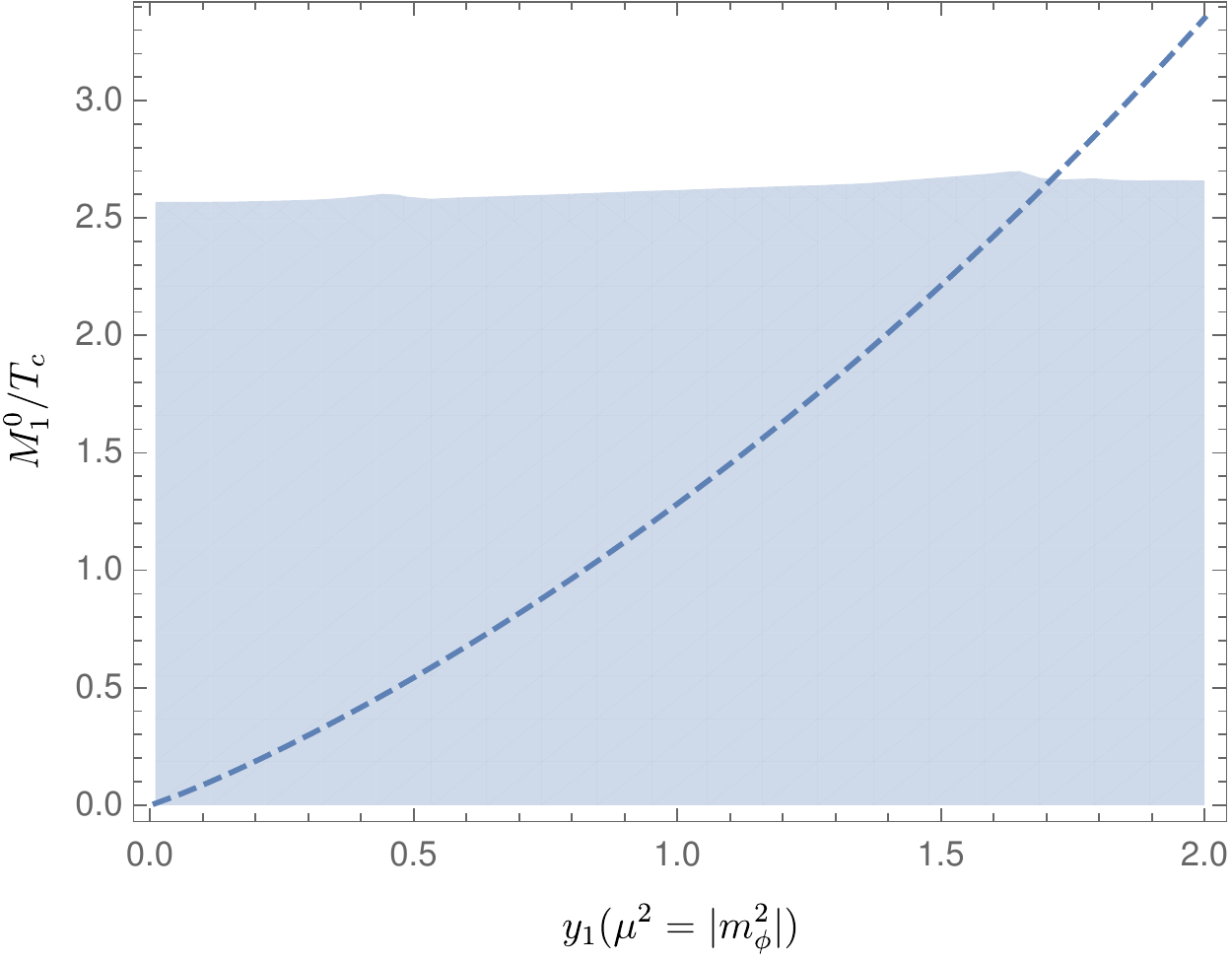}
\includegraphics[width=0.49\textwidth ]{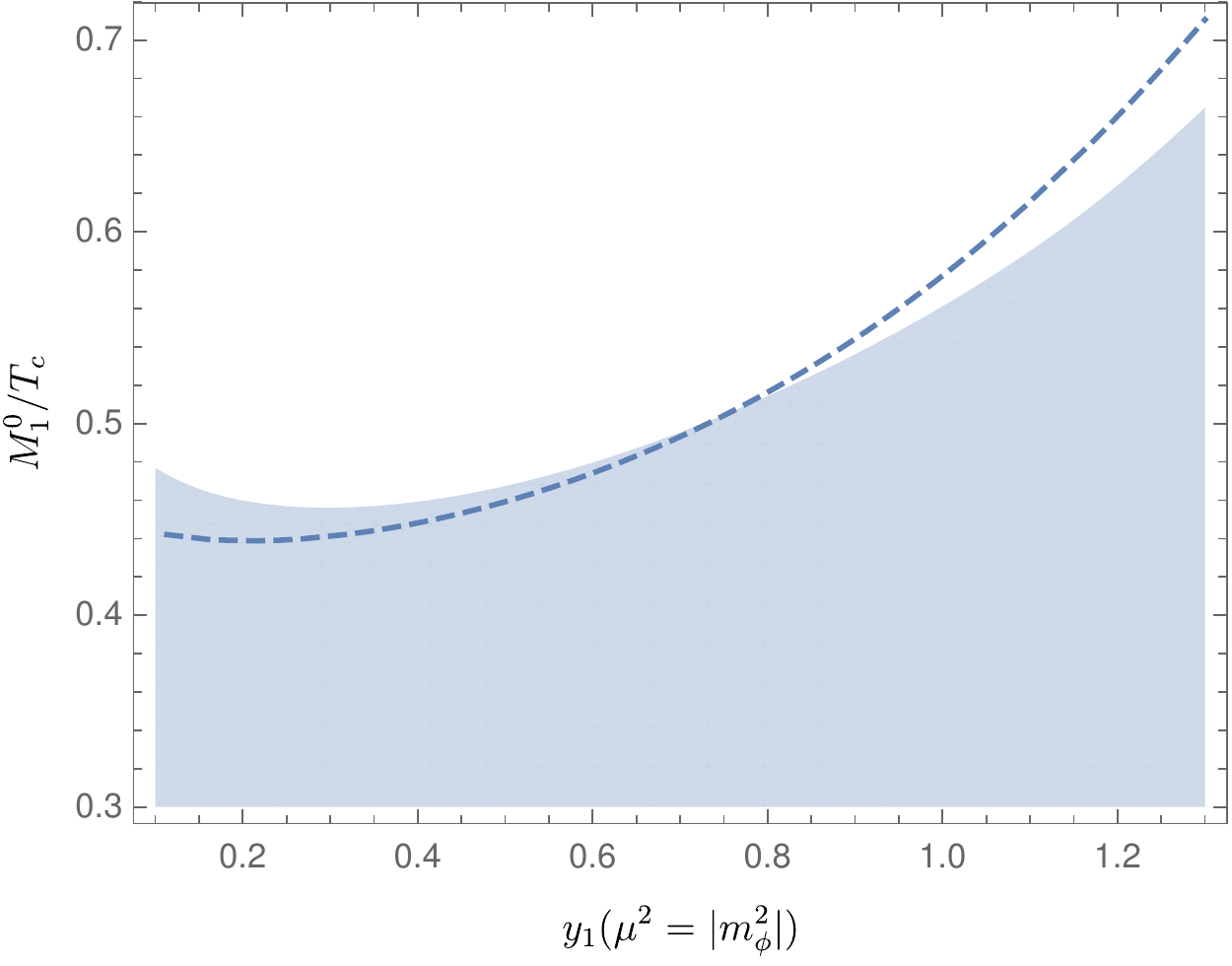}
\caption{Allowed regions for $M^0_I/{T_c}$ in single-scalar models with degenerate $N_I$ and a constant zero-temperature VEV $\langle\Phi\rangle$. The shaded regions show allowed parameters with the RG-improved full one-loop effective potential (including thermal Daisy resummation \cite{Pisarski:1988vd}), while the dashed lines show the  relationship between $M_I^0/T_{\rm c}$ and $y_I$  obtained with the effective potential ignoring zero-temperature loop corrections and keeping only terms in the high-$T$ expansion. Left pane:~upper bounds found by demanding the existence of a local metastable minimum at field values $\sim m_\Phi$, and zero-temperature VEV $\langle\Phi\rangle\sim 6\times 10^9$ GeV. Right pane:~upper bounds obtained by demanding stability of the scalar potential up to the Planck scale, and zero-temperature VEV $\langle\Phi\rangle\sim 0.7\times 10^9$ GeV. }
\label{fig:MTc}
\end{figure}
%%%%%%%%%%%%%%% FIGURE %%%%%%%%%%%%%%%%%%%%%
  
   In the left pane of Fig.~\ref{fig:MTc},  we show the upper bound on $M_I^0/T_{\rm c}$ for the the single-scalar model of Eq.~\eqref{eq:LMaj}  as a function of the Yukawa couplings, $y_I$. For simplicity, we present the results for a quasi-degenerate spectrum of $N_I$, so that all of the $y_I$ are set equal to one another, although similar results hold for hierarchical spectra. To hold the zero-temperature VEV fixed, we set $m^2_\Phi(\mu=m^2_\Phi)=y_I^4  (5\times10^{8})^2\,{\rm GeV}^2$, which gives an approximately $y_I$-independent value of $\langle\Phi\rangle\sim6\times10^9$ GeV in the metastable vacuum (when it exists).  The shaded area on the  plot  shows the bound obtained from the full one-loop potential with thermal corrections including a Daisy resummation (see Appendix~\ref{app:VeffT}) \cite{Pisarski:1988vd}. For comparison, we show with the dashed line the point at which the zero-temperature metastable vacuum disappears in the high-temperature expansion and ignoring any zero-temperature quantum corrections; as is evident, this approximation fails to give the correct upper bound for large values of the Yukawa coupling where higher-order corrections are important to include. 
   
A stricter upper bound on $M^0_I/{T_c}$ can be obtained by requiring a positive effective quartic up to field values of order the Planck scale  (\emph{i.e.,} up to $M_{\rm P}=m_{\rm P}/\sqrt{8\pi}\approx2.4\times10^{18}$ GeV). This gives a minimum value of $\lambda$ ensuring stability, $\lambda_{\rm min}$. Assuming $m^2_\Phi=\lambda_{\rm min}(5\times10^8)^2 {\rm GeV}^2$, which gives a $y_I$-independent  local VEV around $0.7\times10^9$ GeV, the ensuing bound on $M^0_{N_1}/{T_c}$ is illustrated on the right plot of figure \ref{fig:MTc}. The high-temperature approximation works much better at deriving this bound because the requirement of stability up to the Planck scale requires that zero-temperature quantum corrections remain suppressed near the metastable vacuum.  For a second-order phase transition to a local minimum with $\langle\Phi\rangle\sim m_\Phi$, the high-temperature expansion  for all background-dependent masses is expected to give reasonably accurate results.

Even in the most optimistic case in which the local vacuum is not separated  from the unstable region by a barrier (as illustrated in the left pane of Fig.~\ref{fig:MTc}), it is clear that the allowed values of 
 $M^0_I/{T_c}$ are far from the regime in which one expects an enhanced baryon asymmetry due to a rapid transition, $M^0_{I}/{T_{\rm c}}\gtrsim10$. Are there any scenarios in which this conclusion may be evaded? One may expect that larger values of $M^0_I/T_{\rm c}$ could be allowed in theories where radiative corrections are suppressed or cancel among different fields, as is typical in supersymmetric models. For example, if we consider a potential containing additional scalars $S$ possessing mixed quartic couplings to $\Phi$, then loops of $S$ give positive contributions to the running of $\lambda$ that cancel the running due to the Yukawa couplings, $y_I$. An exact cancellation is expected in a models where the various couplings are related by supersymmetry. However, such a scenario poses a new problem:~the new couplings of the bosons $S$ to $\Phi$ induce a non-analytic cubic term in the effective potential, resulting in a thermal barrier between the origin and the metastable vacuum (see Appendix~\ref{app:VeffT}):
 \begin{align}
 \label{eq:VT3}
  V^T(\Phi)\supset -\frac{T}{12\pi} (m^2_S(\Phi))^{3/2}.
 \end{align}
 With such a barrier, the phase transition to the broken vacuum will be first order, rather than second order. As argued in Section \ref{sec:1PT},  first-order phase-transitions with large $\langle\Phi\rangle/T_c$ strongly inhibit, rather than enhance, the  production of a baryon asymmetry through $N_I$ decays.
 
 An exception to the above reasoning is if the $\Phi$-dependent terms in Eq.~\eqref{eq:VT3}  are subdominant to the $\Phi$-independent terms in Eq.~\eqref{eq:VT3} for $\Phi$ values near the metastable vacuum. This occurs if, for instance, $S$ has a large tree-level mass, $m^2_S\gg \lambda_{S\Phi}\langle\Phi\rangle^2$, where $\lambda_{S\Phi}$ is the mixed quartic coupling between $S$ and $\Phi$. In this limit the thermal corrections can be approximated by even powers of $\Phi^\dagger\Phi$, which generate no barrier:
 \begin{align}
(m^2_S(\Phi))^{3/2}=(m^2_S)^{3/2}\left[1+\frac{3}{2}\lambda_{S\Phi}\frac{\Phi^\dagger\Phi}{m^2_S}+\frac{3}{8}\lambda^2_{S\Phi}\left(\frac{\Phi^\dagger\Phi}{m^2_S}\right)^2+{\cal O}\left(\frac{\Phi^\dagger\Phi}{m^2_S}\right)^3\right].
 \end{align}
 However, physically we know that large values of $m_S^2$ correspond to a decoupling of $S$ from the spectrum, and the decoupling theorems ensure that the physical result reproduces the single-scalar limit. The inconsistency is found by noting that the earlier results of the $S$ contribution to the effective potential were derived assuming the dominance of $\log\Phi$ terms in the effective quartic coupling $\lambda_{\rm eff}$, which cease to dominate the potential  when other scales such as $m_S^2$ becomes large. We have confirmed with numerical calculations in a model with additional complex scalars that, indeed, no sizeable enhancement of $M_I^0/T_{\rm c}$ is found with respect to the single-scalar case when one requires a second-order phase transition.

In conclusion, models in which the scalar field, $\Phi$, is the only field undergoing a second-order phase transition do not naturally accommodate large values of $M^0_I/{T_c}$. Thus one cannot obtain sizeable enhancements of the baryon asymmetry with respect to the constant-mass case in a single-field model; instead, typical models  yield an asymmetry  that is at most comparable to that in constant mass models, and may be smaller due to the effects of annihilations.

%%%%%%%%%%%%%%%%%%%%%%%%%%%%%%%%%%%%%%%%%%%%%%%%%%%%%%%%%
%%%%%%%%%%%%%%%%%%%%%%%%%%%%%%%%%%%%%%%%%%%%%%%%%%%%%%%%%
%%%%%%%%%%%%%%%%%%%%%%%%%%%%%%%%%%%%%%%%%%%%%%%%%%%%%%%%%
%%%%%%%%%%%%%%%%%%%%%%%%%%%%%%%%%%%%%%%%%%%%%%%%%%%%%%%%%
%%%%%%%%%%%%%%%%%%%%%%%%%%%%%%%%%%%%%%%%%%%%%%%%%%%%%%%%%

\subsection{\label{subsec:multiple}Two-Field Models}

In the previous section, we assumed that the breaking of the baryon or lepton symmetry was the result of the dynamics of a single scalar field. There is, however, no reason to assume that the breaking of such a symmetry is so simple; in the SM, for example, there exist multiple sources of electroweak symmetry breaking, namely the Higgs scalar and the chiral condensate. In this section, we study the implications of additional symmetry-breaking fields on the phase transition.

In particular, we focus on the scenario of a multi-step phase transition, where the transition to the $\langle\Phi\rangle\neq0$ vacuum proceeds from another minimum in field-space, rather than from the origin. This results in a qualitatively different dependence of the phase transition temperature, $T_{\rm c}$, relative to the single-field case due to the fact that $T_{\rm c}$ no longer results directly from the temperature-dependent effective mass of $\Phi$, but rather depends on the temperature at which the two vacua become degenerate and the transition is allowed. In the case of a single-field model, we found that a delayed phase transition (with post-phase-transition mass large compared to $T_{\rm c}$) required large couplings to increase the finite-temperature corrections, which in turn led to large radiative corrections that destabilized the minimum. For a multi-field model, parametrically small values of $T_c$ may instead result due to small, zero-temperature energy differences between the minima. 

Multi-field transitions have already been  studied in the literature \cite{Land:1992sm,Hammerschmitt:1994fn,Cui:2011qe,Patel:2012pi,Blinov:2015sna}. The previous investigations have focused on scalars responsible for the breaking of electroweak symmetry, but the results can be qualitatively applied to the case of two singlets, one of them ($\Phi$) being responsible for breaking baryon or lepton number. We therefore consider a two-field model consisting of $\Phi$ and a real scalar, $\varphi$. $\Phi$ is the only field directly carrying baryon or lepton number, and so it alone couples to the RH neutrinos, $N_I$, of our simplified model of leptogenesis. The tree-level Lagrangian is
\begin{align}
\label{eq:LMajS}
{\cal L}\supset-\frac{y_I}{2}\,\Phi\, \overline{N}_I^{\rm c} N_I+{\rm h.c.}+m^2_\Phi |\Phi|^2+\frac{m^2_\varphi}{2} \varphi^2-\frac{\lambda_\Phi}{4}|\Phi|^4-\frac{\lambda_\varphi}{4!}\varphi^4-\frac{\lambda_{\Phi \varphi}}{2}|\Phi|^2 \varphi^2.
\end{align}
We are interested in the limit in which there occurs first a phase transition to a vacuum with $(\Phi,\varphi)=(0, v_\varphi)$, which can be either second or first order. This is followed by a second-order phase transition in which $\Phi$ acquires a VEV (and $\varphi$ may or may not have a VEV). At high temperatures, the $(0, v_\varphi)$ vacuum should be preferred over configurations with nonzero $\Phi$, implying  that thermal corrections predominate along the $\Phi$ axis. This follows naturally from the condition that only $\Phi$ couples to the thermal bath of the  $N_I$.

The requirement that the final phase transition is second order implies that there should be no barriers generated between the $(0,v_\varphi)$ minimum and the minimum with nonzero $\Phi$ at the time of the transition. In particular, there should exist no tree-level barrier between the vacua so that the minimum in the $\varphi$ axis should be unstable (a saddle point) at zero temperature. We are further interested in a situation in which the critical temperature of this second transition can go parametrically to zero, ensuring a large value of $M_{I}^0/T_{\rm c} = y_I\langle\Phi_0\rangle/T_{\rm c}$. We consider as an example the case where the metastable and true zero-$T$ vacua are parametrically unrelated:~in this case, the energy splitting between them is arbitrary, and in the case where the vacua are nearly degenerate, the temperature at which the intermediate vacuum becomes unstable can be very small.
This is the case if the $\Phi\neq0$ vacuum 
is aligned with the $\Phi$ axis, since  the energies of the unstable  $(0,v_\varphi)$ and the true $(v_\Phi,0)$ minimum are determined at tree-level by the independent ratios $m^4_\Phi/\lambda_\Phi$ and  $m^4_\varphi/\lambda_\varphi$. 

To realize this scenario, we study tree-level potentials with  an unstable minimum in the $\varphi$ direction and a stable minimum aligned with the $\Phi$ axis, along with a smooth valley of decreasing energy connecting them.  This is ensured if the following conditions are satisfied \cite{Cui:2011qe,Blinov:2015sna}:
\begin{align}
\label{eq:2fieldeq}
\frac{m^2_\varphi}{m^2_\Phi}\frac{\lambda_\Phi}{2}\leq \lambda_{\Phi \varphi}\leq\frac{m^2_\Phi}{m^2_\varphi}\frac{\lambda_\varphi}{3}.
\end{align}
Being lifted by thermal corrections, the $(v_\Phi,0)$ ``vacuum'' has higher energy than that at $(0,v_\varphi)$ at finite temperature. If the vacua are nearly degenerate at $T=0$, their energies  cross at a very small value of $T$, suggesting that one may achieve large values of $M_{I}^0/T_{\rm c}$. A critical temperature for baryogenesis is that at which the $(0,\varphi)$ vacuum destabilizes and the complex scalar $\Phi$ can start developing a VEV ($T_{\rm c}\equiv\tilde T_\varphi$). As shown in Section \ref{sec:1PT}, a first-order phase transition leads to a suppressed asymmetry, so we consider second-order phase transitions. To avoid energy barriers resulting in a first-order phase transition, $\tilde T_\varphi$ must be larger than the temperature $\tilde T_\Phi$ at which the $(v_\Phi,0)$ critical point becomes a stable, local minimum. Using the high-temperature expansion, it can be shown (see Appendix \ref{app:VeffT} for details) that the $\tilde T_\varphi\rightarrow0$ limit, together with the condition   $\tilde T_\varphi>\tilde T_\Phi$ for a second-order phase transition between the $(0,v_\varphi)$  and $(v_\Phi,0)$ minima, is achieved when both of the inequalities in Eq.~\eqref{eq:2fieldeq} are saturated.  As expected, the temperature $\tilde T_\varphi$ at which the lepton-number-breaking field  $\Phi$ acquires a VEV goes to zero parametrically in this degenerate limit.

Crucially, the above arguments do not rely on large Yukawa couplings between $N_I$ and $\Phi$ to realize a fast and delayed second-order phase transitions. The conclusions of the previous section, where radiative corrections destroy the stability of the vacuum, therefore do not apply to multi-field models. However, quantum corrections can still be important in correctly modelling the phase transition:~in the limit where Eq.~\eqref{eq:2fieldeq} are saturated, the shape of the potential between the minima is quite flat at tree level and becomes sensitive to quantum corrections, however small. In particular, such corrections can spoil the existence of a line of decreasing energy connecting the minima. One of two things can occur to spoil the phase transition. First, a zero-temperature barrier may appear, which  prevents a second order phase transition and inhibits asymmetry generation for large $M_I^0/T_{\rm c}$. Second, a new global minimum with concurrent, nonzero values of $\Phi$ and $\varphi$ might develop. In the latter case, the energy of this minimum stops depending on the energy of the  $(0,v_\varphi)$ saddle-point, and consequently the critical temperature of the second phase transition can no longer become parametrically close to zero. Although a second-order phase transition is possible in this case, an upper bound on $M_{I}^0/T_c$ is obtained.

We examine the effects of radiative corrections on phase transitions in the simplest two-field model, Eq.~\eqref{eq:LMajS}. The quantum corrections arise predominantly due to the Yukawa couplings between $\Phi$ and $N_I$. It is illustrative to consider the path of least energy connecting the minima aligned with each of the $\Phi$ and $\varphi$ axes, which is computed by parameterizing the field space in polar coordinates
\begin{align}
\varphi= r\cos\theta, \quad \Phi = r \sin\theta,
\end{align}
and calculating the energy along the minimal path as
\begin{align}
 V_{\rm min}(\theta)=\min_r V(r,\theta).
\end{align}
Considering only the tree-level potential, we consider parameters in Eq.~\eqref{eq:LMajS} giving rise to a smooth path of monotonically decreasing energy at zero temperature between the $(0,v_\varphi)$ and $(v_\Phi,0)$ minima, meaning that $V_{\rm min}(\theta)$ is a monotonically decreasing function of $\theta$ between 0 and $\pi$. We then include radiative corrections that may spoil the monotonic behaviour when $V_{\rm min}(0)$ and  $V_{\rm min}(\pi)$ become sufficiently degenerate. It should be noted that, in our analysis, we do not concern ourselves with issues of fine tuning (which such degeneracies necessarily involve in the absence of a symmetry) and simply ask how radiative corrections would spoil the shape of the potential; we turn later to a discussion on the question of how symmetries may alter this perspective.

%%%%%%%%%%%%%%% FIGURE %%%%%%%%%%%%%%%%%%%%%
\begin{figure}[t]
\centering
\includegraphics[width=0.49\textwidth ]{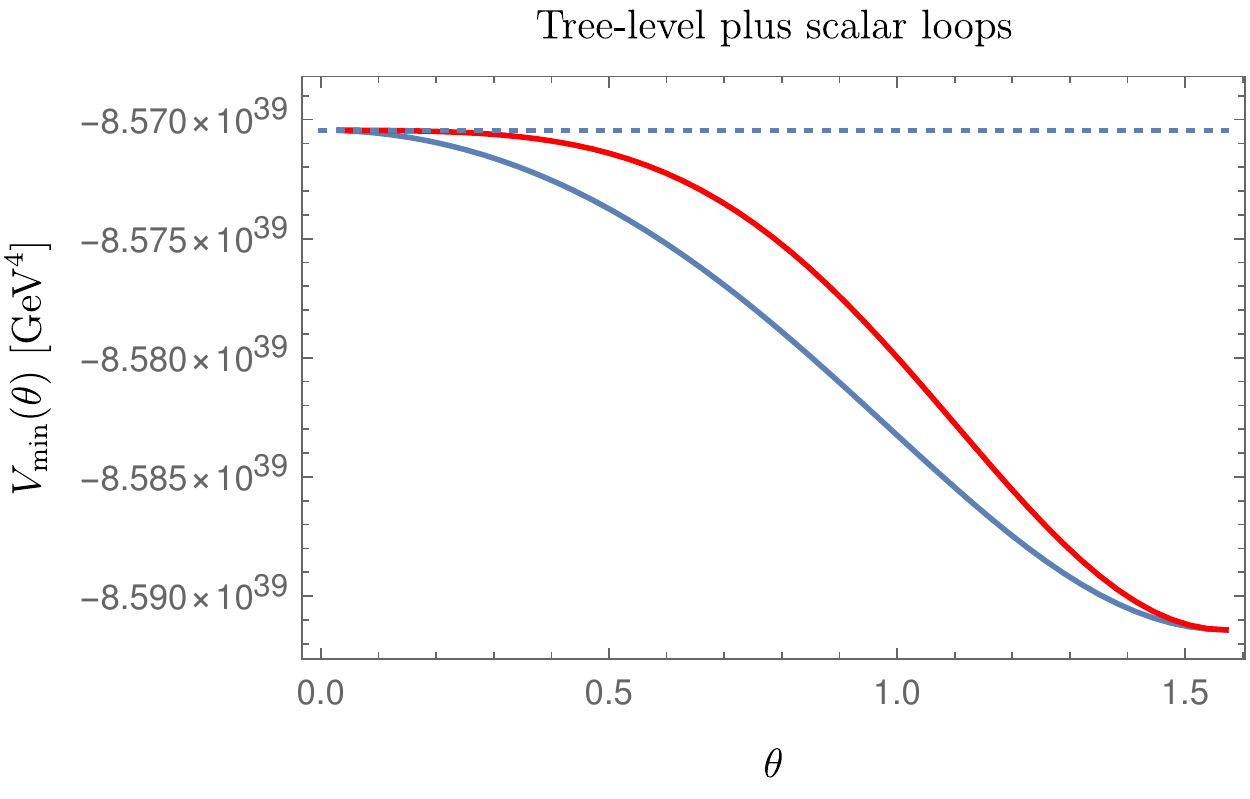}
\includegraphics[width=0.49\textwidth ]{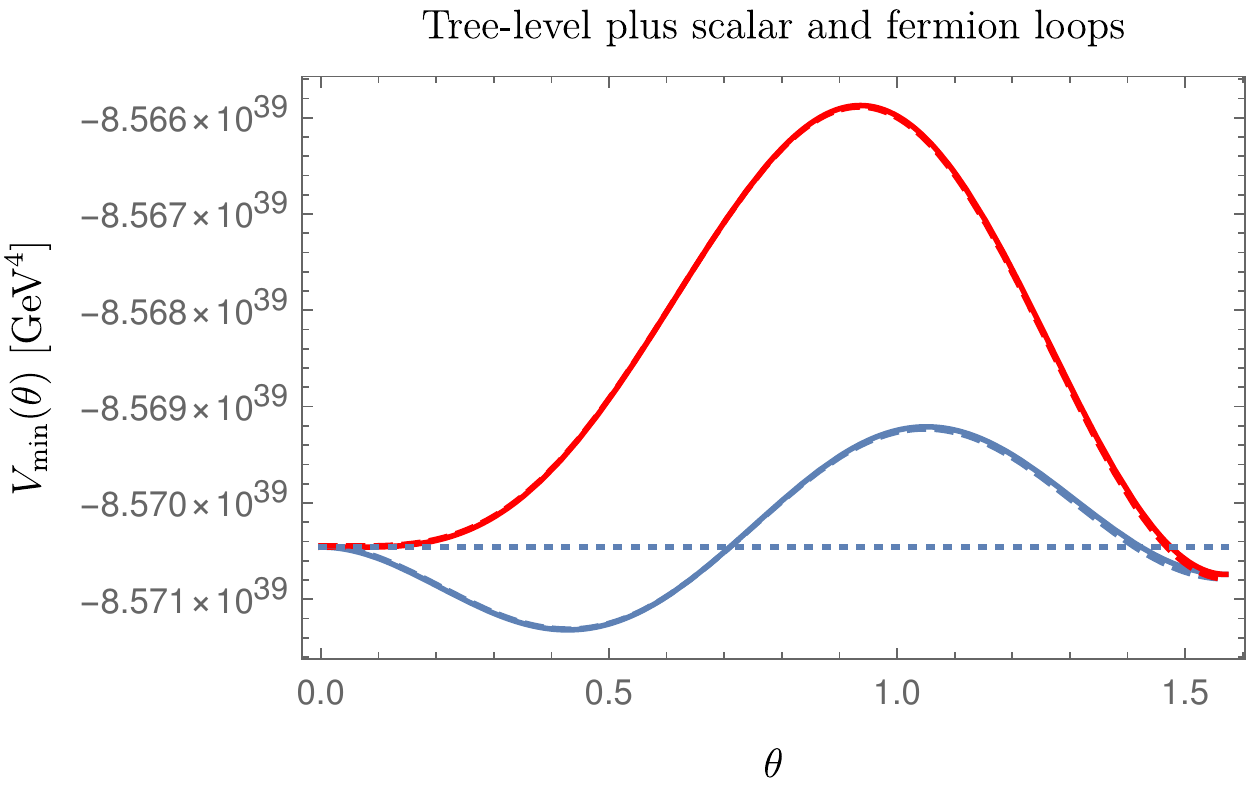}
\caption{Scalar potential $V_{\rm min}(\theta)$ along the lowest-energy path connecting the minima along the $\varphi$ and $\Phi$ axes in the two-field symmetry breaking model of Eq.~\eqref{eq:LMajS}. We have parameterized the path with $\theta\equiv\arctan \Phi/\varphi$. On the left, only scalar corrections are included, while the right plot accounts for both scalar and fermion loops. The blue lines have  $\lambda_{\Phi \varphi}=0.07000$, and the red lines   $\lambda_{\Phi \varphi}=0.06994$; the other parameters are specified in the main text. On the right, the solid lines were obtained with a fixed renormalization scale, and the dashed lines with a field-dependent scale set to the maximum of the effective masses in a given $\Phi,\varphi$ background. }
\label{fig:Vtheta}
\end{figure}
%%%%%%%%%%%%%%% FIGURE %%%%%%%%%%%%%%%%%%%

We illustrate in Fig.~\ref{fig:Vtheta} our results for the effects of radiative corrections on the zero-temperature potential, showing the difference between the potential shape when scalar and fermion radiative corrections are added to the potential. We fix  $\lambda_\Phi=0.115475$, $\lambda_\varphi=0.25400$, $y_1=0.33333$, $y_2=0.35570$, $m^2_\Phi=(5.61270\times10^9{\,\rm GeV})^2$, $m^2_\varphi=(6.17252\times10^9{\,\rm GeV})^2$, and  either $\lambda_{\Phi \varphi}=0.07000$ (red curves) or 0.06994 (blue curves). 
Our calculations show that scalar loops preserve the smooth, monotonic path between vacua, while fermionic interactions either introduce a barrier or a minimum with non-zero VEVs for both fields (\emph{i.e.,} a global minimum at $\theta\neq0,\pi$). The appearance of such features is not an artifact of a truncation of perturbation theory, as the theoretical error of the calculations remains much smaller than the size of the features generated by fermionic loops. To show this explicitly, we calculated the potential both with a fixed renormalization scale $\mu=\sqrt{m^2_\Phi}$, or a field-dependent scale $\mu=\max\{m_i(\Phi,\varphi)\}$ set to the maximum of the scalar and fermionic masses in a given $(\Phi,\varphi)$ background. The  choice of fixed $\mu$ gives the solid lines on the right of Fig.~\ref{fig:Vtheta}, while the field-dependent choice produces the dashed lines. 

In the examples shown in Fig.~\ref{fig:Vtheta}, we obtain $M_1^0/T_{\rm c}\approx8.3$ for $T_c$ defined as the critical temperature at which the minima in the two field directions become degenerate, and $M_{1}^0$ as the field-dependent mass for $N_1$ in the $(v_\Phi,0)$ minimum. Potentials with a barrier (such as the red curve in the right pane of Fig.~\ref{fig:Vtheta}) lead to first-order phase transitions and a suppressed baryon asymmetry, and so we must instead consider parameter regimes such as that leading to the blue curve on the right-hand of Fig.~\ref{fig:Vtheta}. In this case, a global minimum occurs with  simultaneous non-zero $\Phi$ and $\varphi$, and as argued above it is not possible to realize a $T_{\rm c}\rightarrow0$ limit in this case. Furthermore, the value of $\Phi$ at the global minimum is smaller than in the $(v_\Phi,0)$ configuration:~for instance, for the potential with the blue lines in Fig.~\ref{fig:Vtheta} we get $M_1^0/T_{\rm c}=1.7$. It is evident that quantum corrections spoil the region of parameter space that na\"ively realizes a large, parametric enhancement of $M_1^0/T_{\rm c}$ at leading order. We were unable to find any successful benchmark point in the regime of asymmetry enhancement, (\emph{i.e.,} with $M_{N_1}/T_c\gtrsim 10$) as in Fig.~\ref{fig:enhancement}.

It is perhaps unsurprising that radiative corrections ruin the parts of parameter space that give rise to delayed phase transitions in tree-level calculations. The level of tuning of parameters is quite high (see, for instance, how a change at the $\sim10^{-4}$ level in coupling results in the different red and blue curves in Fig.~\ref{fig:Vtheta}), and in the absence of a symmetry, there is no reason that these contributions should cancel. For example, if $\varphi$ and $\Phi$ were embedded in the multiplet of some higher symmetry, then the potential could be exactly flat between the two quasi-minima of the potential. However, the fact that the $N_I$ should only get a mass in the second stage of the phase transition (and therefore only couple appreciably to $\Phi$) resulting in a hard breaking of the symmetry, and leads to the dangerous radiative corrections studied above.

One may consider an alternate possibility, namely that the contribution of the fermionic loops to the scalar potential is cancelled by the contributions of  additional scalars with mixed quartic couplings  related to the Yukawa couplings, $y_I$, as in supersymmetric models. In Section \ref{subsec:single}, we found that this introduces a thermal barrier to the new scalar contributions to the effective potential for $\Phi$ and $\varphi$. This is less of an issue for multi-field models because the barrier height scales with the temperature as in Eq.~\eqref{eq:VT3}, and so if $T_{\rm c}$ is  parametrically close to zero, the barrier could potentially be irrelevant at $T=T_{\rm c}$. We have checked that it is indeed  possible to get larger values of $M_1^0/T_c$ in the two-field model of Eq.~\eqref{eq:LMajS} supplemented with two additional scalars. However, when nearing the asymmetry-enhancing regime with $M_1^0/T_c\gtrsim 10$, the required degeneracy between
the $(0,v_\varphi)$ saddle-point and the $(v_\Phi,0)$ minimum becomes comparable to the theoretical uncertainty of the one-loop computation. This results in a strong dependence of the shape of the potential on the choice of the unphysical renormalization scale, which can affect the appearance (or lack thereof) of barriers and other non-trivial features as in Fig.~\ref{fig:Vtheta}. Resolving the issue would require the precision of a two-loop calculation, which is outside the scope of this work but is likely to involve substantial fine-tuning in the model.

We conclude that models with an enhancement of the asymmetry due to a very fast second order phase transition are in principle possible when multi-field transitions are invoked. In the case of two-step transitions, viable models seem to require 
a carefully arranged particle content with some supersymmetric-like coupling relations, which in the absence of a complete model manifests as a fine tuning. Even still, a definite conclusion requires higher order calculations and presumably a more UV-complete model. Perhaps it is possible to  evade the tuning requirements with a larger number of scalar fields and transitions, in which the fields get trapped in a region with zero $\Phi$ until very low temperatures are reached; as outlined here, the challenge is to achieve this while guaranteeing a barrier-free path towards the lepton-number-breaking $\Phi$ vacuum at both finite and zero temperature.

\subsection{Scalar-Sector Phenomenology}

We  comment briefly on the phenomenology associated with the new sector of scalars giving rise to RHN masses. The primary driving factor for the phenomenology of the scalar sector is whether $\Phi$ is a complex or a real field. If $\Phi$ is real, there is no massless Goldstone mode and the only scalar degree of freedom, $\Phi$, has a typical mass $M_\Phi \sim v_\Phi \sim M_I$. Thus, $\Phi$ may not have any interesting phenomenological consequences if its mass is very high ($\gg10$ TeV), whereas for lower masses it could be produced at colliders via Higgs mixing or other couplings to the SM. In the interesting case $M_\Phi > 2 M_I$, $\Phi$ could decay into $N_I N_I$ when produced via mixing with the SM Higgs. Alternatively, the SM Higgs could decay via its mixing with $\Phi$ into $N_I N_I$, giving rise to a variety of striking multilepton and/or displaced-vertex signatures at high-energy colliders \cite{Pilaftsis:1991ug,Graesser:2007pc,Aranda:2007dq,Shoemaker:2010fg,DiazCruz:2010rs,Maiezza:2015lza,Nemevsek:2016enw,Accomando:2016rpc}. For this scenario to be realized, the mass scale of the RHNs would have to be low and a resonant enhancement of the $CP$ source is needed to obtain the observed baryon asymmetry \cite{Pilaftsis:1997jf,Pilaftsis:2003gt}.

If  $\Phi$ is instead a complex scalar,  there exists an additional scalar degree of freedom. If the lepton-number symmetry is a gauge symmetry, the additional scalar component is eaten by the  gauge boson. This massive $Z'$ boson could be directly produced at the LHC and other colliders and decay into long-lived RHNs or other final states \cite{Huitu:2008gf,Basso:2008iv,AguilarSaavedra:2009ik,Perez:2009mu,Batell:2016zod}. As with the $\Phi$-induced signatures, this scenario is only interesting if the new vector mass is within reach of the LHC, $M_{Z'} \lesssim $ TeV.

Finally, if the lepton-number symmetry is a global symmetry,  there exists a Goldstone boson whose mass is protected by a shift symmetry, and it can therefore be substantially lighter than the RHN or radial scalar mode. This particle is known as the Majoron \cite{Chikashige:1980ui} and can be detected in several ways. For example, one or two Majorons could be emitted in double-beta decay processes, and its interactions are therefore constrained by experiments such as EXO-200 \cite{Albert:2014fya} and KamLAND-Zen \cite{Gando:2012pj}. Neutrino scattering in the early universe would also lead to Majoron production, and there exist cosmological constraints on the existence of Majorons due to overclosure \cite{Cline:1993ht}. In some models, the Majoron itself can constitute a substantial fraction of dark matter \cite{Berezinsky:1993fm}; however, Majoron decays into SM neutrinos during the Cosmic Microwave Background (CMB) epoch are strongly constrained \cite{Lattanzi:2007ux}, and Majoron-mediated neutrino self-interactions can also be constrained by the CMB \cite{Bell:2005dr,Forastieri:2015paa}. Present-day Majoron relics can decay at tree level into pairs of SM neutrinos \cite{Dudas:2014bca} or radiatively into $\gamma\gamma$ \cite{Berezinsky:1993fm}, giving rise to constraints from searches for monochromatic neutrinos \cite{Dudas:2014bca} or   photons \cite{Bazzocchi:2008fh}. Finally, we note that  constraints and signals on Majoron models are possible due to Majoron production in SM neutrino decays during supernovae \cite{Berezhiani:1989za,Choi:1989hi,Kachelriess:2000qc}.

%%%%%%%%%%%%%%%%%%%%%%%%%%%%%%%%%%%%%%%%%%%%%%%%%%%%%%%%%
%%%%%%%%%%%%%%%%%%%%%%%%%%%%%%%%%%%%%%%%%%%%%%%%%%%%%%%%%
%%%%%%%%%%%%%%%%%%%%%%%%%%%%%%%%%%%%%%%%%%%%%%%%%%%%%%%%%
%%%%%%%%%%%%%%%%%%%%%%%%%%%%%%%%%%%%%%%%%%%%%%%%%%%%%%%%%
%%%%%%%%%%%%%%%%%%%%%%%%%%%%%%%%%%%%%%%%%%%%%%%%%%%%%%%%%

\section{Conclusions}

 We have systematically studied the effects of a phase transition on the baryon asymmetry generated via out-of-equilibrium decays. In particular, we have focused on the scenario in which the parent particle responsible for baryogenesis obtains its mass via spontaneous symmetry breaking, and phase transitions in the early universe therefore give rise to a time-dependent mass. This in principle allows for the possibility of an enhanced departure from thermal equilibrium, leading to deviations in the usual predictions for the baryon asymmetry.
 
The change in the baryon asymmetry due to a time-varying mass for the parent particle depends strongly on the nature of the phase transition. We have found the following:
\benum
\item {\bf First-order phase transition:}~While an enhancement of the baryon asymmetry due to suppression of washout effects is, in principle, possible with a first-order phase transition, realistic models lead to a reflection of the parent particle at the bubble wall during the phase transition, resulting in an exponential suppression of the asymmetry (see Fig.~\ref{fig:PT_strength}).

\item {\bf Second-order phase transition:}~If the mass of the parent particle changes rapidly during the phase transition (\emph{i.e.,} on time scales shorter than the Hubble expansion time), a suppression of baryon-number-violating inverse decays can lead to an enhanced baryon asymmetry. The requirement is that the ratio of the zero temperature mass of the parent relative to the temperature of the phase transition should approximately exceed $10$ (see Fig.~\ref{fig:enhancement}).

\item {\bf Slowly varying mass:}~If the mass of the parent particle changes very slowly with time (\emph{i.e.,} on time scales longer than the Hubble expansion time), there is no change in the efficacy of baryogenesis. However, since the mass was different in the early universe relative to today, this changes the relationship among parameters since the time of leptogenesis. In particular, in models of leptogenesis where the right-handed neutrino mass changes slowly with time, then the time scale at which leptogenesis occurs was different in the early universe than would be expected from current neutrino oscillation data. We find that an enhancement of the asymmetry requires larger RH neutrino masses in the early universe, and thus higher reheating temperatures,  which is contrary to existing statements in the literature. 

\eenum 

With a time-dependent mass due to a varying background scalar field, the asymmetry can also be modified due to annihilations of the parent particle into quanta of the same scalar field. We found that the role of these annihilations is to damp the asymmetry by providing new, baryon-number-conserving modes for depleting the parent abundance. However, the importance of the annihilations depends on the parameters of the model, and they tend to become more important for delayed or fast transitions.

Finally, we considered realistic models of  phase transitions driven by the dynamics of single or multiple scalars. We found that obtaining fast and delayed phase transitions, which are necessary to enhance the asymmetry via a second-order phase transition, typically requires large Yukawa couplings that induce radiative corrections spoiling the lateness of the phase transition, or its second-order nature. We discussed possible cancellations among quantum corrections that could lead to a delayed  asymmetry tied to a second-order phase transition, concluding that while such phase transitions are possible in principle, they typically rely on some type of fine tuning or the contributions of higher-order corrections that are beyond the scope of this paper. This fine-tuning is indicative of the challenge of realizing an adiabatic second-order phase transition that is delayed but does not lead to a barrier in the free energy.  Therefore, we conclude that an enhancement of the asymmetry from out-of-equilibrium decays is unlikely in a generic model, but may be realized in specific UV-complete theories.\\

  \noindent {\bf Acknowledgements:}~We are grateful to Peter Ballett, Nikita Blinov, David Morrissey, Michael Peskin, Natalia Toro, Jessica Turner and Ye-Ling Zhou for helpful discussions. BS is supported by the United States Department of Energy under contract DE-AC02-76SF00515.

\appendix

\section{$CP$-Violating Source in the Resonant Regime} \label{app:resonant}

In the regime of resonant leptogenesis with near-degenerate $N_I$, $M_I\sim M_J$ \cite{Pilaftsis:1997jf,Pilaftsis:2003gt}, a resummation of self-energy effects is necessary to resolve the  $x=1$ singularity in Eq.~\eqref{eq:XIJ}. For two flavours of $N_I$ particles, the flavoured $CP$-sources can be written as \cite{Pilaftsis:2003gt}
\begin{equation}
\label{eq:CP_source_resonant}
 \begin{aligned}
  \varepsilon^\alpha_I=\,\frac{1}{8\pi}\,\frac{1}{(F^\dagger F)_{II}}\,\sum_{J\neq I}\,&\Bigg\{\mathrm{Im}[F^\dagger_{I\alpha} F_{\alpha J}(F^\dagger F)_{IJ}]g'(x_{IJ})\\
  %%%%%%
  &\left.\,\,+(M^2_I-M^2_J)\,\frac{\mathrm{Im}\left[F^\dagger_{I\alpha} F_{\alpha J}(M^2_I (F^\dagger F)_{JI}+M_I M_J (F^\dagger F)_{IJ}) \right]}{(M^2_I-M^2_J)^2+\Delta_{IJ}^2}\right\}.
 \end{aligned}
\end{equation}
The $x_{IJ}$ are defined in Eq.~\eqref{eq:XIJ}, while the function $g'(x)$ is given by the non-singular part of $g(x)$ in the same equation, 
\begin{equation}
 g'(x) = \sqrt{x}\left[1-(1+x)\log\left(\frac{1+x}{x}\right)\right].
\end{equation}
The $\Delta_{IJ}$ regulate the singularity in the degenerate limit of the sources in Eq.~\eqref{eq:CP_source}, and have been the subject of discussion in the literature; here we choose $\Delta_{12}=\Delta_{21}=M_1 \Gamma_{N_1} +M_2 \Gamma_{N_2}$, which allows the semi-classical Boltzmann equations to better capture coherent quantum effects \cite{Garny:2011hg}. In the limit $\Delta_{IJ}\rightarrow0$, one recovers Eq.~\eqref{eq:CP_source} from Eq.~\eqref{eq:CP_source_resonant}.

\section{\label{app:Tav}Thermally Averaged Rates}
In this Appendix, we establish conventions for our calculations, collecting  formulae for the thermally averaged rates. These are constructed from the corresponding reaction densities, which for a process of the form $a+b+\dots \rightarrow i+j+\dots$ are defined as
\begin{equation}
\label{eq:rateT}
 \gamma(a+b+\dots \rightarrow i+j+\dots)\equiv \frac{1}{\mathcal{S}}\int d\Pi_a f_a \,d\Pi_b  f_b \dots  |M(a+b+\dots \rightarrow i+j+\dots)|^2 \,\overline{\delta}\,d\Pi_i d\Pi_j\dots.
\end{equation}
In the previous equation, we have ignored Bose-enhancement and Pauli-blocking effects, while the squared-matrix element is summed over all external polarizations and $\mathcal{S}$ is a symmetry factor accounting for identical-particle phase-space integration in the initial or final state. $d\Pi_a=d^3p_a/[(2\pi)^3 2E_a]$, while $\overline\delta\equiv (2\pi)^4 \delta^{(4)}(\sum p_a-\sum p_i)$. $f_a$ denote the equilibrium number densities, which we  take as given by the Maxwell-Boltzmann distribution with zero chemical potential,
\begin{equation}
 f_a(p)=e^{-M_a/T}.
\end{equation}
For a decay process of a particle $N_I$, such as the  $N_I\rightarrow L_\alpha H$ processes considered in this paper, the thermally averaged decay rate is defined as
\be
\label{eq:GammaNav}
\langle \Gamma_{N_I}\rangle \equiv \frac{\langle \gamma(N_I\rightarrow \dots)\rangle}{n^{\rm eq}_{N_I}},
\ee
where $n^{\rm eq}_{N_I}$ is the total equilibrium number density. For a particle $a$ with $g_a$ degrees of freedom,
\be
\label{eq:nb}
n^{\rm eq}_{a}=\frac{g_a}{(2\pi)^3}\int d^3 p f_{N_I}(p)=\frac{g_aT^3}{2\pi^2}\left(\frac{M_a}{T}\right)^2 K_2\left[\frac{M_a}{T}\right],\quad \quad z^2 K_2[z]=\int_z^\infty x e^{-x}\sqrt{x^2-z^2}.
\ee
where the function $K_2$ behaves asymptotically as
\be
\begin{aligned}
\label{eq:K2asympt}
z^2 K_2[z]=& 2+{\cal O}(z), & z^2 K_2[z]=&\left(\frac{15}{8}+z\right)\sqrt{\frac{\pi z}{2}}e^{-z}+{\cal O}\left(\frac{1}{z}\right).
\end{aligned}
\ee
Using Eqs.~\eqref{eq:rateT}, \eqref{eq:GammaNav} and \eqref{eq:nb}, for a momentum-independent two-body decay rate such as $\Gamma_{N_I\rightarrow XY}$ (see Eq.~\eqref{eq:Gamma}) one has
\begin{align}
\label{eq:GammaIav}
 \langle \Gamma_{N_I}\rangle = \frac{K_1(z_I)}{K_2(z_I)} \Gamma_{N_I}, \quad \quad z K_1[z]=\int_z^\infty  e^{-x}\sqrt{x^2-z^2}, 
\end{align}
where $z_I = M_I/T$ and
\be
\begin{aligned}
\label{eq:K1asympt}
z K_1[z]=& 1+{\cal O}(z), & z K_1[z]=&\sqrt{\frac{\pi z}{2}}e^{-z}+{\cal O}\left(\frac{1}{z}\right).
\end{aligned}
\ee
Averaged annihilation cross sections for a reaction $N_I+N_I\rightarrow \dots$ are defined as 
\be
\label{eq:GammaAav}
\langle \sigma_{N_I N_I\rightarrow \dots}v\rangle \equiv \frac{\langle \gamma(N_I N_I\rightarrow \dots)\rangle}{(n^{\rm eq}_{N_I})^2}.
\ee
Using Eqs.~\eqref{eq:rateT} and \eqref{eq:nb}, in the case of a two-body annihilation $N_I N_I\rightarrow XY$ one can write
\begin{align}
\label{eq:sigmaav}
 \langle  \sigma_{N_I N_I\rightarrow XY}v\rangle = \frac{1}{16T^5(z_I^2 K_2[z_I])^2} \int ds s^{3/2} K_1\left[\frac{\sqrt{s}}{T}\right]\left(1-\frac{4 M^2_I}{s}\right)\sigma_{N_I N_I\rightarrow XY}(s),
\end{align}
with $\sigma_{N_I N_I\rightarrow XY}(s)$ the usual annihilation cross section averaged over the spins of the $N_I$, and where we have included an additional factor of $1/2$ for the symmetry factor associated with the initial-state phase-space integration.

%%%%%%%%%%%%%%%%%%%%%%%%%%%%%%%%%%%%%%%%%%%%%%%%%%%%%%%%%%%%%%%%%%%%%%%
%%%%%%%%%%%%%%%%%%%%%%%%%%%%%%%%%%%%%%%%%%%%%%%%%%%%%%%%%%%%%%%%%%%%%%%
%%%%%%%%%%%%%%%%%%%%%%%%%%%%%%%%%%%%%%%%%%%%%%%%%%%%%%%%%%%%%%%%%%%%%%%
%%%%%%%%%%%%%%%%%%%%%%%%%%%%%%%%%%%%%%%%%%%%%%%%%%%%%%%%%%%%%%%%%%%%%%%
%%%%%%%%%%%%%%%%%%%%%%%%%%%%%%%%%%%%%%%%%%%%%%%%%%%%%%%%%%%%%%%%%%%%%%%
%%%%%%%%%%%%%%%%%%%%%%%%%%%%%%%%%%%%%%%%%%%%%%%%%%%%%%%%%%%%%%%%%%%%%%%

\section{\label{app:VeffT}Effective Potential at Zero and Finite Temperature}

We give an overview of the construction of the effective potential at finite temperature for a theory with fermions and scalars in the background of scalars fields $\phi_i$; details can be found in many reviews (such as Ref.~\cite{Quiros:1999jp}).
At one-loop order, the potential may be expressed as $V=V^0+V^T$, with $V^0$
the zero-temperature contribution including one-loop Coleman-Weinberg corrections, and with $V^T$ designating the finite-temperature correction. In the  $\overline{\rm MS}$ scheme, $V^0$ is given by
\begin{align}
 \label{eq:V0}V^0=V^{{\rm tree}}+\frac{1}{64\pi^2}&\left[\sum_S m^4_S(\phi_i)\left(\log\frac{m^2_S(\phi_i)}{\mu^2}-\frac{3}{2}\right)
 -2\sum_F m^4_F(\phi_i)\left(\log\frac{m^2_F(\phi_i)}{\mu^2}-\frac{3}{2}\right)\right],
\end{align}
where $V^{{\rm tree}}$ is the tree-level potential, $m_X(\phi_i)$ is the mass of the field $X$ in a background configuration $\phi_i$, and the labels $S, F$ corresponds to scalars and Weyl fermions (the factor of 2 is from each spin contribution for a given Weyl fermion). On the other hand,  the one-loop
thermal corrections go as
\begin{align}
 \label{eq:VT}
  V^T=\frac{T^4}{2\pi^2}\left[\sum_B J_B\left(\frac{m^2_S(\phi_i)}{T^2}\right)-2\sum_F J_F\left(\frac{m^2_F(\phi_i)}{T^2}\right)\right].
 \end{align}
 The thermal functions $J_S$ and $J_F$ are given by
 \begin{align}
  J_S(x)=\int_0^\infty dy\, y^2\log\left[1-\exp(-\sqrt{x^2+y^2})\right],\quad J_F(x)=\int_0^\infty dy\, y^2\log\left[1+\exp(-\sqrt{x^2+y^2})\right].
 \end{align}
 In the large temperature limit, $T\gg m_X(\phi_i)$, the above functions may be approximated as
 \begin{align}
 \label{eq:exp}
  J_S=-\frac{\pi^4}{45}+\frac{\pi^2}{12}x-\frac{\pi}{6}\,x^{3/2}+{\cal O}(x)^4),\quad J_F=\frac{7\pi^4}{360}-\frac{\pi^2}{24}x+{\cal O}(x^{2}).
 \end{align}
 The above expansions, together with Eq.~\eqref{eq:VT}, imply that the leading effect of a thermal bath is to introduce positive $T^2$ corrections to the quadratic terms in the potential, stabilizing the scalar fields by enhancing
 their masses. A single scalar field $\phi$ undergoes a second-order phase transition if, at some critical temperature $T_{\rm c}$, its effective mass crosses zero and becomes negative, triggering the development of a nonzero VEV. The transition remains of the second order as long as no barrier is generated, which happens if the thermal corrections do not generate terms that are cubic in the background fields. Such terms come from the non-analytic ${\cal O}(x^{3/2})$ terms in Eq.~\eqref{eq:exp}; no $\phi^3$ barrier is generated if these contributions are suppressed, or if for the range of interest of $\phi$ the bosonic masses that generate the barrier are dominated by their $\phi$-independent parts. Whenever $\phi^3$ interactions can be neglected, one may use the leading terms to approximate the temperature dependence of the VEV $\langle\phi(T)\rangle$
 and the masses of the excitations above the background. This gives the temperature dependence of Eq.~\eqref{eq:SOPT}. 
 
 The non-analyticity of $J_S(x)$ for small $x$ in Eq.~\eqref{eq:exp} signals a singular behaviour for zero bosonic masses. In fact, small bosonic masses can be seen to lead to a breakdown of perturbation theory, which can be avoided by resumming thermal corrections to the scalar two-point function \cite{Pisarski:1988vd}, a procedure known as ``Daisy resummation''. Practically, this can be implemented by substituting $m^2_S(\Phi_i)$ in Eq.~\eqref{eq:VT} with the thermally corrected masses.

 %%%%%%%%%%%%%%%%%%%%%%%%%%%%%%%%%%%%%%%%%%%%%%%%%%%%%%%%%%%%%%%%%%%%%%%%%%%%%%%
 %%%%%%%%%%%%%%%%%%%%%%%%%%%%%%%%%%%%%%%%%%%%%%%%%%%%%%%%%%%%%%%%%%%%%%%%%%%%%%%
 %%%%%%%%%%%%%%%%%%%%%%%%%%%%%%%%%%%%%%%%%%%%%%%%%%%%%%%%%%%%%%%%%%%%%%%%%%%%%%%
 %%%%%%%%%%%%%%%%%%%%%%%%%%%%%%%%%%%%%%%%%%%%%%%%%%%%%%%%%%%%%%%%%%%%%%%%%%%%%%%
 
 \subsection{\label{subapp:ann:single}Single-Field Model}
 
 For the model with interaction as in Eq.~\eqref{eq:LMaj}, consisting of a scalar field $\Phi$ that breaks the baryon number symmetry and gives a mass to right-handed neutrinos $N_I$, the masses of real scalar fields and Weyl fermions in the background $\Phi$ field are:
 \begin{equation}
  \begin{aligned}
   m^2_{S,1}=&-m^2 +\frac{\lambda  \Phi^2}{2}, &   m^2_{S,2}=&-m^2 +\frac{3\lambda  \Phi^2}{2}  & m_{F,1}=&y_1\Phi, & m_{F,2}=&y_2\Phi,
  \end{aligned}
 \end{equation}
 The one-loop beta functions are,  defining $\kappa\equiv (16\pi^2)^{-1}$:
 \begin{equation}
  \begin{aligned}
   \beta_\lambda=&\kappa\left(5 \lambda ^2+2 \lambda  y_1^2+2 \lambda  y_2^2-4 y_1^4-4 y_2^4\right)+{\cal O}(\kappa^2), & \beta_{y_1}=&\frac{\kappa}{2} y_1 \left(3 y_1^2+y_2^2\right) +{\cal O}(\kappa^2),\\
   %%%%%%
   \beta_{y_2}=&\frac{\kappa}{2} y_2 \left(y_1^2+3 y_2^2\right)+{\cal O}(\kappa^2), & \beta_{m^2}=&\kappa m^2 \left(2 \lambda +y_1^2+y_2^2\right) +{\cal O}(\kappa^2).
  \end{aligned}
 \end{equation}

 The leading thermal corrections induce the following correction to the $\Phi$ mass parameter (with the conventions of Eq.~\eqref{eq:LMaj}):
 \begin{align}
 \label{eq:MT}
  -m^2\rightarrow-m^2+\frac{T^2}{24}\left(2\lambda+\sum_I y^2_I\right),
 \end{align}
which gives the following estimate of the critical temperature:
 \begin{align}
 \label{eq:TcMaj}
 T^2_c=\frac{12\lambda\Phi^2_0}{2\lambda+\sum_I y^2_I},
 \end{align}
where $\Phi_0$ is the zero-temperature VEV, $\Phi^2_0=2m^2/\lambda$ . In the broken phase, the temperature-dependent minimum lies at a VEV given by equation \eqref{eq:SOPT} and $T_{\rm c}$ given in Eq.~\eqref{eq:TcMaj}. The masses of scalar excitations around the temperature-dependent minimum,
 relevant for computing annihilation cross sections as in Section \ref{sec:damping}, scale as $m^2_{|\Phi|}(T)=\lambda \langle\Phi(T)\rangle^2$ and zero,  as Goldstone's theorem also applies at finite temperature. The field-dependent masses of the $N_I$ are given by Eq.~\eqref{eq:Majorana_SSB}.

 %%%%%%%%%%%%%%%%%%%%%%%%%%%%%%%%%%%%%%%%%%%%%%%%%%%%%%%%%%%%%%%%%%%%%%%%%%%%%%%
 %%%%%%%%%%%%%%%%%%%%%%%%%%%%%%%%%%%%%%%%%%%%%%%%%%%%%%%%%%%%%%%%%%%%%%%%%%%%%%%
 %%%%%%%%%%%%%%%%%%%%%%%%%%%%%%%%%%%%%%%%%%%%%%%%%%%%%%%%%%%%%%%%%%%%%%%%%%%%%%%
 %%%%%%%%%%%%%%%%%%%%%%%%%%%%%%%%%%%%%%%%%%%%%%%%%%%%%%%%%%%%%%%%%%%%%%%%%%%%%%%
 
  \subsection{Two-Field ($N_I$-$\varphi$-$\Phi$) Model}
  
  This model is characterized by the interactions in Eq.~\eqref{eq:LMajS}. The scalar and fermion masses in an arbitrary $\Phi,\varphi$ background, including the leading finite temperature corrections (as needed for the Daisy resummation) are given by
  \begin{equation}
   \begin{aligned}
    m^2_{S,1}=&\frac{1}{24} \left(12 \Phi ^2 \lambda _{\Phi }-24 m^2_{\Phi }+12 \varphi^2 \lambda _{\Phi\varphi}\right)+\frac{ T^2}{24} \left(2 \lambda _{\Phi }+\lambda _{\Phi\varphi}+y_1^2+y_2^2\right),\\
    %%%%
    m^2_{S,2}=&\frac{1}{48} \left\{24 \Phi ^2 \lambda _{\Phi\varphi}+36 \Phi ^2 \lambda _{\Phi }
    -\Big[\left(36 \Phi ^2 \lambda _{\Phi }-24 m^2_\varphi-24 m^2_{\Phi }+12 \varphi^2 \left(\lambda _{\Phi\varphi}+\lambda _\varphi\right)+T^2 \lambda _\varphi+3 \lambda _{\Phi\varphi} \left(T^2+8 \Phi ^2\right)\right.\right.\\
      %%%%
    &\left.+2 T^2 \lambda _{\Phi }+T^2 y_1^2+T^2 y_2^2\right)^2-4 \left(\left(12 \varphi^2+T^2\right) \lambda _\varphi-24 m^2_\varphi\right) \left(-24 m^2_{\Phi }+2 \lambda _{\Phi } \left(T^2+18 \Phi ^2\right)+T^2 \left(y_1^2+y_2^2\right)\right)\\
    %%%%%
    &-4 \lambda _{\Phi\varphi} \left(-24 m^2_\varphi \left(12 \varphi^2+T^2\right)-48 m^2_{\Phi } \left(T^2+12 \Phi ^2\right)+144 \varphi^4 \lambda _\varphi+T^2 \left(\left(24 \varphi^2+T^2\right) \lambda _\varphi\right.\right.\\
    &\left.\left.+2 \left(y_1^2+y_2^2\right) \left(T^2+12 \Phi ^2\right)\right)+4 \lambda _{\Phi } \left(T^4+30 T^2 \Phi ^2+216 \Phi ^4\right)\right)-8 \lambda _{\Phi\varphi}^2 \left(12 \varphi^2 \left(T^2-36 \Phi ^2\right)\right.\\
    %%%%
    & \left.\left. +T^4+12 T^2 \Phi ^2\right)\Big]^{1/2}
    -24 m^2_\varphi-24 m^2_{\Phi }+12 \varphi^2 \lambda _{\Phi\varphi}+12 \varphi^2 \lambda _\varphi\right\}+\frac{T^2}{48}  \left(2 \lambda _{\Phi }+3 \lambda _{\Phi\varphi}+\lambda _\varphi+y_1^2+y_2^2\right),\\
    %%%
    m^2_{S,3}=&\frac{1}{48} \left\{24 \Phi ^2 \lambda _{\Phi\varphi}+36 \Phi ^2 \lambda _{\Phi }
    +\Big[\left(36 \Phi ^2 \lambda _{\Phi }-24 m^2_\varphi-24 m^2_{\Phi }+12 \varphi^2 \left(\lambda _{\Phi\varphi}+\lambda _\varphi\right)+T^2 \lambda _\varphi+3 \lambda _{\Phi\varphi} \left(T^2+8 \Phi ^2\right)\right.\right.\\
    %%%%%
    &\left.+2 T^2 \lambda _{\Phi }+T^2 y_1^2+T^2 y_2^2\right)^2-4 \left(\left(12 \varphi^2+T^2\right) \lambda _\varphi-24 m^2_\varphi\right) \left(-24 m^2_{\Phi }+2 \lambda _{\Phi } \left(T^2+18 \Phi ^2\right)+T^2 \left(y_1^2+y_2^2\right)\right)\\
    %%%%
     &-4 \lambda _{\Phi\varphi} \left(-24 m^2_\varphi \left(12 \varphi^2+T^2\right)-48 m^2_{\Phi } \left(T^2+12 \Phi ^2\right)+144 \varphi^4 \lambda _\varphi+T^2 \left(\left(24 \varphi^2+T^2\right) \lambda _\varphi\right.\right.\\
    %%%%
    &\left.\left.+2 \left(y_1^2+y_2^2\right) \left(T^2+12 \Phi ^2\right)\right)+4 \lambda _{\Phi } \left(T^4+30 T^2 \Phi ^2+216 \Phi ^4\right)\right)-8 \lambda _{\Phi\varphi}^2 \left(12 \varphi^2 \left(T^2-36 \Phi ^2\right)+T^4\right.\\
    %%%%%%
    &\left.\left.+12 T^2 \Phi ^2\right)\Big]^{1/2}
    -24 m^2_\varphi-24 m^2_{\Phi }+12 \varphi^2 \lambda _{\Phi\varphi}+12 \varphi^2 \lambda _\varphi\right\}+\frac{T^2 }{48} \left(2 \lambda _{\Phi }+3 \lambda _{\Phi\varphi}+\lambda _\varphi+y_1^2+y_2^2\right).
   \end{aligned}
  \end{equation}
The one-loop beta functions (used to implement the one-loop potential with a field-dependent renormalization scale) are in this case
 \begin{equation}
  \begin{aligned}
   \beta_{\lambda_\Phi}=&\,\kappa\left(5 \lambda _{\Phi }^2+2 \lambda _{\Phi \varphi }^2+2 \lambda _{\Phi }(y_1^2 + y_2^2 )-4 y_1^4-4 y_2^4\right)+{\cal O}(\kappa^2), &
   %%%%
    \beta_{\lambda_{\varphi}}=&\,3\kappa \left(\lambda _{\varphi }^2+2 \lambda _{\Phi \varphi }^2\right)+{\cal O}(\kappa^2),\\
   %%%%
   \beta_{\lambda_{\Phi\varphi}}=&\,\kappa\lambda _{\Phi \varphi } \left(\lambda _{\varphi }+2 \lambda _{\Phi }+4 \lambda _{\Phi \varphi }+y_1^2+y_2^2\right)+{\cal O}(\kappa^2),&
   %%%%%%
    \beta_{y_1}=&\,\frac{\kappa}{2} y_1 \left(3 y_1^2+y_2^2\right) +{\cal O}(\kappa^2),\\
   %%%%%%
   \beta_{m^2_\Phi}=&\,\kappa \left(\lambda _{\Phi \varphi } m^2_{\varphi }+m^2_{\Phi } \left(2 \lambda _{\Phi }+y_1^2+y_2^2\right)\right) +{\cal O}(\kappa^2),&
   %%%%
   \beta_{y_2}=&\,\frac{\kappa}{2} y_2 \left(y_1^2+3 y_2^2\right)+{\cal O}(\kappa^2),\\
   %%%
   \beta_{m^2_\varphi}=&\,\kappa\left(\lambda _{\varphi } m^2{}_{\varphi }+2 \lambda _{\Phi \varphi } m^2{}_{\Phi }\right)+{\cal O}(\kappa^2).
  \end{aligned}
 \end{equation}
 The temperature-induced mass corrections in the high-temperature approximation are:
 \begin{equation}\begin{aligned}
  -m^2_\Phi\rightarrow&\,-m^2_\Phi+\frac{T^2}{24}\left(2\lambda_\Phi+\sum_I y^2_I+\xi \right),
  \quad\quad -m^2_\varphi\rightarrow&\,-m^2_\varphi+\frac{T^2}{24}\left(\lambda_S+2\xi \right).
 \end{aligned}\end{equation}
 As elaborated in the main text, we are interested in scenarios in which at high temperatures there exists a vacuum in the $(0,\varphi)$ direction, which becomes unstable at lower temperatures and gives rise
 to a second-order phase transition to the the $(\Phi,0)$ minimum. During the phase transition, the minimum  moves from one field direction to the other without encountering any barrier. The fact that such scenarios are, in principle, feasible can be seen with the tree-level potential plus the dominant $T^2$  corrections to the mass parameters (the $T^4$ terms are field-independent and thus do not affect phase transitions). One may define the following temperatures:
 \begin{itemize}
 \item $T_\varphi$: Temperature at which a minimum degenerate with the origin appears in the $(0,\varphi)$ direction.
 
 \item $T_\Phi$: Temperature at which a minimum degenerate with the origin appears in the $(\Phi,0)$ direction.
 
 \item $\tilde T_\varphi$: Temperature at which the minimum in the $(0,\varphi)$ direction becomes unstable.
 
 \item $\tilde T_\Phi$: Temperature at which the minimum in the $(\Phi,0)$ direction becomes stable.
\end{itemize}
 In the scenarios of interest, we want first a transition to a  $(0,v_\varphi)$ vacuum, which should become unstable before the  $(v_\Phi,0)$ minimum in the $\Phi$ axis is stabilized, as otherwise the latter would lie behind  an energy barrier. This means
 $T_\varphi>T_\Phi>\tilde T_\varphi>\tilde T_\Phi$.  In this case $\tilde T_\varphi$  will be the critical temperature at which $\Phi$ starts acquiring a VEV in a second-order phase transition, and in scenarios with an enhanced asymmetry we want to ensure $M_{N_I}/\tilde T_\varphi=y_I\langle\Phi\rangle/\tilde T_\varphi\gg1$.
Parametrizing mass scales in terms of the zero-temperature VEVs of the minima aligned with the field directions,
\begin{align}
\label{eq:VEVs}
 \langle\Phi\rangle^2\equiv \frac{2 m^2_\Phi}{\lambda_\Phi},\quad  \langle\varphi\rangle^2\equiv\frac{6m^2_\varphi}{\lambda_S},
\end{align}
then in these approximations the previously defined temperatures are given by:
\begin{equation}
\label{eq:critTs}
 \begin{aligned}
  T^2_\varphi=&\,\frac{4\lambda_\varphi\langle \varphi\rangle^2}{\lambda_\varphi+2\xi},&
  %%%%%%
  T^2_\Phi=&\,\frac{12\lambda_\Phi\langle\Phi\rangle^2}{2\lambda_\Phi+\xi+\tilde y^2},\\
  %%%%%%
 \tilde  T^2_\varphi=&\,\frac{12\lambda_\varphi(\lambda_\Phi \langle\Phi\rangle^2-\xi \langle \varphi\rangle^2)}{2\lambda_\Phi\lambda_\varphi-6\xi^2+\lambda_\varphi(\tilde y^2-2\xi)}, &
  %%%%%%
 \tilde T^2_\Phi=&\,\frac{4\lambda_\Phi(6\xi\langle\Phi\rangle^2-\lambda_\varphi \langle \varphi\rangle^2)}{2\xi(\xi+\tilde y^2)-\lambda_\Phi(\lambda_\varphi-2\xi)},
  %%%%%%
 \end{aligned}
\end{equation}
where we defined $\tilde y^2\equiv\sum_I y_I^2$. In the scenarios of interest with $\tilde T_\varphi>\tilde T_\Phi$, and furthermore $\tilde T_\varphi\rightarrow0$ in order to get an enhancement of the asymmetry, one has
\begin{align}
\label{eq:Tsineq}
\begin{aligned}
&\lambda_\Phi -\xi\frac{\langle \varphi\rangle^2}{\langle\Phi\rangle^2}\gtrsim0,&
%%%
&\lambda_\varphi-6\xi \frac{\langle\Phi\rangle^2}{\langle \varphi\rangle^2}\lesssim0,
\end{aligned}
\end{align}
which by virtue of equation \eqref{eq:VEVs} enforces the same inequalities as the conditions  Eq.~\eqref{eq:2fieldeq} ensuring the absence of a barrier at zero temperature. Within this tree-level, high-temperature expansion, the transition from the $(0,v_\varphi)$ to the $(v_\Phi)$ minimum is indeed second-order, and one may follow  how the minimum moves from the $\varphi$ to the $\Phi$ axis as a function of temperature. For $\tilde T_\varphi > T > \tilde T_\Phi$, the minimum lies in between the two field directions, and then settles on the $\Phi$ axis for $T\leq \tilde T_\Phi$:
\begin{align}
 &\left.\begin{aligned}
 \langle\varphi\rangle^2(T)=&\,\frac{(T^2-\tilde T_\Phi^2)(2\xi(\xi+\tilde y_X^2)-\lambda_\Phi(\lambda_\varphi-2\xi))}{4(\lambda_\Phi\lambda_\varphi-6\xi^2)}\\
 %%%%%%
 \langle\Phi\rangle^2(T)=&\,\frac{(\tilde T^2_\varphi-T^2)(2\lambda_\Phi\lambda_\varphi-6\xi^2+\lambda_\varphi(\tilde y_X^2-2\xi))}{12(\lambda_\Phi\lambda_\varphi-6\xi^2)}
 \end{aligned}\right\}
 &\tilde T_s \geq T \geq \tilde T_\phi.\\
 %%%%%
 &\left.\begin{aligned}
  \langle\varphi\rangle^2(T)=&0\\
  %%%%
 \langle\Phi\rangle^2(T)=&\,\frac{(T^2_\Phi-T^2)\langle\Phi\rangle^2}{T^2_\phi}
 %%%%%%
 \end{aligned}\right\}
 & \tilde T<\tilde T_\phi.
\end{align}

%%%%%%%%%%%%%%%%%%%%%%%%%%%%%%%%%%%%%%%%%%%%%%%%%%%%%%%%%%%%%%%%%%%%%%%
%%%%%%%%%%%%%%%%%%%%%%%%%%%%%%%%%%%%%%%%%%%%%%%%%%%%%%%%%%%%%%%%%%%%%%%
%%%%%%%%%%%%%%%%%%%%%%%%%%%%%%%%%%%%%%%%%%%%%%%%%%%%%%%%%%%%%%%%%%%%%%%
%%%%%%%%%%%%%%%%%%%%%%%%%%%%%%%%%%%%%%%%%%%%%%%%%%%%%%%%%%%%%%%%%%%%%%%
%%%%%%%%%%%%%%%%%%%%%%%%%%%%%%%%%%%%%%%%%%%%%%%%%%%%%%%%%%%%%%%%%%%%%%%
%%%%%%%%%%%%%%%%%%%%%%%%%%%%%%%%%%%%%%%%%%%%%%%%%%%%%%%%%%%%%%%%%%%%%%%

\section{\label{app:XS}Annihilation Cross Sections}

In this section we consider the annihilation cross sections of $N_I$ particles into the components of a complex scalar field $\Phi$, in a model with the interactions of Eq.~\eqref{eq:LMaj}, distinguishing the broken and unbroken symmetry cases. The cross sections are averaged over the spins of the initial particles.

\subsection{\label{subapp:XS:broken}Broken phase}

We will assume that the field $\Phi$ acquires a background value aligned with its real part; for a $CP$-invariant evolution of $\Phi$, this can be ensured with an appropriate 
rotation of the field. In order to calculate annihilations of $N_I$ into the degrees of freedom in $\Phi$, we redefine the latter as
\begin{align}
 \Phi=\frac{1}{\sqrt{2}}(v_\phi+\phi_r)e^{i \phi_i/v_\phi},
\end{align}
where $v_\phi$ parameterizes the background. The field $\phi_i$ can be interpreted as the (pseudo-)Goldstone of the (approximate) symmetry that forbids the tree-level mass term for the $N_I$. The $N_I$ fields can be redefined to absorb the $\phi_i$ phase, which appears then only in the kinetic terms. The annihilation $NN\rightarrow\phi_r\phi_r$ scalars proceeds through diagrams as in Fig.~\ref{fig:annihilation}. The interactions in Eq.~\eqref{eq:LMaj}
generate a cubic term for $\phi_r$ of the form ${\cal L}\supset \kappa/6 \phi_r^3$, with $\kappa=3\lambda_\phi v_\phi/2$. The annihilation rate is then given in terms of the masses $M$ and $m$ 
of the $N_I$ particles and $\phi_r$ in the $v_\phi$ background as:

\begin{equation}
\begin{aligned}
 \sigma_{N_I N_I\rightarrow\phi_r\phi_r}(s)=&\,\frac{1}{64 \pi  s \left(s-4 M^2\right)}\left\{\sqrt{s-4 m^2} \sqrt{s-4 M^2} \left[\frac{\kappa ^2 \left(s-4 M^2\right)}{m^2 \Gamma _{\phi }^2+\left(m^2-s\right)^2}+\frac{8 \kappa  M y_I \left(m^2-s\right)}{m^2 \Gamma _{\phi }^2+\left(m^2-s\right)^2}\right.\right.\\
 %%%%%%
 &\left.-4 y_I^2 \left(M^4 \left(\frac{4}{m^4+2 m^2 \left(s-2 M^2\right)+M^2 s}+\frac{4}{m^4-4 m^2 M^2+M^2 s}\right)+1\right)\right]\\
 %%%%
 &+\frac{y_I}{s \left(m^2 \Gamma _{\phi }^2+\left(m^2-s\right)^2\right)}\left[2 \kappa  M s \left(m^2-s\right) \left(s-8 M^2\right)+y_I \left(s^2+16 M^2 s-32 M^4\right) \left(m^2 \Gamma _{\phi }^2\right.\right.\\
 %%%%
 &\left.\left.+\left(m^2-s\right)^2\right)\right]\left[\log \left(m^2-\frac{1}{2} \sqrt{s-4 m^2} \sqrt{s-4 M^2}-\frac{s}{2}\right)\right.\\
 %%%%
 &\left.\left.-\log \left(m^2+\frac{1}{2} \sqrt{s-4 m^2} \sqrt{s-4 M^2}-\frac{s}{2}\right)+2 \coth ^{-1}\left(\frac{2 m^2+s}{\sqrt{s-4 m^2} \sqrt{s-4 M^2}}\right)\right] \right\}.
\end{aligned}
\end{equation}
In the above equation, $s$ is the centre-of-mass energy, and we allowed for a scalar width  $\Gamma_\phi$. 

For the annihilation cross section into $\phi_i$, whose  mass is taken as negligible as would be expected for a (pseudo-)Goldstone boson, the process arises again from diagrams as in 
Fig.~\ref{fig:annihilation} but with vertices involving derivative interactions. The result is:
\begin{equation}
 \begin{aligned}
 \sigma_{N_I N_I\rightarrow\phi_i\phi_i}(s)=& \frac{y_I^4}{256 \pi  M^2 \left(4 M^2-s\right) \left(m^2 \Gamma _{\phi }^2+\left(m^2-s\right)^2\right)} \Bigg\{-m^4 \sqrt{s \left(s-4 M^2\right)}-m^2 \Gamma _{\phi }^2 \sqrt{s \left(s-4 M^2\right)}\\
 %%%
& +M^2 \left(m^4+m^2 \Gamma _{\phi }^2-s^2\right) \left[\log \left(-\sqrt{s-4 M^2}-\sqrt{s}\right)-\log \left(\sqrt{s-4 M^2}-\sqrt{s}\right)\right.\\
&\left.\left.+2 \tanh ^{-1}\left(\sqrt{1-\frac{4 M^2}{s}}\right)\right]+4 M^2 s^{3/2} \sqrt{s-4 M^2}\right\}.
 \end{aligned}
\end{equation}

\subsection{\label{subapp:XS:unbroken}Unbroken phase}

In this case the complex $\Phi$ field creates and destroys particles and antiparticles with equal mass. In the absence of a $\Phi$ background there is no induced cubic term in the potential, and so 
contributing diagrams are as in the left pane of Fig.~\ref{fig:annihilation} with $\Phi\Phi^*$ final states. The chiral symmetry under which both $N_I$ and $\Phi$ are charged forbids a tree-level mass for the $N_I$ and is not broken by thermal thermal effects. The total annihilation cross section is (for scalar mass $m$),
\begin{equation}
 \begin{aligned}
  \sigma_{N_I N_I\rightarrow{\rm \Phi\Phi^*}}(s)=& \frac{y^4_I}{32 \pi  s^2 } \left\{\left(s-2 m^2\right) \log \left[\frac{2 m^2-s-\sqrt{s \left(s-4 m^2\right)}}{2 m^2-s+\sqrt{s \left(s-4 m^2\right)}}\right]-2 \sqrt{s \left(s-4 m^2\right)}\right\}.
 \end{aligned}
\end{equation}

%%%%%%%%%%%%%%%%%%%%%%%%%%%%%%%%%%%%%%%%%%%%%%%%%%%%%%%%%%%%%%%%%%%%%%%%
%%%%%%%%%%%%%%%%%%%%%%%%%%%%%%%%%%%%%%%%%%%%%%%%%%%%%%%%%%%%%%%%%%%%%%%%
%%%%%%%%%%%%%%%%%%%%%%%%%%%%%%%%%%%%%%%%%%%%%%%%%%%%%%%%%%%%%%%%%%%%%%%%
%%%%%%%%%%%%%%%%%%%%%%%%%%%%%%%%%%%%%%%%%%%%%%%%%%%%%%%%%%%%%%%%%%%%%%%%
%%%%%%%%%%%%%%%%%%%%%%%%%%%%%%%%%%%%%%%%%%%%%%%%%%%%%%%%%%%%%%%%%%%%%%%%

%%%%%%%%%%%%%%%%%%%%%%%%%%%%%%%%%%%%%%%%%%%%%%%%%%%%%%%%%%%%%%%%%%%%%%%
%%%%%%%%%%%%%%%%%%%%%%%%%%%%%%%%%%%%%%%%%%%%%%%%%%%%%%%%%%%%%%%%%%%%%%%
%%%%%%%%%%%%%%%%%%%%%%%%%%%%%%%%%%%%%%%%%%%%%%%%%%%%%%%%%%%%%%%%%%%%%%%
%%%%%%%%%%%%%%%%%%%%%%%%%%%%%%%%%%%%%%%%%%%%%%%%%%%%%%%%%%%%%%%%%%%%%%%
%%%%%%%%%%%%%%%%%%%%%%%%%%%%%%%%%%%%%%%%%%%%%%%%%%%%%%%%%%%%%%%%%%%%%%%
%%%%%%%%%%%%%%%%%%%%%%%%%%%%%%%%%%%%%%%%%%%%%%%%%%%%%%%%%%%%%%%%%%%%%%%

\section{\label{app:Fs}Select Yukawa Matrices Used in Calculations}

We give in this Appendix the  Yukawa matrices used for the calculations leading to Figures \ref{fig:PT_strength}, \ref{fig:enhancement}, \ref{fig:slowevolution}, and \ref{fig:annihilationplots}. To ensure compatibility with neutrino oscillation measurements, we fixed the light neutrino masses and 
mixing matrix $U$ using the global fits to oscillation experiments of Ref.~\cite{deSalas:2017kay}, assuming a normal hierarchy and taking the  lightest SM neutrino to be massless. Such choices do not completely fix the Yukawa
matrices $F$ of the right-handed neutrinos, which depend on the masses of the heavy neutrinos and additional complex mixing parameters. This can be made manifest with the Casas-Ibarra parameterization \cite{Casas:2001sr}, which may be written as
\begin{align}
 F=\frac{1}{v}U^*D_{\sqrt{m}}OD_{\sqrt{M}}.
\end{align}
In the previous equation $v$ is the Higgs VEV, $D_{\sqrt{m}}$ and $D_{\sqrt{M}}$ denote
diagonal matrices containing the light and heavy masses, respectively, and $O$ is a complex orthogonal matrix.
Throughout the paper we consider two right-handed neutrinos (which justifies the choice of a massless lightest neutrino), and the orthogonal matrix can be then parameterized in terms of a single complex parameter $\theta$:
\begin{align}
 O=\left[
\begin{array}{ccc}
 0,& \cos\theta,& \sin\theta\\
 %%%%%
 0,&-\sin\theta,& \cos\theta
\end{array}
\right].
\end{align}

\subsection*{Figure \ref{fig:PT_strength}}

% \begin{align}
% \label{eq:Yuk_fig1}
% F=\left[
% \begin{array}{cc}
%  0.0044448\, -0.00781103 i, & -0.000585946+0.000711637 i \\
%  0.00166355\, -0.00412115 i, & -0.0066961+0.00202516 i 
% \end{array}
% \right].
% \end{align}
\begin{align}
\label{eq:Yuk_fig2_solid}\theta=&0.980813 + 0.334919 i,\quad M=\{10^9 {\rm GeV},2\times10^9 {\rm GeV}\},\\
 %%%
 \nonumber F=&\left[
 \begin{array}{cc}
 0.00015934 - 0.000254585 i, & -0.000493917 - 
   0.000207035 i \\
  0.000928904 + 0.0000455962 i, & 
   0.000263774 - 0.000393313 i \\
  0.000673622 + 0.000243432 i, & 
   0.00113985 - 0.000272048 i \\
 \end{array}
\right].
\end{align}

\subsection*{Figure \ref{fig:enhancement}}
\subsubsection*{\hskip1cm Solid line}
Same as in equation \eqref{eq:Yuk_fig2_solid}.

\subsubsection*{\hskip1cm Dashed line}
\begin{align}
\label{eq:Yuk_fig2_dashed}
\theta=&1.500813 - 0.434919 i,\quad M=\{10^9 {\rm GeV},2\times10^9 {\rm GeV}\},\\
%%%%%
\nonumber
 F=&\left[
\begin{array}{cc}
-0.0000466965 - 0.0000608621 i,& -0.000356602 - 
   0.0000433057 i,\\
   0.000954627 + 0.000124872 i,& -0.000437452 + 
   0.000588519 i,\\
   %%%%
   0.00104749 - 0.000157692 i,& 
  0.000537639 + 0.000647699 i
\end{array}
\right].
\end{align}

\subsection*{Figure \ref{fig:slowevolution}}

\hskip1cm Same as Eq.~\eqref{eq:Yuk_fig2_solid}.

\subsection*{Figure \ref{fig:annihilationplots}}

\hskip1cm As detailed in the text, our choice of Yukawa matrices is based upon two reference matrices, $F^h_7$ for hierarchical scenarios, and $F^d_7$ for degenerate ones, with:
\subsubsection*{\hskip1cm Hierarchical scenarios}
\begin{align}
 \label{eq:Yuk_fig5}
 \theta=&-5.76361 + 0.773428 i,\quad M=\{10^7 {\rm GeV},2\times10^7 {\rm GeV}\},\\
 %%%
 \nonumber F^h_7&=\left[
\begin{array}{cc}
  0.0000437894 - 0.0000288214 i, & -0.000047967 - 
  0.000057027 i, \\
  %%%
  0.0000925538 + 0.0000443457 i,& 
 0.000100645 - 0.0000834525 i,\\
 %%%%
 0.0000297338 + 0.0000795733 i,& 
 0.000178024 - 0.0000255772 i
\end{array}
\right].
\end{align}
\subsubsection*{\hskip1cm Degenerate scenarios}
\begin{align}
 \label{eq:Yuk_fig5deg}
 \theta=&-5.76361 + 0.773428 i,\quad M=\{10^7 {\rm GeV},1.000001\times10^7 {\rm GeV}\},\\
 %%%
 \nonumber F^d_7&=\left[
\begin{array}{cc}
  0.0000437894 - 0.0000288214 i,& -0.0000339178 - 
   0.0000403242 i,\\
   %%%
   0.0000925538 + 0.0000443457 i, &
  0.0000711669 - 0.0000590098 i,\\
  %%%
  0.0000297338 + 0.0000795733 i, &
  0.000125882 - 0.0000180858 i
\end{array}
\right].
\end{align}

\subsection*{Figure \ref{fig:2ndorderannihilationplots}}

\hskip1.2cm The left plot uses the Yukawa matrix of Eq.~\eqref{eq:Yuk_fig5deg}, while the right plot uses Eq.~\eqref{eq:Yuk_fig2_solid}.

\bibliographystyle{JHEP}
\bibliography{PT_baryogenesis_ref}

\providecommand{\href}[2]{#2}\begingroup\raggedright\begin{thebibliography}{10}

\bibitem{Glashow:1961tr}
S.~L. Glashow, \emph{{Partial Symmetries of Weak Interactions}},
  \href{http://dx.doi.org/10.1016/0029-5582(61)90469-2}{\emph{Nucl. Phys.}
  {\bfseries 22} (1961) 579--588}.

\bibitem{Weinberg:1967tq}
S.~Weinberg, \emph{{A Model of Leptons}},
  \href{http://dx.doi.org/10.1103/PhysRevLett.19.1264}{\emph{Phys. Rev. Lett.}
  {\bfseries 19} (1967) 1264--1266}.

\bibitem{Salam:1968rm}
A.~Salam, \emph{{Weak and Electromagnetic Interactions}}, {\emph{Conf. Proc.}
  {\bfseries C680519} (1968) 367--377}.

\bibitem{ALEPH:2005ab}
{\scshape SLD Electroweak Group, DELPHI, ALEPH, SLD, SLD Heavy Flavour Group,
  OPAL, LEP Electroweak Working Group, L3} collaboration, S.~Schael et~al.,
  \emph{{Precision electroweak measurements on the $Z$ resonance}},
  \href{http://dx.doi.org/10.1016/j.physrep.2005.12.006}{\emph{Phys. Rept.}
  {\bfseries 427} (2006) 257--454},
  [\href{https://arxiv.org/abs/hep-ex/0509008}{{\ttfamily hep-ex/0509008}}].

\bibitem{LEP-2}
{The ALEPH, DELPHI, L3, OPAL Collaborations, the LEP Electroweak Working
  Group}, \emph{{Electroweak Measurements in Electron-Positron Collisions at
  W-Boson-Pair Energies at LEP}}, {\emph{Phys. Rept.} {\bfseries 532} (2013)
  119}, [\href{https://arxiv.org/abs/1302.3415}{{\ttfamily 1302.3415}}].

\bibitem{Aad:2012tfa}
{\scshape ATLAS} collaboration, G.~Aad et~al., \emph{{Observation of a new
  particle in the search for the Standard Model Higgs boson with the ATLAS
  detector at the LHC}},
  \href{http://dx.doi.org/10.1016/j.physletb.2012.08.020}{\emph{Phys. Lett.}
  {\bfseries B716} (2012) 1--29},
  [\href{https://arxiv.org/abs/1207.7214}{{\ttfamily 1207.7214}}].

\bibitem{Chatrchyan:2012xdj}
{\scshape CMS} collaboration, S.~Chatrchyan et~al., \emph{{Observation of a new
  boson at a mass of 125 GeV with the CMS experiment at the LHC}},
  \href{http://dx.doi.org/10.1016/j.physletb.2012.08.021}{\emph{Phys. Lett.}
  {\bfseries B716} (2012) 30--61},
  [\href{https://arxiv.org/abs/1207.7235}{{\ttfamily 1207.7235}}].

\bibitem{Kirzhnits:1972iw}
D.~A. Kirzhnits, \emph{{Weinberg model in the hot universe}}, {\emph{JETP
  Lett.} {\bfseries 15} (1972) 529--531}.

\bibitem{Dolan:1973qd}
L.~Dolan and R.~Jackiw, \emph{{Symmetry Behavior at Finite Temperature}},
  \href{http://dx.doi.org/10.1103/PhysRevD.9.3320}{\emph{Phys. Rev.} {\bfseries
  D9} (1974) 3320--3341}.

\bibitem{Weinberg:1974hy}
S.~Weinberg, \emph{{Gauge and Global Symmetries at High Temperature}},
  \href{http://dx.doi.org/10.1103/PhysRevD.9.3357}{\emph{Phys. Rev.} {\bfseries
  D9} (1974) 3357--3378}.

\bibitem{Sakharov:1967dj}
A.~D. Sakharov, \emph{{Violation of CP Invariance, c Asymmetry, and Baryon
  Asymmetry of the Universe}},
  \href{http://dx.doi.org/10.1070/PU1991v034n05ABEH002497}{\emph{Pisma Zh.
  Eksp. Teor. Fiz.} {\bfseries 5} (1967) 32--35}.

\bibitem{Kuzmin:1985mm}
V.~A. Kuzmin, V.~A. Rubakov and M.~E. Shaposhnikov, \emph{{On the Anomalous
  Electroweak Baryon Number Nonconservation in the Early Universe}},
  \href{http://dx.doi.org/10.1016/0370-2693(85)91028-7}{\emph{Phys. Lett.}
  {\bfseries B155} (1985) 36}.

\bibitem{Shaposhnikov:1986jp}
M.~E. Shaposhnikov, \emph{{Possible Appearance of the Baryon Asymmetry of the
  Universe in an Electroweak Theory}}, {\emph{JETP Lett.} {\bfseries 44} (1986)
  465--468}.

\bibitem{Shaposhnikov:1987tw}
M.~E. Shaposhnikov, \emph{{Baryon Asymmetry of the Universe in Standard
  Electroweak Theory}},
  \href{http://dx.doi.org/10.1016/0550-3213(87)90127-1}{\emph{Nucl. Phys.}
  {\bfseries B287} (1987) 757--775}.

\bibitem{Bochkarev:1987wf}
A.~I. Bochkarev and M.~E. Shaposhnikov, \emph{{Electroweak Production of Baryon
  Asymmetry and Upper Bounds on the Higgs and Top Masses}},
  \href{http://dx.doi.org/10.1142/S0217732387000537}{\emph{Mod. Phys. Lett.}
  {\bfseries A2} (1987) 417}.

\bibitem{Kajantie:1996mn}
K.~Kajantie, M.~Laine, K.~Rummukainen and M.~E. Shaposhnikov, \emph{{Is there a
  hot electroweak phase transition at m(H) larger or equal to m(W)?}},
  \href{http://dx.doi.org/10.1103/PhysRevLett.77.2887}{\emph{Phys. Rev. Lett.}
  {\bfseries 77} (1996) 2887--2890},
  [\href{https://arxiv.org/abs/hep-ph/9605288}{{\ttfamily hep-ph/9605288}}].

\bibitem{Morrissey:2012db}
D.~E. Morrissey and M.~J. Ramsey-Musolf, \emph{{Electroweak baryogenesis}},
  \href{http://dx.doi.org/10.1088/1367-2630/14/12/125003}{\emph{New J. Phys.}
  {\bfseries 14} (2012) 125003},
  [\href{https://arxiv.org/abs/1206.2942}{{\ttfamily 1206.2942}}].

\bibitem{Weinberg:1979bt}
S.~Weinberg, \emph{{Cosmological Production of Baryons}},
  \href{http://dx.doi.org/10.1103/PhysRevLett.42.850}{\emph{Phys. Rev. Lett.}
  {\bfseries 42} (1979) 850--853}.

\bibitem{Fukugita:1986hr}
M.~Fukugita and T.~Yanagida, \emph{{Baryogenesis Without Grand Unification}},
  \href{http://dx.doi.org/10.1016/0370-2693(86)91126-3}{\emph{Phys. Lett.}
  {\bfseries B174} (1986) 45--47}.

\bibitem{Minkowski:1977sc}
P.~Minkowski, \emph{{$\mu \to e\gamma$ at a Rate of One Out of $10^{9}$ Muon
  Decays?}}, \href{http://dx.doi.org/10.1016/0370-2693(77)90435-X}{\emph{Phys.
  Lett.} {\bfseries B67} (1977) 421--428}.

\bibitem{Yoshimura:1978ex}
M.~Yoshimura, \emph{{Unified Gauge Theories and the Baryon Number of the
  Universe}}, \href{http://dx.doi.org/10.1103/PhysRevLett.41.281}{\emph{Phys.
  Rev. Lett.} {\bfseries 41} (1978) 281--284}.

\bibitem{Ignatiev:1978uf}
A.~{\relax Yu}. Ignatiev, N.~V. Krasnikov, V.~A. Kuzmin and A.~N. Tavkhelidze,
  \emph{{Universal CP Noninvariant Superweak Interaction and Baryon Asymmetry
  of the Universe}},
  \href{http://dx.doi.org/10.1016/0370-2693(78)90900-0}{\emph{Phys. Lett.}
  {\bfseries B76} (1978) 436--438}.

\bibitem{Khoze:2016zfi}
V.~V. Khoze and A.~D. Plascencia, \emph{{Dark Matter and Leptogenesis Linked by
  Classical Scale Invariance}},
  \href{http://dx.doi.org/10.1007/JHEP11(2016)025}{\emph{JHEP} {\bfseries 11}
  (2016) 025}, [\href{https://arxiv.org/abs/1605.06834}{{\ttfamily
  1605.06834}}].

\bibitem{Mohapatra:1986aw}
R.~N. Mohapatra, \emph{{Mechanism for Understanding Small Neutrino Mass in
  Superstring Theories}},
  \href{http://dx.doi.org/10.1103/PhysRevLett.56.561}{\emph{Phys. Rev. Lett.}
  {\bfseries 56} (1986) 561--563}.

\bibitem{Mohapatra:1986bd}
R.~N. Mohapatra and J.~W.~F. Valle, \emph{{Neutrino Mass and Baryon Number
  Nonconservation in Superstring Models}},
  \href{http://dx.doi.org/10.1103/PhysRevD.34.1642}{\emph{Phys. Rev.}
  {\bfseries D34} (1986) 1642}.

\bibitem{Pilaftsis:2005rv}
A.~Pilaftsis and T.~E.~J. Underwood, \emph{{Electroweak-scale resonant
  leptogenesis}},
  \href{http://dx.doi.org/10.1103/PhysRevD.72.113001}{\emph{Phys. Rev.}
  {\bfseries D72} (2005) 113001},
  [\href{https://arxiv.org/abs/hep-ph/0506107}{{\ttfamily hep-ph/0506107}}].

\bibitem{Pilaftsis:2008qt}
A.~Pilaftsis, \emph{{Electroweak Resonant Leptogenesis in the Singlet Majoron
  Model}}, \href{http://dx.doi.org/10.1103/PhysRevD.78.013008}{\emph{Phys.
  Rev.} {\bfseries D78} (2008) 013008},
  [\href{https://arxiv.org/abs/0805.1677}{{\ttfamily 0805.1677}}].

\bibitem{Gu:2009hn}
P.-H. Gu and U.~Sarkar, \emph{{Leptogenesis Bound on Spontaneous Symmetry
  Breaking of Global Lepton Number}},
  \href{http://dx.doi.org/10.1140/epjc/s10052-011-1560-2}{\emph{Eur. Phys. J.}
  {\bfseries C71} (2011) 1560},
  [\href{https://arxiv.org/abs/0909.5468}{{\ttfamily 0909.5468}}].

\bibitem{Sierra:2014sta}
D.~Aristizabal~Sierra, M.~Tortola, J.~W.~F. Valle and A.~Vicente,
  \emph{{Leptogenesis with a dynamical seesaw scale}},
  \href{http://dx.doi.org/10.1088/1475-7516/2014/07/052}{\emph{JCAP} {\bfseries
  1407} (2014) 052}, [\href{https://arxiv.org/abs/1405.4706}{{\ttfamily
  1405.4706}}].

\bibitem{Ganguly:1994pi}
A.~Ganguly, J.~C. Parikh and U.~Sarkar, \emph{{Low-energy leptogenesis in
  left-right symmetric models}},
  \href{http://dx.doi.org/10.1016/0370-2693(96)00892-1}{\emph{Phys. Lett.}
  {\bfseries B385} (1996) 175--180},
  [\href{https://arxiv.org/abs/hep-ph/9408271}{{\ttfamily hep-ph/9408271}}].

\bibitem{McLerran:1990zh}
L.~D. McLerran, M.~E. Shaposhnikov, N.~Turok and M.~B. Voloshin, \emph{{Why the
  baryon asymmetry of the universe is approximately 10**-10}},
  \href{http://dx.doi.org/10.1016/0370-2693(91)91794-V}{\emph{Phys. Lett.}
  {\bfseries B256} (1991) 451--456}.

\bibitem{Bi:2003yr}
X.-J. Bi, P.-h. Gu, X.-l. Wang and X.-m. Zhang, \emph{{Thermal leptogenesis in
  a model with mass varying neutrinos}},
  \href{http://dx.doi.org/10.1103/PhysRevD.69.113007}{\emph{Phys. Rev.}
  {\bfseries D69} (2004) 113007},
  [\href{https://arxiv.org/abs/hep-ph/0311022}{{\ttfamily hep-ph/0311022}}].

\bibitem{Pascoli:2016gkf}
S.~Pascoli, J.~Turner and Y.-L. Zhou, \emph{{Baryogenesis via leptonic
  CP-violating phase transition}},
  \href{https://arxiv.org/abs/1609.07969}{{\ttfamily 1609.07969}}.

\bibitem{Cohen:1990it}
A.~G. Cohen, D.~B. Kaplan and A.~E. Nelson, \emph{{Baryogenesis at the weak
  phase transition}},
  \href{http://dx.doi.org/10.1016/0550-3213(91)90395-E}{\emph{Nucl. Phys.}
  {\bfseries B349} (1991) 727--742}.

\bibitem{Fornal:2017owa}
B.~Fornal, Y.~Shirman, T.~M.~P. Tait and J.~R. West, \emph{{Asymmetric Dark
  Matter and Baryogenesis from $SU(2)$-Lepton}},
  \href{https://arxiv.org/abs/1703.00199}{{\ttfamily 1703.00199}}.

\bibitem{Long:2017rdo}
A.~J. Long, A.~Tesi and L.-T. Wang, \emph{{Baryogenesis at a
  Lepton-Number-Breaking Phase Transition}},
  \href{https://arxiv.org/abs/1703.04902}{{\ttfamily 1703.04902}}.

\bibitem{Katz:2016adq}
A.~Katz and A.~Riotto, \emph{{Baryogenesis and Gravitational Waves from Runaway
  Bubble Collisions}},
  \href{http://dx.doi.org/10.1088/1475-7516/2016/11/011}{\emph{JCAP} {\bfseries
  1611} (2016) 011}, [\href{https://arxiv.org/abs/1608.00583}{{\ttfamily
  1608.00583}}].

\bibitem{Cohen:2008nb}
T.~Cohen, D.~E. Morrissey and A.~Pierce, \emph{{Changes in Dark Matter
  Properties After Freeze-Out}},
  \href{http://dx.doi.org/10.1103/PhysRevD.78.111701}{\emph{Phys. Rev.}
  {\bfseries D78} (2008) 111701},
  [\href{https://arxiv.org/abs/0808.3994}{{\ttfamily 0808.3994}}].

\bibitem{Baldes:2016rqn}
I.~Baldes, T.~Konstandin and G.~Servant, \emph{{A First-Order Electroweak Phase
  Transition in the Standard Model from Varying Yukawas}},
  \href{https://arxiv.org/abs/1604.04526}{{\ttfamily 1604.04526}}.

\bibitem{Davidson:2008bu}
S.~Davidson, E.~Nardi and Y.~Nir, \emph{{Leptogenesis}},
  \href{http://dx.doi.org/10.1016/j.physrep.2008.06.002}{\emph{Phys. Rept.}
  {\bfseries 466} (2008) 105--177},
  [\href{https://arxiv.org/abs/0802.2962}{{\ttfamily 0802.2962}}].

\bibitem{Covi:1996wh}
L.~Covi, E.~Roulet and F.~Vissani, \emph{{CP violating decays in leptogenesis
  scenarios}},
  \href{http://dx.doi.org/10.1016/0370-2693(96)00817-9}{\emph{Phys. Lett.}
  {\bfseries B384} (1996) 169--174},
  [\href{https://arxiv.org/abs/hep-ph/9605319}{{\ttfamily hep-ph/9605319}}].

\bibitem{Pilaftsis:1997jf}
A.~Pilaftsis, \emph{{CP violation and baryogenesis due to heavy Majorana
  neutrinos}}, \href{http://dx.doi.org/10.1103/PhysRevD.56.5431}{\emph{Phys.
  Rev.} {\bfseries D56} (1997) 5431--5451},
  [\href{https://arxiv.org/abs/hep-ph/9707235}{{\ttfamily hep-ph/9707235}}].

\bibitem{Pilaftsis:2003gt}
A.~Pilaftsis and T.~E.~J. Underwood, \emph{{Resonant leptogenesis}},
  \href{http://dx.doi.org/10.1016/j.nuclphysb.2004.05.029}{\emph{Nucl. Phys.}
  {\bfseries B692} (2004) 303--345},
  [\href{https://arxiv.org/abs/hep-ph/0309342}{{\ttfamily hep-ph/0309342}}].

\bibitem{Kolb:1990vq}
E.~W. Kolb and M.~S. Turner, \emph{{The Early Universe}}, {\emph{Front. Phys.}
  {\bfseries 69} (1990) 1--547}.

\bibitem{Abada:2006ea}
A.~Abada, S.~Davidson, A.~Ibarra, F.~X. Josse-Michaux, M.~Losada and A.~Riotto,
  \emph{{Flavour Matters in Leptogenesis}},
  \href{http://dx.doi.org/10.1088/1126-6708/2006/09/010}{\emph{JHEP} {\bfseries
  09} (2006) 010}, [\href{https://arxiv.org/abs/hep-ph/0605281}{{\ttfamily
  hep-ph/0605281}}].

\bibitem{Gu:2003er}
P.~Gu, X.~Wang and X.~Zhang, \emph{{Dark energy and neutrino mass limits from
  baryogenesis}},
  \href{http://dx.doi.org/10.1103/PhysRevD.68.087301}{\emph{Phys. Rev.}
  {\bfseries D68} (2003) 087301},
  [\href{https://arxiv.org/abs/hep-ph/0307148}{{\ttfamily hep-ph/0307148}}].

\bibitem{Buchmuller:2000nd}
W.~Buchmuller and S.~Fredenhagen, \emph{{Quantum mechanics of baryogenesis}},
  \href{http://dx.doi.org/10.1016/S0370-2693(00)00573-6}{\emph{Phys. Lett.}
  {\bfseries B483} (2000) 217--224},
  [\href{https://arxiv.org/abs/hep-ph/0004145}{{\ttfamily hep-ph/0004145}}].

\bibitem{DeSimone:2007gkc}
A.~De~Simone and A.~Riotto, \emph{{Quantum Boltzmann Equations and
  Leptogenesis}},
  \href{http://dx.doi.org/10.1088/1475-7516/2007/08/002}{\emph{JCAP} {\bfseries
  0708} (2007) 002}, [\href{https://arxiv.org/abs/hep-ph/0703175}{{\ttfamily
  hep-ph/0703175}}].

\bibitem{Nelson:1991ab}
A.~E. Nelson, D.~B. Kaplan and A.~G. Cohen, \emph{{Why there is something
  rather than nothing: Matter from weak interactions}},
  \href{http://dx.doi.org/10.1016/0550-3213(92)90440-M}{\emph{Nucl. Phys.}
  {\bfseries B373} (1992) 453--478}.

\bibitem{deSalas:2017kay}
P.~F. de~Salas, D.~V. Forero, C.~A. Ternes, M.~Tortola and J.~W.~F. Valle,
  \emph{{Status of neutrino oscillations 2017}},
  \href{https://arxiv.org/abs/1708.01186}{{\ttfamily 1708.01186}}.

\bibitem{Li:2001st}
M.-z. Li, X.-l. Wang, B.~Feng and X.-m. Zhang, \emph{{Quintessence and
  spontaneous leptogenesis}},
  \href{http://dx.doi.org/10.1103/PhysRevD.65.103511}{\emph{Phys. Rev.}
  {\bfseries D65} (2002) 103511},
  [\href{https://arxiv.org/abs/hep-ph/0112069}{{\ttfamily hep-ph/0112069}}].

\bibitem{Davidson:2002qv}
S.~Davidson and A.~Ibarra, \emph{{A Lower bound on the right-handed neutrino
  mass from leptogenesis}},
  \href{http://dx.doi.org/10.1016/S0370-2693(02)01735-5}{\emph{Phys. Lett.}
  {\bfseries B535} (2002) 25--32},
  [\href{https://arxiv.org/abs/hep-ph/0202239}{{\ttfamily hep-ph/0202239}}].

\bibitem{Ellis:1984eq}
J.~R. Ellis, J.~E. Kim and D.~V. Nanopoulos, \emph{{Cosmological Gravitino
  Regeneration and Decay}},
  \href{http://dx.doi.org/10.1016/0370-2693(84)90334-4}{\emph{Phys. Lett.}
  {\bfseries B145} (1984) 181--186}.

\bibitem{Ade:2015xua}
{\scshape Planck} collaboration, P.~A.~R. Ade et~al., \emph{{Planck 2015
  results. XIII. Cosmological parameters}},
  \href{http://dx.doi.org/10.1051/0004-6361/201525830}{\emph{Astron.
  Astrophys.} {\bfseries 594} (2016) A13},
  [\href{https://arxiv.org/abs/1502.01589}{{\ttfamily 1502.01589}}].

\bibitem{Esteban:2016qun}
I.~Esteban, M.~C. Gonzalez-Garcia, M.~Maltoni, I.~Martinez-Soler and
  T.~Schwetz, \emph{{Updated fit to three neutrino mixing: exploring the
  accelerator-reactor complementarity}},
  \href{http://dx.doi.org/10.1007/JHEP01(2017)087}{\emph{JHEP} {\bfseries 01}
  (2017) 087}, [\href{https://arxiv.org/abs/1611.01514}{{\ttfamily
  1611.01514}}].

\bibitem{Casas:2001sr}
J.~A. Casas and A.~Ibarra, \emph{{Oscillating neutrinos and muon ---> e,
  gamma}}, \href{http://dx.doi.org/10.1016/S0550-3213(01)00475-8}{\emph{Nucl.
  Phys.} {\bfseries B618} (2001) 171--204},
  [\href{https://arxiv.org/abs/hep-ph/0103065}{{\ttfamily hep-ph/0103065}}].

\bibitem{Degrassi:2012ry}
G.~Degrassi, S.~Di~Vita, J.~Elias-Miro, J.~R. Espinosa, G.~F. Giudice,
  G.~Isidori et~al., \emph{{Higgs mass and vacuum stability in the Standard
  Model at NNLO}}, \href{http://dx.doi.org/10.1007/JHEP08(2012)098}{\emph{JHEP}
  {\bfseries 08} (2012) 098},
  [\href{https://arxiv.org/abs/1205.6497}{{\ttfamily 1205.6497}}].

\bibitem{Buttazzo:2013uya}
D.~Buttazzo, G.~Degrassi, P.~P. Giardino, G.~F. Giudice, F.~Sala, A.~Salvio
  et~al., \emph{{Investigating the near-criticality of the Higgs boson}},
  \href{http://dx.doi.org/10.1007/JHEP12(2013)089}{\emph{JHEP} {\bfseries 12}
  (2013) 089}, [\href{https://arxiv.org/abs/1307.3536}{{\ttfamily 1307.3536}}].

\bibitem{Metaxas:1995ab}
D.~Metaxas and E.~J. Weinberg, \emph{{Gauge independence of the bubble
  nucleation rate in theories with radiative symmetry breaking}},
  \href{http://dx.doi.org/10.1103/PhysRevD.53.836}{\emph{Phys. Rev.} {\bfseries
  D53} (1996) 836--843},
  [\href{https://arxiv.org/abs/hep-ph/9507381}{{\ttfamily hep-ph/9507381}}].

\bibitem{Plascencia:2015pga}
A.~D. Plascencia and C.~Tamarit, \emph{{Convexity, gauge-dependence and
  tunneling rates}},
  \href{http://dx.doi.org/10.1007/JHEP10(2016)099}{\emph{JHEP} {\bfseries 10}
  (2016) 099}, [\href{https://arxiv.org/abs/1510.07613}{{\ttfamily
  1510.07613}}].

\bibitem{Ford:1992mv}
C.~Ford, D.~R.~T. Jones, P.~W. Stephenson and M.~B. Einhorn, \emph{{The
  Effective potential and the renormalization group}},
  \href{http://dx.doi.org/10.1016/0550-3213(93)90206-5}{\emph{Nucl. Phys.}
  {\bfseries B395} (1993) 17--34},
  [\href{https://arxiv.org/abs/hep-lat/9210033}{{\ttfamily hep-lat/9210033}}].

\bibitem{Espinosa:2007qp}
J.~R. Espinosa, G.~F. Giudice and A.~Riotto, \emph{{Cosmological implications
  of the Higgs mass measurement}},
  \href{http://dx.doi.org/10.1088/1475-7516/2008/05/002}{\emph{JCAP} {\bfseries
  0805} (2008) 002}, [\href{https://arxiv.org/abs/0710.2484}{{\ttfamily
  0710.2484}}].

\bibitem{Pisarski:1988vd}
R.~D. Pisarski, \emph{{Scattering Amplitudes in Hot Gauge Theories}},
  \href{http://dx.doi.org/10.1103/PhysRevLett.63.1129}{\emph{Phys. Rev. Lett.}
  {\bfseries 63} (1989) 1129}.

\bibitem{Land:1992sm}
D.~Land and E.~D. Carlson, \emph{{Two stage phase transition in two Higgs
  models}}, \href{http://dx.doi.org/10.1016/0370-2693(92)90616-C}{\emph{Phys.
  Lett.} {\bfseries B292} (1992) 107--112},
  [\href{https://arxiv.org/abs/hep-ph/9208227}{{\ttfamily hep-ph/9208227}}].

\bibitem{Hammerschmitt:1994fn}
A.~Hammerschmitt, J.~Kripfganz and M.~G. Schmidt, \emph{{Baryon asymmetry from
  a two stage electroweak phase transition?}},
  \href{http://dx.doi.org/10.1007/BF01557241}{\emph{Z. Phys.} {\bfseries C64}
  (1994) 105--110}, [\href{https://arxiv.org/abs/hep-ph/9404272}{{\ttfamily
  hep-ph/9404272}}].

\bibitem{Cui:2011qe}
Y.~Cui, L.~Randall and B.~Shuve, \emph{{Emergent Dark Matter, Baryon, and
  Lepton Numbers}},
  \href{http://dx.doi.org/10.1007/JHEP08(2011)073}{\emph{JHEP} {\bfseries 08}
  (2011) 073}, [\href{https://arxiv.org/abs/1106.4834}{{\ttfamily 1106.4834}}].

\bibitem{Patel:2012pi}
H.~H. Patel and M.~J. Ramsey-Musolf, \emph{{Stepping Into Electroweak Symmetry
  Breaking: Phase Transitions and Higgs Phenomenology}},
  \href{http://dx.doi.org/10.1103/PhysRevD.88.035013}{\emph{Phys. Rev.}
  {\bfseries D88} (2013) 035013},
  [\href{https://arxiv.org/abs/1212.5652}{{\ttfamily 1212.5652}}].

\bibitem{Blinov:2015sna}
N.~Blinov, J.~Kozaczuk, D.~E. Morrissey and C.~Tamarit, \emph{{Electroweak
  Baryogenesis from Exotic Electroweak Symmetry Breaking}},
  \href{http://dx.doi.org/10.1103/PhysRevD.92.035012}{\emph{Phys. Rev.}
  {\bfseries D92} (2015) 035012},
  [\href{https://arxiv.org/abs/1504.05195}{{\ttfamily 1504.05195}}].

\bibitem{Pilaftsis:1991ug}
A.~Pilaftsis, \emph{{Radiatively induced neutrino masses and large Higgs
  neutrino couplings in the standard model with Majorana fields}},
  \href{http://dx.doi.org/10.1007/BF01482590}{\emph{Z. Phys.} {\bfseries C55}
  (1992) 275--282}, [\href{https://arxiv.org/abs/hep-ph/9901206}{{\ttfamily
  hep-ph/9901206}}].

\bibitem{Graesser:2007pc}
M.~L. Graesser, \emph{{Experimental Constraints on Higgs Boson Decays to
  TeV-scale Right-Handed Neutrinos}},
  \href{https://arxiv.org/abs/0705.2190}{{\ttfamily 0705.2190}}.

\bibitem{Aranda:2007dq}
A.~Aranda, O.~Blanno and J.~Lorenzo Diaz-Cruz, \emph{{A Model of Neutrino and
  Higgs Physics at the Electroweak Scale}},
  \href{http://dx.doi.org/10.1016/j.physletb.2007.12.026}{\emph{Phys. Lett.}
  {\bfseries B660} (2008) 62--66},
  [\href{https://arxiv.org/abs/0707.3662}{{\ttfamily 0707.3662}}].

\bibitem{Shoemaker:2010fg}
I.~M. Shoemaker, K.~Petraki and A.~Kusenko, \emph{{Collider signatures of
  sterile neutrinos in models with a gauge-singlet Higgs}},
  \href{http://dx.doi.org/10.1007/JHEP09(2010)060}{\emph{JHEP} {\bfseries 09}
  (2010) 060}, [\href{https://arxiv.org/abs/1006.5458}{{\ttfamily 1006.5458}}].

\bibitem{DiazCruz:2010rs}
J.~L. Diaz-Cruz, O.~Felix-Beltran, A.~Rosado and S.~Rosado-Navarro,
  \emph{{Electroweak right-handed neutrinos and new signals at the LHC}},
  \href{http://dx.doi.org/10.1142/S0217751X11053675}{\emph{Int. J. Mod. Phys.}
  {\bfseries A26} (2011) 2865--2880},
  [\href{https://arxiv.org/abs/1007.2134}{{\ttfamily 1007.2134}}].

\bibitem{Maiezza:2015lza}
A.~Maiezza, M.~Nemev¨ek and F.~Nesti, \emph{{Lepton Number Violation in Higgs
  Decay at LHC}},
  \href{http://dx.doi.org/10.1103/PhysRevLett.115.081802}{\emph{Phys. Rev.
  Lett.} {\bfseries 115} (2015) 081802},
  [\href{https://arxiv.org/abs/1503.06834}{{\ttfamily 1503.06834}}].

\bibitem{Nemevsek:2016enw}
M.~Nemev¨ek, F.~Nesti and J.~C. Vasquez, \emph{{Majorana Higgses at
  colliders}}, \href{http://dx.doi.org/10.1007/JHEP04(2017)114}{\emph{JHEP}
  {\bfseries 04} (2017) 114},
  [\href{https://arxiv.org/abs/1612.06840}{{\ttfamily 1612.06840}}].

\bibitem{Accomando:2016rpc}
E.~Accomando, L.~Delle~Rose, S.~Moretti, E.~Olaiya and C.~H.
  Shepherd-Themistocleous, \emph{{Novel SM-like Higgs decay into displaced
  heavy neutrino pairs in U(1)? models}},
  \href{http://dx.doi.org/10.1007/JHEP04(2017)081}{\emph{JHEP} {\bfseries 04}
  (2017) 081}, [\href{https://arxiv.org/abs/1612.05977}{{\ttfamily
  1612.05977}}].

\bibitem{Huitu:2008gf}
K.~Huitu, S.~Khalil, H.~Okada and S.~K. Rai, \emph{{Signatures for right-handed
  neutrinos at the Large Hadron Collider}},
  \href{http://dx.doi.org/10.1103/PhysRevLett.101.181802}{\emph{Phys. Rev.
  Lett.} {\bfseries 101} (2008) 181802},
  [\href{https://arxiv.org/abs/0803.2799}{{\ttfamily 0803.2799}}].

\bibitem{Basso:2008iv}
L.~Basso, A.~Belyaev, S.~Moretti and C.~H. Shepherd-Themistocleous,
  \emph{{Phenomenology of the minimal B-L extension of the Standard model: Z'
  and neutrinos}},
  \href{http://dx.doi.org/10.1103/PhysRevD.80.055030}{\emph{Phys. Rev.}
  {\bfseries D80} (2009) 055030},
  [\href{https://arxiv.org/abs/0812.4313}{{\ttfamily 0812.4313}}].

\bibitem{AguilarSaavedra:2009ik}
J.~A. Aguilar-Saavedra, \emph{{Heavy lepton pair production at LHC: Model
  discrimination with multi-lepton signals}},
  \href{http://dx.doi.org/10.1016/j.nuclphysb.2009.11.021}{\emph{Nucl. Phys.}
  {\bfseries B828} (2010) 289--316},
  [\href{https://arxiv.org/abs/0905.2221}{{\ttfamily 0905.2221}}].

\bibitem{Perez:2009mu}
P.~Fileviez~Perez, T.~Han and T.~Li, \emph{{Testability of Type I Seesaw at the
  CERN LHC: Revealing the Existence of the B-L Symmetry}},
  \href{http://dx.doi.org/10.1103/PhysRevD.80.073015}{\emph{Phys. Rev.}
  {\bfseries D80} (2009) 073015},
  [\href{https://arxiv.org/abs/0907.4186}{{\ttfamily 0907.4186}}].

\bibitem{Batell:2016zod}
B.~Batell, M.~Pospelov and B.~Shuve, \emph{{Shedding Light on Neutrino Masses
  with Dark Forces}},
  \href{http://dx.doi.org/10.1007/JHEP08(2016)052}{\emph{JHEP} {\bfseries 08}
  (2016) 052}, [\href{https://arxiv.org/abs/1604.06099}{{\ttfamily
  1604.06099}}].

\bibitem{Chikashige:1980ui}
Y.~Chikashige, R.~N. Mohapatra and R.~D. Peccei, \emph{{Are There Real
  Goldstone Bosons Associated with Broken Lepton Number?}},
  \href{http://dx.doi.org/10.1016/0370-2693(81)90011-3}{\emph{Phys. Lett.}
  {\bfseries 98B} (1981) 265--268}.

\bibitem{Albert:2014fya}
{\scshape EXO-200} collaboration, J.~B. Albert et~al., \emph{{Search for
  Majoron-emitting modes of double-beta decay of $^{136}$Xe with EXO-200}},
  \href{http://dx.doi.org/10.1103/PhysRevD.90.092004}{\emph{Phys. Rev.}
  {\bfseries D90} (2014) 092004},
  [\href{https://arxiv.org/abs/1409.6829}{{\ttfamily 1409.6829}}].

\bibitem{Gando:2012pj}
{\scshape KamLAND-Zen} collaboration, A.~Gando et~al., \emph{{Limits on
  Majoron-emitting double-beta decays of Xe-136 in the KamLAND-Zen
  experiment}}, \href{http://dx.doi.org/10.1103/PhysRevC.86.021601}{\emph{Phys.
  Rev.} {\bfseries C86} (2012) 021601},
  [\href{https://arxiv.org/abs/1205.6372}{{\ttfamily 1205.6372}}].

\bibitem{Cline:1993ht}
J.~M. Cline, K.~Kainulainen and K.~A. Olive, \emph{{Constraints on majoron
  models, neutrino masses and baryogenesis}},
  \href{http://dx.doi.org/10.1016/0927-6505(93)90005-X}{\emph{Astropart. Phys.}
  {\bfseries 1} (1993) 387--398},
  [\href{https://arxiv.org/abs/hep-ph/9304229}{{\ttfamily hep-ph/9304229}}].

\bibitem{Berezinsky:1993fm}
V.~Berezinsky and J.~W.~F. Valle, \emph{{The KeV majoron as a dark matter
  particle}}, \href{http://dx.doi.org/10.1016/0370-2693(93)90140-D}{\emph{Phys.
  Lett.} {\bfseries B318} (1993) 360--366},
  [\href{https://arxiv.org/abs/hep-ph/9309214}{{\ttfamily hep-ph/9309214}}].

\bibitem{Lattanzi:2007ux}
M.~Lattanzi and J.~W.~F. Valle, \emph{{Decaying warm dark matter and neutrino
  masses}}, \href{http://dx.doi.org/10.1103/PhysRevLett.99.121301}{\emph{Phys.
  Rev. Lett.} {\bfseries 99} (2007) 121301},
  [\href{https://arxiv.org/abs/0705.2406}{{\ttfamily 0705.2406}}].

\bibitem{Bell:2005dr}
N.~F. Bell, E.~Pierpaoli and K.~Sigurdson, \emph{{Cosmological signatures of
  interacting neutrinos}},
  \href{http://dx.doi.org/10.1103/PhysRevD.73.063523}{\emph{Phys. Rev.}
  {\bfseries D73} (2006) 063523},
  [\href{https://arxiv.org/abs/astro-ph/0511410}{{\ttfamily
  astro-ph/0511410}}].

\bibitem{Forastieri:2015paa}
F.~Forastieri, M.~Lattanzi and P.~Natoli, \emph{{Constraints on secret neutrino
  interactions after Planck}},
  \href{http://dx.doi.org/10.1088/1475-7516/2015/07/014}{\emph{JCAP} {\bfseries
  1507} (2015) 014}, [\href{https://arxiv.org/abs/1504.04999}{{\ttfamily
  1504.04999}}].

\bibitem{Dudas:2014bca}
E.~Dudas, Y.~Mambrini and K.~A. Olive, \emph{{Monochromatic neutrinos generated
  by dark matter and the seesaw mechanism}},
  \href{http://dx.doi.org/10.1103/PhysRevD.91.075001}{\emph{Phys. Rev.}
  {\bfseries D91} (2015) 075001},
  [\href{https://arxiv.org/abs/1412.3459}{{\ttfamily 1412.3459}}].

\bibitem{Bazzocchi:2008fh}
F.~Bazzocchi, M.~Lattanzi, S.~Riemer-Sorensen and J.~W.~F. Valle, \emph{{X-ray
  photons from late-decaying majoron dark matter}},
  \href{http://dx.doi.org/10.1088/1475-7516/2008/08/013}{\emph{JCAP} {\bfseries
  0808} (2008) 013}, [\href{https://arxiv.org/abs/0805.2372}{{\ttfamily
  0805.2372}}].

\bibitem{Berezhiani:1989za}
Z.~G. Berezhiani and A.~{\relax Yu}. Smirnov, \emph{{Matter Induced Neutrino
  Decay and Supernova {SN1987A}}},
  \href{http://dx.doi.org/10.1016/0370-2693(89)90052-X}{\emph{Phys. Lett.}
  {\bfseries B220} (1989) 279--284}.

\bibitem{Choi:1989hi}
K.~Choi and A.~Santamaria, \emph{{Majorons and Supernova Cooling}},
  \href{http://dx.doi.org/10.1103/PhysRevD.42.293}{\emph{Phys. Rev.} {\bfseries
  D42} (1990) 293--306}.

\bibitem{Kachelriess:2000qc}
M.~Kachelriess, R.~Tomas and J.~W.~F. Valle, \emph{{Supernova bounds on Majoron
  emitting decays of light neutrinos}},
  \href{http://dx.doi.org/10.1103/PhysRevD.62.023004}{\emph{Phys. Rev.}
  {\bfseries D62} (2000) 023004},
  [\href{https://arxiv.org/abs/hep-ph/0001039}{{\ttfamily hep-ph/0001039}}].

\bibitem{Garny:2011hg}
M.~Garny, A.~Kartavtsev and A.~Hohenegger, \emph{{Leptogenesis from first
  principles in the resonant regime}},
  \href{http://dx.doi.org/10.1016/j.aop.2012.10.007}{\emph{Annals Phys.}
  {\bfseries 328} (2013) 26--63},
  [\href{https://arxiv.org/abs/1112.6428}{{\ttfamily 1112.6428}}].

\bibitem{Quiros:1999jp}
M.~Quiros, \emph{{Finite temperature field theory and phase transitions}},  in
  \emph{{Proceedings, Summer School in High-energy physics and cosmology:
  Trieste, Italy, June 29-July 17, 1998}}, pp.~187--259, 1999.
\newblock \href{https://arxiv.org/abs/hep-ph/9901312}{{\ttfamily
  hep-ph/9901312}}.

\end{thebibliography}\endgroup

\end{document}